\documentclass[aps,pra,reprint,superscriptaddress,longbibliography,nobibnotes]{revtex4-2}
\usepackage{CJK}
\usepackage{lineno}
\usepackage{graphicx}
\usepackage{amsmath}
\usepackage{bbold}
\usepackage{amssymb}
\usepackage[english]{babel}
\usepackage{color}
\usepackage[version=4]{mhchem}
\usepackage{dsfont}
\usepackage{placeins}
\usepackage{mathtools}
\usepackage[normalem]{ulem}
\usepackage{lipsum}
\usepackage{array}
\usepackage{comment}
\usepackage{physics}

\usepackage{soul}
\newcommand{\Caltech}{California Institute of Technology, Pasadena, CA 91125, USA}

\usepackage{caption}

\DeclareCaptionLabelSeparator{bar}{ \textbf{\textbar}~}
\usepackage{enumitem}
\setlist{nolistsep}
\usepackage[hidelinks]{hyperref}
\hypersetup{colorlinks=true,citecolor=blue,linkcolor=blue,urlcolor=blue}

\usepackage{bibunits}

\begin{document}


\title{Gate-based Readout and Cooling of Neutral Atoms}

\author{Richard Bing-Shiun Tsai}
\altaffiliation{These authors contributed equally to this work.}
\author{Lewis R. B. Picard}
\altaffiliation{These authors contributed equally to this work.}
\author{Xiangkai Sun}
\altaffiliation{These authors contributed equally to this work.}
\author{Yuan Le}
\author{Kon H. Leung}
\author{Manuel Endres}
\email{mendres@caltech.edu}
\affiliation{\Caltech}

\begin{abstract}
Neutral atom arrays have seen tremendous progress in quantum simulation, quantum metrology, and fault-tolerant quantum computing. However, hardware constraints such as atom loss and heating remain significant challenges. In this work, we introduce a comprehensive ancilla-based toolbox for optical tweezer experiments that utilizes high-fidelity Rydberg entangling gates and ancilla atoms to mitigate these physical limitations. First, we demonstrate repeated ancilla-based atom readout, achieving improved detection fidelity over multiple rounds with minimal perturbation to data atoms. Second, leveraging the quantized motional states in tweezer-trapped strontium atoms, we transduce quantum information from the electronic to the motional manifold. This enables us to perform mid-circuit ancilla-based atom loss detection in a coherence-preserving fashion. Finally, we demonstrate algorithmic cooling, a circuit-based sequence that deterministically cools data atoms by transferring their motional entropy to the electronic states of ancilla atoms. We observe a marked reduction in the atomic temperature of data atoms. These tools offer a pathway to continuous operation in tweezer clocks and complement recent developments in continuous reloading experiments.
\end{abstract}

\maketitle

\textit{Introduction.}---Neutral atom arrays have rapidly developed as a compelling platform for quantum science, leading to numerous advances in quantum simulation \cite{Shaw2024C,qiao2025realization,evered2025probing}, quantum metrology~\cite{Finkelstein2024,Cao2024,bornet2023scalable}, and fault-tolerant quantum computing~\cite{Bluvstein2026,sales2025experimental}. However, inherent to optical trapping are several underlying error mechanisms, such as heating from trapping light \cite{savard1997laser} and atom loss from background gas collisions \cite{Manetsch2025,zhang2025high,schymik2021single}, which become increasingly deleterious as computational tasks are extended to longer durations. While recent work has shown that loss-aware decoding may enable fault-tolerant quantum computing with neutral atoms even in the presence of atom loss~\cite{Bluvstein2026}, the performance of such schemes remains ultimately limited by the underlying physical atom loss rates and mid-circuit measurement fidelities. Moreover, direct detection of atom presence often results in heating via photon scattering, necessitating additional cooling for subsequent protocols to mitigate decoherence and atom loss~\cite{Covey2019A,radnaev2025universal}.  

Ancilla-based detection schemes provide a powerful means to circumvent the invasiveness of direct detection~\cite{Finkelstein2024,Bluvstein2024,Anand2024,Muniz2025,machu2025non}. In these protocols, information about the data atoms (either atom presence or quantum state) is mapped to the ancilla atoms and subsequently retrieved during mid-circuit readout, leaving the data atoms minimally perturbed. However, the effectiveness of such ancilla-based schemes is often limited by the mid-circuit readout fidelity~\cite{Norcia2023A,Deist2022_midcircuit,Graham2023} and the mapping process involving entangling gates. Critically, such schemes must also preserve data qubit coherence. While it is well established in trapped ion platforms~\cite{Stricker2020}, coherence-preserving atom loss detection was only recently demonstrated in neutral atom experiments for hyperfine qubits; albeit without mid-circuit readout~\cite{Chow2024}. 

\begin{figure}[ht!]
    \centering
    \includegraphics[width=\columnwidth]{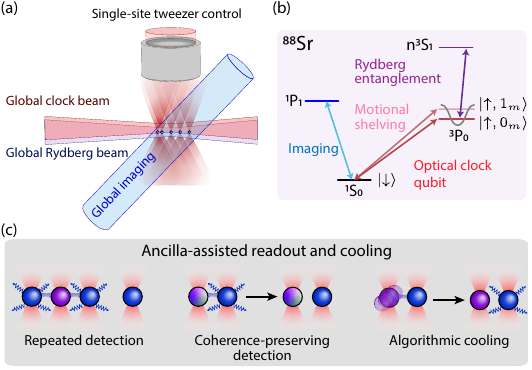}
    \caption{\textbf{Overview of ancilla-based toolbox for tweezer array experiments.}
    (a) Sketch of our experimental setup. The one-dimensional tweezer array, generated by an acousto-optical deflector (not shown), supports in-situ dynamical array reconfiguration and local arbitrary $Z$~gate implementation via single-site atom movements. Global clock and Rydberg laser beams propagate co-linearly along the array, implementing single-qubit gates and entangling gates, respectively. Global imaging is achieved with a pair of counter-propagating angled beams overlapping with the whole array.
    (b) Relevant electronic states of atomic strontium. We encode quantum information in $^3\text{P}_0$ (the clock state) and $^1\text{S}_0$ (the ground state), which we denote $\ket{\uparrow}$ and $\ket{\downarrow}$, respectively. Motional Fock states in the tweezer are denoted by $\ket{n_m}$, with the first two $\ket{1_m}$ and $\ket{0_m}$ employed during mid-circuit readout. Collectively, these states constitute a motional \textit{omg}-architecture, adapted for tweezer-trapped $^{88}\text{Sr}$ atoms.
    (c) The three main ancilla-based applications demonstrated with ancilla atoms (blue) and data atoms (purple) in this work: repeated atom readout, coherence-preserving atom loss detection, ancilla-assisted removal of motional excitation (i.e., algorithmic cooling).
    }
    \vspace{-0.7cm}
    \label{fig:overview}
\end{figure}

\begin{figure*}[ht!]
    \centering
    \includegraphics[width=\textwidth]{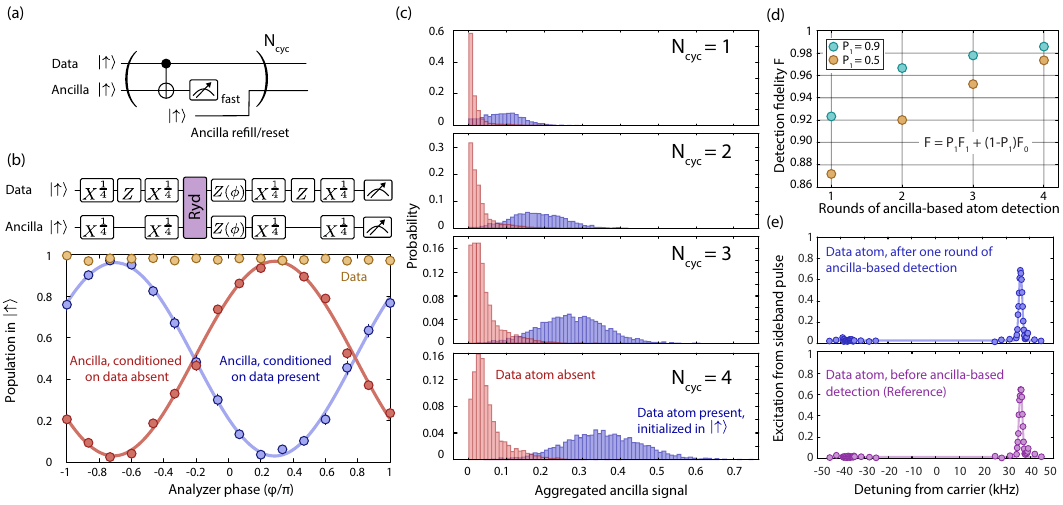}
    \caption{\textbf{Repeated ancilla-based atom readout.}
    (a) Circuit representation of the scheme. Data atoms, if present, are initialized in the electronic excited state $\ket{\uparrow}$. Under ideal conditions, the data atom remains unperturbed and a new ancilla atom is brought in via dynamical array reconfiguration after each round of ancilla-based atom detection. 
    (b) Decomposition of CNOT gate (shown in (a)) and calibration of the single-atom phase $\phi$ after the Rydberg pulse. We maximize the ancilla atom population in $\ket{\downarrow}$ ($\ket{\uparrow}$) if data atom is present (absent). The electronic state of the data atom is insensitive to this phase $\phi$ thanks to the site-selective $Z$~gate implemented via local atom movement. Solid lines are sinusoidal fits.
    (c) Histograms of ancilla signal obtained from repeated mid-circuit readout for two distinct configurations: data atom absent (red), and data atom present (blue). The ancilla signals up to $N_{\text{cyc}}$ cycles are aggregated by an unweighted, cumulative sum. The initial overlap of the two distributions for $N_{\text{cyc}} = 1$ is limited by the fast imaging fidelity on the ancilla atom and CNOT gate fidelity. As $N_{\text{cyc}}$ increases, the distributions separate further, demonstrating improved detection with repeated ancilla-based readout.
    (d) From the histograms, we compute the detection fidelity $F$ with data atom presence prior $P_1$ and classification fidelity $F_1$ ($F_0$) in the presence (absence) of the data atom. For each $P_1$ and $N_{\text{cyc}}$, a classification threshold needs to be determined to obtain $F_1$ and $F_0$. This threshold is optimized against the overall detection fidelity $F$. As $N_{\text{cyc}}$ increases, we see a gradual improvement in $F$, a quantitative trend supporting the visual separation of the two distributions in (c).
    (e) From sideband spectroscopy, we infer an average motional occupation number of $0.002^{+5}_{-2}$ for the data atoms after erasure-cooling, before ancilla-based detection (lower). This number is measured to be $0.010^{+7}_{-7}$ after one round of ancilla-based detection (upper). This shows that data atoms are minimally perturbed in this ancilla-based detection scheme. Error bars represent 1$\sigma$ confidence interval and are typically smaller than the markers.
    } 
    \vspace{-0.3cm}
    \label{fig:repeated_QND}
\end{figure*}

In this Letter, we realize a comprehensive toolbox for ancilla-based atom readout and loss detection, using our experimental setup consisting of individually trapped strontium atoms equipped with high-fidelity entangling gates [Fig.~\ref{fig:overview}]. We first show how ancilla-based atom readout, when performed in a repeated fashion, enables improved detection fidelity while preserving the benefits of fast imaging and flexibility for erasure conversion~\cite{Scholl2023A,Muzi_Falconi_2025,Ma2025}. Then, we demonstrate how this ancilla-based protocol could be implemented in a coherence-preserving manner to achieve pure atom loss detection, making use of the motional \textit{omg}-architecture~\cite{Lis2023A,Ma2022_Yb,Chen2022} adapted for $^{88}\text{Sr}$ atoms (see Fig.~\ref{fig:overview}(b) and Supplemental Material~\cite{SM}).

We further address the heating problem using our ancilla toolbox. The central idea hinges on coherently transferring motional excitations from the data atoms to the ancilla atoms, a process called \textit{algorithmic cooling}. Algorithmic cooling was initially formulated as a method to purify the spin of an ensemble of qubits in an NMR quantum computer by coupling them to a spin bath using controlled quantum gates \cite{boykinAlgorithmicCoolingScalable2002}. Recently, algorithmic cooling has been implemented in trapped ion systems to cool the motional state of ions that cannot be efficiently laser-cooled by coupling them to co-trapped cold ions ~\cite{King2021}. To the best of our knowledge, an experimental realization of this method in neutral atom arrays has not yet been demonstrated. In this work, we employ a circuit-based sequence that deterministically cools data atoms by transferring their motional entropy to the electronic states of ancilla atoms. We observe a marked increase in motional ground state occupation of data atoms using this technique.

\medskip
\textit{Repeated ancilla-based atom readout.}---Detecting atom presence, or reading out a given atomic state, conventionally involves collecting photons scattered by driving the atom with (near-)resonant light. However, ensuring high survival probability entails millisecond-scale imaging times and concomitant cooling \cite{Covey2019A,blodgett2023imaging,Manetsch2025}. Conversely, fast, microsecond-scale imaging with high scattering rates often results in low survival~\cite{Scholl2023A,Senoo2025,Su2025} or, at the very least, significant heating~\cite{Muzi_Falconi_2025}. Ancilla-based readout addresses these tradeoffs by mapping the desired information onto ancilla atoms for fast readout, leaving the data atoms unperturbed. This effectively realizes a quantum nondemolition measurement of the state of the data atoms~\cite{braginsky1980quantum,braginsky1996quantum}, which we leverage to determine atom presence.

We demonstrate improved detection fidelity with repeated consecutive rounds of ancilla-based readout, overcoming the limitations set by mid-circuit readout and the imperfect entangling gates used to map information in single-round schemes~\cite{Finkelstein2024,Shaw2025_erasure_cooling}. To this end, we employ a circuit-based sequence to map the electronic state of a data atom to the state of the ancilla atom [Fig.~\ref{fig:repeated_QND}(a)]. The mapping part is essentially a CNOT gate, consisting of local $Z$~gates (performed with local atom movement~\cite{Shaw2024}), single-qubit gates, and Rydberg entangling gates~\cite{Levine2019,Graham2019,Evered2023,Ma2023,Tsai2025}. To compensate for the single-atom phase induced by the Rydberg evolution~\cite{Levine2019,Tsai2025}, we scan the phase of the $X^{\frac{1}{2}}$ gate applied after the Rydberg pulse, in effect applying a virtual $Z$-rotation in the single-qubit frame [Fig.~\ref{fig:repeated_QND}(b)]. After this sequence of gates, fast imaging is applied to the whole array~\cite{Scholl2023A}. In this specific demonstration, both ancilla atoms and data atoms (if present) are initialized to $\ket{\uparrow}$, which is dark to the imaging light. Therefore, only the ancilla atom can scatter photons in the case where the paired data atom is present and flips the ancilla to $\ket{\downarrow}$. A fresh ancilla atom, kept in $\ket{\uparrow}$, is brought in with the aid of dynamical array reconfiguration~\cite{Bluvstein2022,Finkelstein2024} after each round of ancilla-based detection, up to $N_\text{cyc}$ cycles (largest $N_\text{cyc}=4$ in this work). Each round takes less than 800 $\mu$s, with dynamical array reconfiguration and single-qubit gates taking up the majority of the duration. 

We initialize the tweezer array into two distinct configurations: ancillae with a paired data atom (i.e., data atom present), and unpaired ancillae (i.e., data atom absent). The histograms for the combined signal~\cite{SM} at ancilla sites are shown in Fig.~\ref{fig:repeated_QND}(c), where we observe a clear separation between the two configurations, which becomes more pronounced as $N_\text{cyc}$ increases. To characterize this separation quantitatively, we adopt the conventional definition of detection fidelity $F = P_1\times F_1+(1-P_1)\times F_0$~\cite{Madjarovthesis,Covey2019A}, where $F_1$ ($F_0$) is the probability of correctly classifying the presence (absence) of a data atom from the ancilla signal, and $P_1$ is the data atom presence prior. To illustrate the results, we plot $F$ as a function of $N_\text{cyc}$ with two characteristic $P_1$ [Fig.~\ref{fig:repeated_QND}(d)]: $P_1 = 0.5$ corresponds to the typical initial tweezer loading probability; $P_1 = 0.9$ corresponds to a representative mid-sequence atom survival probability where atom loss is not negligible. In both cases, the detection fidelity $F$ is initially ${\sim}0.90$ (limited by fast imaging fidelity at ${\sim}0.1$-mK trap depth) and increases with the number of rounds, reaching close to 0.99.

We further show that data atoms are not significantly heated. To this end, we quantify changes in the data atoms' temperature with sideband spectroscopy [Fig.~\ref{fig:repeated_QND}(e)]. We first initialize the atoms to close to motional ground state, $\bar{n} = 0.002^{+5}_{-2}$, with erasure cooling~\cite{Shaw2025_erasure_cooling}. After one round of ancilla-based atom presence detection, the data atoms are measured to have an average motional occupation number of $\bar{n} = 0.010^{+7}_{-7}$, indicating that there is effectively zero heating of the data atoms during a single round. These low atomic temperature values and their uncertainties are obtained with a maximum-likelihood analysis of the measured sideband spectra~\cite{SM}.

\begin{figure}[ht!]
    \centering
    \includegraphics[width=\columnwidth]{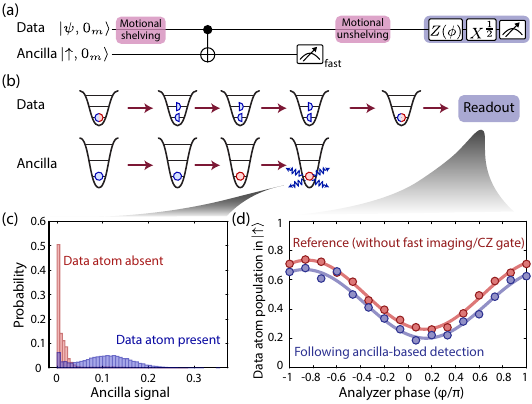}
    \caption{\textbf{Coherence-preserving ancilla-based atom loss detection.}
    (a) Circuit representation of the detection scheme. The detection scheme starts with motional shelving of the data atom, which transduces the quantum information $\ket{\psi}$ into the motional state manifold and prepares it into $\ket{\uparrow}$. A subsequent CNOT gate flips the ancilla to $\ket{\downarrow}$ contingent on the data atom's presence; otherwise, the ancilla remains in $\ket{\uparrow}$. This is followed by mid-circuit fast imaging of the ancilla. Finally, motional unshelving on the data atom restores the initial internal superposition $\ket{\psi}$, followed by a direct readout (purple).
    (b) Diagram representation. We site-selectively address a data atom with the shelving pulse by engineering different trap frequencies for the data and ancilla atoms. Fill colors denote electronic states: blue, $\ket{\uparrow}$; red, $\ket{\downarrow}$. Semicircles represent state superpositions, either in the electronic or motional manifold. In the depicted scenario, the data atom is present and initialized in $(\ket{\uparrow}+ \ket{\downarrow})/\sqrt{2}$. Consequently, the ancilla atom is flipped to $\ket{\downarrow}$ by the entangling gate and detected via fast imaging.
    (c) Ancilla signal histograms. The two distributions, conditioning on the presence of the data atom, have clear peak separations. Assuming $P_1 = 0.5$, the detection fidelity is $F=0.88$ (see text for discussion on technical limitations).
    (d) Post-detection coherence, limited by single-qubit coherence time. Blue circles: Data atom population directly read out with an analyzer pulse of phase $\phi$, following one round of ancilla-based protocol. Red circles: Reference experiment using an identical sequence, but with the Rydberg and fast imaging light blocked. The relative offset reduction between the Ramsey fringes is attributed to shelving errors (see text). Solid lines are sinusoidal fits. Error bars represent $1\sigma$ confidence interval and are typically smaller than the markers.
    } 
    \vspace{-0.3cm}
    \label{fig:spin_preserving_QND}
\end{figure}

\medskip
\textit{Coherence-preserving ancilla-based atom loss detection.}---Now, we turn to demonstrating ancilla-based atom loss detection while preserving the quantum information stored in the data atoms [Fig.~\ref{fig:spin_preserving_QND}], similar to recent work on leakage detection for neutral atom experiments~\cite{Chow2024}. However, instead of using two entangling gates for state-independent action on the ancilla, we exploit the metastable qubit manifold (see motional \textit{omg}-architecture in Supplemental Material~\cite{SM}) to directly map data atom presence to the state of the ancilla with \textit{a single entangling gate} [Fig.~\ref{fig:spin_preserving_QND}(a,b)], and we also show subsequent mid-circuit readout of the ancilla atom. More specifically, before the entangling gate, we apply a sideband pulse on the clock transition and transfer the data atom, if present, to $\ket{\uparrow}$ regardless of the initial electronic state. As a result, the state $\ket{\psi}\otimes\ket{0_m} = \alpha \ket{\uparrow, 0_m}+\beta \ket{\downarrow, 0_m}$ is transduced into the motional states as $\alpha \ket{\uparrow, 0_m}+\beta \ket{\uparrow, 1_m}=\ket{\uparrow}\otimes(\alpha \ket{0_m}+\beta \ket{1_m})$~\cite{Shaw2025_erasure_cooling,Finkelstein2024}. A CNOT gate then maps this information of data atom presence onto an ancilla atom, flipping its electronic state to $\ket{\downarrow}$, followed by mid-circuit readout of the ancilla atom with fast imaging. In the case of data atom absence, the ancilla atom electronic state remains in $\ket{\uparrow}$, which is dark to the subsequent fast imaging. This protocol is related to the collective-bit-flip approach, as demonstrated on a trapped ion platform~\cite{Stricker2020}, where only one single entangling pulse is explicitly needed. 

We characterize the detection fidelity in the described protocol using data and ancilla atoms prepared in the motional ground state with erasure cooling~\cite{Shaw2025_erasure_cooling}, with the ancilla initialized to $\ket{\uparrow}$. Our scheme works for arbitrary superpositions $\ket{\psi}$ of data atom electronic states. Here, we initialize $\ket{\psi}=\ket{+} \equiv (\ket{\uparrow}+\ket{\downarrow})/\sqrt{2}$. As before, the resulting histograms of ancilla signals display two cleanly separated distributions corresponding to the presence or absence of a data atom [Fig.~\ref{fig:spin_preserving_QND}(c)]. Adopting the same definition for detection fidelity $F$ as in Fig.~\ref{fig:repeated_QND}(d), we obtain $F=0.88$ assuming a prior $P_1 = 0.5$, indicating successful data atom presence detection via mid-circuit readout of ancilla atoms. Notably, the histogram also reveals a small peak near zero signal. This could be due to imperfections during motional shelving (transduction) to the subspace $\{\ket{\uparrow,0_m}, \ket{\uparrow,1_m}\}$; such errors leave some residual data atom population in $\ket{\downarrow}$ prior to the CNOT gate, consequently leaving the paired ancilla unflipped in $\ket{\uparrow}$~\cite{SM}. Future upgrades addressing technical limitations, such as the transduction fidelity and the fast imaging fidelity at ${\sim}0.1 $-mK trap depth, should further improve the detection fidelity $F$. 

To see how well the data atom coherence is preserved during the whole ancilla-based detection procedure, we transduce the information back to the electronic states after fast imaging and perform a readout with a $\pi/2$-pulse with varying phase [Fig.~\ref{fig:spin_preserving_QND}(d)]. To isolate the effects of idle time in the sequence and single-qubit coherence time between $\ket{\uparrow}$ and $\ket{\downarrow}$, mainly limited by our clock laser~\cite{Finkelstein2024,Madjarov2019}, we perform the same sequence without the entangling gate and fast imaging, as the reference curve (red). The oscillation curve obtained after the full ancilla-based detection (blue) shows clear preservation of the qubit coherence, with a small global population offset reduction due to unsuccessful shelving events, leading to atom loss induced by the fast imaging beam. This suggests that the main limitation of this ancilla-based detection scheme is the motional shelving fidelity, of which we discuss the technical limitations in Supplemental Material~\cite{SM}.

\medskip
\textit{Algorithmic cooling.}---Neutral atoms have been cooled to close to the motional ground state of an optical tweezer using sub-Doppler cooling~\cite{chin2017polarization,Brown2019,blodgett2023imaging}---including schemes that leverage state-dependent trapping~\cite{Cooper2018,urech2022narrow,holzl2023motional,blodgett2025narrow}---as well as through resolved or Raman sideband cooling~\cite{Kaufman2012,Thompson2013,yu2018motional,Norcia2018,Cooper2018,Jenkins2022}. These methods rely fundamentally on spontaneous emission by an excited atom to reset its electronic state and remove motional entropy from the system. The heating induced by this spontaneous emission step generally sets the lower limit on the temperature which any given cooling method can achieve~\cite{Wineland1979,Kaufman2012,yu2018motional,phatak2024generalized}. Alternatively, an ensemble can be cooled by selectively removing hot atoms in a measurement-based scheme~\cite{Shaw2025_erasure_cooling}. In contrast to the aforementioned techniques, algorithmic cooling introduces a fundamentally different paradigm in which a quantum circuit is used to coherently map motional entropy of a data qubit onto the internal state of an ancilla qubit, which is then reset or discarded.

\begin{figure}[ht!]
    \centering
    \includegraphics[width=\columnwidth]{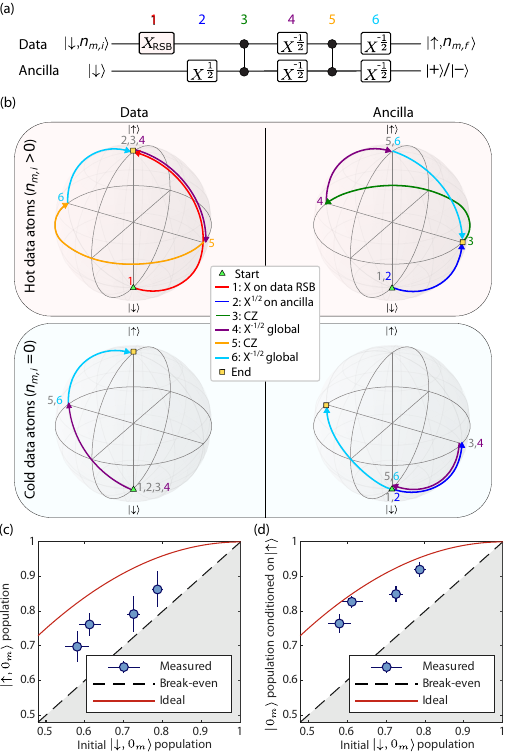}
    \caption{\textbf{Algorithmic cooling of atoms using entangling gates with ancilla atoms.} 
    (a) Algorithmic cooling quantum circuit, consisting of six gates. At the end of the circuit, data atoms are prepared in electronic state $\ket{\uparrow}$ and motional state $\ket{n_{m,f}}$. If the cooling round has been successful,  $n_{m,f} = \mathrm{max}(n_{m,i} - 1,0)$. Ancilla atoms are left in the $\ket{+}$ or $\ket{-}$ state, depending on whether the data atom was in the motional ground state to begin with.
    (b) Bloch sphere representations of the rotations performed on the data and ancilla atoms at each step. The top row shows the trajectories when the data atoms are initially not in the motional ground state, and the bottom row shows the trajectories when they are.
    (c) Fraction of total atom population, after one round of algorithmic cooling, which is measured in both the motional ground state and the correct electronic state, as a function of initial motional ground state population. This represents the combined success of cooling and state preparation in the circuit. Dashed line: break-even threshold for total usable population. Solid red line: theoretical maximum, corresponding to the perfect removal of one motional quantum from an initial Boltzmann distribution.
    (d) Motional ground state population, post-selected on the correct electronic state ($\ket{\uparrow}$) at the end of the cooling round. Error bars represent $1\sigma$ confidence intervals.
    } 
    \vspace{-0.3cm}
    \label{fig:algorithmic_cooling}
\end{figure}

We implement algorithmic cooling using neutral atoms in optical tweezers. In the context of ions, the data and ancilla atoms are often different atomic or molecular species. In our case, the data and ancilla are both neutral $^{88}\text{Sr}$ atoms, which are addressed selectively using local tweezer movements and depth ramps~\cite{Shaw2024,Shaw2023A}. The circuit we use for algorithmic cooling is illustrated in Fig.~\ref{fig:algorithmic_cooling}(a). Initially, both data and ancilla are in the $\ket{\downarrow}$ state, and the data atom may be either in the motional ground state or a motional excited state. The coupling between motional and electronic degrees of freedom is provided by the first step, which is a red sideband $\pi$-pulse (RSB) on the clock transition, driving data atoms to the $\ket{\uparrow}$ state and removing one motional quantum only if the atom was initially not in the motional ground state \cite{Shaw2025_erasure_cooling}. This is followed by a $\pi/2$-pulse applied selectively to the ancilla atoms using local tweezer movements. The remaining steps consist of a pair of CZ gates and $\pi/2$-pulses applied globally, which serve to swap the $Z$-basis state of the data atom to the $X$-basis of the ancilla. The action of this pulse sequence for both initially hot and initially cold data atoms is illustrated on the Bloch sphere in Fig.~\ref{fig:algorithmic_cooling}(b). In both cases, the data atoms are in $\ket{\uparrow}$ at the end of the circuit, but in the case where they were initially hot, their motional state has also been reduced by one quantum. This procedure thus achieves fully coherent cooling of the motion of the data atom. The ancilla will end in either the $\ket{+}$ or $\ket{-}$ state, depending on the initial motional state of the data atom. In this work we run only a single cycle of algorithmic cooling. However, if the ancilla atom were replaced or optically pumped to reset its electronic state then this cooling cycle could be repeated many times. We note that this circuit in principle does not rely on the ancilla atom being initially motionally cold, but only requires that its electronic state be initialized with a high degree of accuracy. This distinguishes the approach from sympathetic cooling via Rydberg interactions, as proposed in Ref.~\cite{Belyansky2019}, which relies on cooling data atoms by direct coupling to the motion of a cold bath of ancilla atoms. In practice we do employ motionally cold ancilla atoms, prepared using erasure cooling~\cite{Shaw2025_erasure_cooling}, in order to maximize the fidelity of the motionally-sensitive single-qubit clock pulses~\cite{Finkelstein2024}. The circuit is also shown in Fig.S1 of Supplemental Material~\cite{SM} with the motional levels of the atoms explicitly illustrated.

The primary limit on the performance of the cooling sequence is the finite single-qubit gate fidelity on the clock transition due to clock laser decoherence. This results in imperfect state preparation of the data atoms in the $\ket{\uparrow}$ state at the end of the sequence. To evaluate the performance of the algorithmic cooling sequence, in Fig.~\ref{fig:algorithmic_cooling}(c), we plot the fraction of the total data atom population which is in both the correct electronic state and the motional ground state after one cooling cycle as a function of the initial motional ground state fraction. This can be thought of as the total amount of ``useful" cooled atom population at the end of a cooling cycle. The initial ground state fraction is varied by applying site-selective laser heating to the data atoms for between 30 and 150 $\mu$s. We compare the measured results against a simple model where one motional quantum is removed from an initial Boltzmann distribution of motional states (solid red lines in Fig.~\ref{fig:algorithmic_cooling}(c-d)), corresponding to the ideal performance limit. A detailed description of the model is given in the Supplemental Material~\cite{SM}.

We find that one round of algorithmic cooling provides an increase in the fraction of usable motional ground state atoms for all initial temperatures, despite the imperfect electronic state preparation. On average we detect 7.3(2)\% of the atoms in the wrong electronic state at the end of the sequence. The motional ground state fractions and electronic state fractions are determined from the sideband thermometry spectra shown in Fig.S2 of Supplemental Material~\cite{SM}. We analyze the sideband spectra in the conventional way, using the assumption of a thermal distribution of population across motional states \cite{Monroe1995}, which we note may not strictly hold after algorithmic cooling. In the Supplemental Material we determine bounds on the possible deviation between the measured and actual ground state populations, which we find are always smaller than the statistical uncertainty in this work~\cite{SM}.
In Fig.~\ref{fig:algorithmic_cooling}(d) we examine the motional degree of freedom in isolation, plotting the fraction of atoms in the motional ground state, conditioned on the atoms being in $\ket{\uparrow}$. This represents the final motional ground state fraction that would be achieved by discarding atoms in the incorrect electronic state, as is done in erasure cooling \cite{Shaw2025_erasure_cooling}. The population in the incorrect electronic state could also be recovered via optical pumping. However, the repumping process in strontium atoms involves scattering via the relatively long-lived intermediate state $^3\text{P}_1$, which introduces significant heating due to the differential trap depth at our current tweezer wavelength and polarization. Alkali atoms, by contrast, can be optically pumped into particular hyperfine states with high fidelity and minimal heating~\cite{Kaufman2012}, making this algorithmic cooling approach a powerful alternative for species without narrow optical transitions for sub-Doppler cooling.

\medskip
\textit{Summary and outlook.}---In this work, we have demonstrated improved detection fidelity in ancilla-based atom readout by performing multiple rounds of detection while keeping the data atoms minimally perturbed. In this specific demonstration, ancilla atoms are discarded and not reused. However, the resource overhead of heated ancilla atoms can be circumvented using a zone architecture~\cite{Bluvstein2024,Bluvstein2026,Manetsch2025}, where ancilla atoms could be imaged and cooled in a separate zone, or continuous reloading experiments~\cite{Gyger2024,Li2025,Chiu2025}, where fresh ancilla atoms could be brought in continuously. Repeated ancilla-based detection could further see enhanced performance with adaptive strategies~\cite{Hu2025,Zhao2024_repetitive}, for instance, where imaging time is adapted in real-time depending on the number of photons collected, or even based on the information extracted from previous rounds of detection.

We have also shown that our atom loss detection scheme can be done in a coherence-preserving fashion by transducing the initial quantum state from the optical qubit manifold to the motional qubit manifold. As demonstrated, coherence is indeed preserved, with the overall coherence mainly limited by clock laser noise~\cite{Finkelstein2024}. As for the information extracted from ancilla atoms, current readout fidelity is primarily limited by the efficiency of motional shelving operations. We anticipate significant improvements with further laser upgrades~\cite{Shaw2025_erasure_cooling}.

To our knowledge, this work represents the first realization of algorithmic cooling in neutral atom arrays. We observe a significant improvement in motional ground-state populations in the correct electronic state over a broad range of initial temperatures. The primary limitation to the electronic state preparation fidelity is currently single-qubit gate errors in longer sequences. This limitation could be mitigated in strontium by increasing the single-qubit gate fidelity via a higher clock transition Rabi frequency. The algorithmic cooling technique shown here could also be adapted to alkali atoms, and may be particularly well suited to dual-species atom array architectures, which natively support selective single-qubit gates, detection and reset of data and ancilla qubits~\cite{Anand2024}.

Demonstrated on a tweezer clock platform~\cite{Madjarov2019,Norcia2019}, this ancilla-based toolbox establishes a critical pathway toward continuous clock operation~\cite{Katori_2021,Liu2025_zerodeadtimeclock} and quantum-computing-enhanced metrology~\cite{allen2025quantumcomputingenhancedsensing,Direkci2026}. Moreover, recent proposals~\cite{dudinets2025} suggest a realistic scheme where mobile ancilla atoms mediate entanglement between distant atoms, which could be leveraged to create entangled probe states across a large array of clock qubits. In the future, combining these tools with ancilla reuse~\cite{Muniz2025,Bluvstein2026}, continuous reloading ~\cite{Gyger2024,Li2025,Chiu2025}, or dual-species architectures~\cite{Anand2024}, featuring designated processing atoms (e.g., rubidium) and probing atoms (e.g., strontium), will allow us to leverage a continuous stream of fresh ancillae. This approach effectively removes key interruptions associated with readout and cooling, facilitating true continuous operation for optical tweezer clocks.

\section*{Acknowledgments}
We thank Elie Bataille, Nelson Darkwah Oppong, Zunqi Li, Hannah Manetsch, Gyohei Nomura, and Jeff Thompson for fruitful discussions and their feedback on this work. We also thank Ran Finkelstein and Adam Shaw for early contributions related to this work. We acknowledge support from the Army Research Office TINA QC program (W911NF2410388), the NSF QLCI program (2016245), the Army Research Office MURI program (W911NF2010136), the DOE (DESC0021951), the AFOSR (FA9550-23-1-0625), the Institute for Quantum Information and Matter, an NSF Physics Frontiers Center (NSF Grant PHY-2317110), the Heising-Simons Foundation (2024-4852), and the Weston Havens Foundation. R.B.S.T. acknowledges support from the Taiwan-Caltech Fellowship. L.R.B.P. acknowledges support from the David and Ellen Lee Postdoctoral Prize Fellowship. Y.L. and K.H.L. acknowledge support from the AWS Quantum postdoctoral fellowship. Competing interest: M.E. is co-founder and shareholder of and L.R.B.P. is an employee of Oratomic, Inc.

\bibliography{library-endreslab}

\begin{thebibliography}{74}%
\makeatletter
\providecommand \@ifxundefined [1]{%
 \@ifx{#1\undefined}
}%
\providecommand \@ifnum [1]{%
 \ifnum #1\expandafter \@firstoftwo
 \else \expandafter \@secondoftwo
 \fi
}%
\providecommand \@ifx [1]{%
 \ifx #1\expandafter \@firstoftwo
 \else \expandafter \@secondoftwo
 \fi
}%
\providecommand \natexlab [1]{#1}%
\providecommand \enquote  [1]{``#1''}%
\providecommand \bibnamefont  [1]{#1}%
\providecommand \bibfnamefont [1]{#1}%
\providecommand \citenamefont [1]{#1}%
\providecommand \href@noop [0]{\@secondoftwo}%
\providecommand \href [0]{\begingroup \@sanitize@url \@href}%
\providecommand \@href[1]{\@@startlink{#1}\@@href}%
\providecommand \@@href[1]{\endgroup#1\@@endlink}%
\providecommand \@sanitize@url [0]{\catcode `\\12\catcode `\$12\catcode `\&12\catcode `\#12\catcode `\^12\catcode `\_12\catcode `\%12\relax}%
\providecommand \@@startlink[1]{}%
\providecommand \@@endlink[0]{}%
\providecommand \url  [0]{\begingroup\@sanitize@url \@url }%
\providecommand \@url [1]{\endgroup\@href {#1}{\urlprefix }}%
\providecommand \urlprefix  [0]{URL }%
\providecommand \Eprint [0]{\href }%
\providecommand \doibase [0]{https://doi.org/}%
\providecommand \selectlanguage [0]{\@gobble}%
\providecommand \bibinfo  [0]{\@secondoftwo}%
\providecommand \bibfield  [0]{\@secondoftwo}%
\providecommand \translation [1]{[#1]}%
\providecommand \BibitemOpen [0]{}%
\providecommand \bibitemStop [0]{}%
\providecommand \bibitemNoStop [0]{.\EOS\space}%
\providecommand \EOS [0]{\spacefactor3000\relax}%
\providecommand \BibitemShut  [1]{\csname bibitem#1\endcsname}%
\let\auto@bib@innerbib\@empty
\bibitem [{\citenamefont {Shaw}\ \emph {et~al.}(2024{\natexlab{a}})\citenamefont {Shaw}, \citenamefont {Chen}, \citenamefont {Choi}, \citenamefont {Mark}, \citenamefont {Scholl}, \citenamefont {Finkelstein}, \citenamefont {Elben}, \citenamefont {Choi},\ and\ \citenamefont {Endres}}]{Shaw2024C}%
  \BibitemOpen
  \bibfield  {author} {\bibinfo {author} {\bibfnamefont {A.~L.}\ \bibnamefont {Shaw}}, \bibinfo {author} {\bibfnamefont {Z.}~\bibnamefont {Chen}}, \bibinfo {author} {\bibfnamefont {J.}~\bibnamefont {Choi}}, \bibinfo {author} {\bibfnamefont {D.~K.}\ \bibnamefont {Mark}}, \bibinfo {author} {\bibfnamefont {P.}~\bibnamefont {Scholl}}, \bibinfo {author} {\bibfnamefont {R.}~\bibnamefont {Finkelstein}}, \bibinfo {author} {\bibfnamefont {A.}~\bibnamefont {Elben}}, \bibinfo {author} {\bibfnamefont {S.}~\bibnamefont {Choi}},\ and\ \bibinfo {author} {\bibfnamefont {M.}~\bibnamefont {Endres}},\ }\bibfield  {title} {\bibinfo {title} {Benchmarking highly entangled states on a 60-atom analogue quantum simulator},\ }\href {https://doi.org/10.1038/s41586-024-07173-x} {\bibfield  {journal} {\bibinfo  {journal} {Nature}\ }\textbf {\bibinfo {volume} {628}},\ \bibinfo {pages} {71} (\bibinfo {year} {2024}{\natexlab{a}})}\BibitemShut {NoStop}%
\bibitem [{\citenamefont {Qiao}\ \emph {et~al.}(2025)\citenamefont {Qiao}, \citenamefont {Emperauger}, \citenamefont {Chen}, \citenamefont {Homeier}, \citenamefont {Hollerith}, \citenamefont {Bornet}, \citenamefont {Martin}, \citenamefont {G{\'e}ly}, \citenamefont {Klein}, \citenamefont {Barredo}, \citenamefont {Geier}, \citenamefont {Chiu}, \citenamefont {Grusdt}, \citenamefont {Bohrdt}, \citenamefont {Lahaye},\ and\ \citenamefont {Browaeys}}]{qiao2025realization}%
  \BibitemOpen
  \bibfield  {author} {\bibinfo {author} {\bibfnamefont {M.}~\bibnamefont {Qiao}}, \bibinfo {author} {\bibfnamefont {G.}~\bibnamefont {Emperauger}}, \bibinfo {author} {\bibfnamefont {C.}~\bibnamefont {Chen}}, \bibinfo {author} {\bibfnamefont {L.}~\bibnamefont {Homeier}}, \bibinfo {author} {\bibfnamefont {S.}~\bibnamefont {Hollerith}}, \bibinfo {author} {\bibfnamefont {G.}~\bibnamefont {Bornet}}, \bibinfo {author} {\bibfnamefont {R.}~\bibnamefont {Martin}}, \bibinfo {author} {\bibfnamefont {B.}~\bibnamefont {G{\'e}ly}}, \bibinfo {author} {\bibfnamefont {L.}~\bibnamefont {Klein}}, \bibinfo {author} {\bibfnamefont {D.}~\bibnamefont {Barredo}}, \bibinfo {author} {\bibfnamefont {S.}~\bibnamefont {Geier}}, \bibinfo {author} {\bibfnamefont {N.-C.}\ \bibnamefont {Chiu}}, \bibinfo {author} {\bibfnamefont {F.}~\bibnamefont {Grusdt}}, \bibinfo {author} {\bibfnamefont {A.}~\bibnamefont {Bohrdt}}, \bibinfo {author} {\bibfnamefont {T.}~\bibnamefont {Lahaye}},\ and\ \bibinfo {author} {\bibfnamefont {A.}~\bibnamefont
  {Browaeys}},\ }\bibfield  {title} {\bibinfo {title} {Realization of a doped quantum antiferromagnet in a {Rydberg} tweezer array},\ }\href {https://doi.org/10.1038/s41586-025-09377-1} {\bibfield  {journal} {\bibinfo  {journal} {Nature}\ }\textbf {\bibinfo {volume} {644}},\ \bibinfo {pages} {889} (\bibinfo {year} {2025})}\BibitemShut {NoStop}%
\bibitem [{\citenamefont {Evered}\ \emph {et~al.}(2025)\citenamefont {Evered}, \citenamefont {Kalinowski}, \citenamefont {Geim}, \citenamefont {Manovitz}, \citenamefont {Bluvstein}, \citenamefont {Li}, \citenamefont {Maskara}, \citenamefont {Zhou}, \citenamefont {Ebadi}, \citenamefont {Xu}, \citenamefont {Campo}, \citenamefont {Cain}, \citenamefont {Ostermann}, \citenamefont {Yelin}, \citenamefont {Sachdev}, \citenamefont {Greiner}, \citenamefont {Vuleti{\'{c}}},\ and\ \citenamefont {Lukin}}]{evered2025probing}%
  \BibitemOpen
  \bibfield  {author} {\bibinfo {author} {\bibfnamefont {S.~J.}\ \bibnamefont {Evered}}, \bibinfo {author} {\bibfnamefont {M.}~\bibnamefont {Kalinowski}}, \bibinfo {author} {\bibfnamefont {A.~A.}\ \bibnamefont {Geim}}, \bibinfo {author} {\bibfnamefont {T.}~\bibnamefont {Manovitz}}, \bibinfo {author} {\bibfnamefont {D.}~\bibnamefont {Bluvstein}}, \bibinfo {author} {\bibfnamefont {S.~H.}\ \bibnamefont {Li}}, \bibinfo {author} {\bibfnamefont {N.}~\bibnamefont {Maskara}}, \bibinfo {author} {\bibfnamefont {H.}~\bibnamefont {Zhou}}, \bibinfo {author} {\bibfnamefont {S.}~\bibnamefont {Ebadi}}, \bibinfo {author} {\bibfnamefont {M.}~\bibnamefont {Xu}}, \bibinfo {author} {\bibfnamefont {J.}~\bibnamefont {Campo}}, \bibinfo {author} {\bibfnamefont {M.}~\bibnamefont {Cain}}, \bibinfo {author} {\bibfnamefont {S.}~\bibnamefont {Ostermann}}, \bibinfo {author} {\bibfnamefont {S.~F.}\ \bibnamefont {Yelin}}, \bibinfo {author} {\bibfnamefont {S.}~\bibnamefont {Sachdev}}, \bibinfo {author} {\bibfnamefont {M.}~\bibnamefont
  {Greiner}}, \bibinfo {author} {\bibfnamefont {V.}~\bibnamefont {Vuleti{\'{c}}}},\ and\ \bibinfo {author} {\bibfnamefont {M.~D.}\ \bibnamefont {Lukin}},\ }\bibfield  {title} {\bibinfo {title} {Probing the {Kitaev} honeycomb model on a neutral-atom quantum computer},\ }\href {https://doi.org/10.1038/s41586-025-09475-0} {\bibfield  {journal} {\bibinfo  {journal} {Nature}\ }\textbf {\bibinfo {volume} {645}},\ \bibinfo {pages} {341} (\bibinfo {year} {2025})}\BibitemShut {NoStop}%
\bibitem [{\citenamefont {Finkelstein}\ \emph {et~al.}(2024)\citenamefont {Finkelstein}, \citenamefont {Tsai}, \citenamefont {Sun}, \citenamefont {Scholl}, \citenamefont {Direkci}, \citenamefont {Gefen}, \citenamefont {Choi}, \citenamefont {Shaw},\ and\ \citenamefont {Endres}}]{Finkelstein2024}%
  \BibitemOpen
  \bibfield  {author} {\bibinfo {author} {\bibfnamefont {R.}~\bibnamefont {Finkelstein}}, \bibinfo {author} {\bibfnamefont {R.~B.-S.}\ \bibnamefont {Tsai}}, \bibinfo {author} {\bibfnamefont {X.}~\bibnamefont {Sun}}, \bibinfo {author} {\bibfnamefont {P.}~\bibnamefont {Scholl}}, \bibinfo {author} {\bibfnamefont {S.}~\bibnamefont {Direkci}}, \bibinfo {author} {\bibfnamefont {T.}~\bibnamefont {Gefen}}, \bibinfo {author} {\bibfnamefont {J.}~\bibnamefont {Choi}}, \bibinfo {author} {\bibfnamefont {A.~L.}\ \bibnamefont {Shaw}},\ and\ \bibinfo {author} {\bibfnamefont {M.}~\bibnamefont {Endres}},\ }\bibfield  {title} {\bibinfo {title} {Universal quantum operations and ancilla-based read-out for tweezer clocks},\ }\href {https://doi.org/10.1038/s41586-024-08005-8} {\bibfield  {journal} {\bibinfo  {journal} {Nature}\ }\textbf {\bibinfo {volume} {634}},\ \bibinfo {pages} {321} (\bibinfo {year} {2024})}\BibitemShut {NoStop}%
\bibitem [{\citenamefont {Cao}\ \emph {et~al.}(2024)\citenamefont {Cao}, \citenamefont {Eckner}, \citenamefont {Lukin~Yelin}, \citenamefont {Young}, \citenamefont {Jandura}, \citenamefont {Yan}, \citenamefont {Kim}, \citenamefont {Pupillo}, \citenamefont {Ye}, \citenamefont {Darkwah~Oppong},\ and\ \citenamefont {Kaufman}}]{Cao2024}%
  \BibitemOpen
  \bibfield  {author} {\bibinfo {author} {\bibfnamefont {A.}~\bibnamefont {Cao}}, \bibinfo {author} {\bibfnamefont {W.~J.}\ \bibnamefont {Eckner}}, \bibinfo {author} {\bibfnamefont {T.}~\bibnamefont {Lukin~Yelin}}, \bibinfo {author} {\bibfnamefont {A.~W.}\ \bibnamefont {Young}}, \bibinfo {author} {\bibfnamefont {S.}~\bibnamefont {Jandura}}, \bibinfo {author} {\bibfnamefont {L.}~\bibnamefont {Yan}}, \bibinfo {author} {\bibfnamefont {K.}~\bibnamefont {Kim}}, \bibinfo {author} {\bibfnamefont {G.}~\bibnamefont {Pupillo}}, \bibinfo {author} {\bibfnamefont {J.}~\bibnamefont {Ye}}, \bibinfo {author} {\bibfnamefont {N.}~\bibnamefont {Darkwah~Oppong}},\ and\ \bibinfo {author} {\bibfnamefont {A.~M.}\ \bibnamefont {Kaufman}},\ }\bibfield  {title} {\bibinfo {title} {Multi-qubit gates and {Schr{\"o}dinger} cat states in an optical clock},\ }\href {https://doi.org/10.1038/s41586-024-07913-z} {\bibfield  {journal} {\bibinfo  {journal} {Nature}\ }\textbf {\bibinfo {volume} {634}},\ \bibinfo {pages} {315} (\bibinfo {year}
  {2024})}\BibitemShut {NoStop}%
\bibitem [{\citenamefont {Bornet}\ \emph {et~al.}(2023)\citenamefont {Bornet}, \citenamefont {Emperauger}, \citenamefont {Chen}, \citenamefont {Ye}, \citenamefont {Block}, \citenamefont {Bintz}, \citenamefont {Boyd}, \citenamefont {Barredo}, \citenamefont {Comparin}, \citenamefont {Mezzacapo}, \citenamefont {Roscilde}, \citenamefont {Lahaye}, \citenamefont {Yao},\ and\ \citenamefont {Browaeys}}]{bornet2023scalable}%
  \BibitemOpen
  \bibfield  {author} {\bibinfo {author} {\bibfnamefont {G.}~\bibnamefont {Bornet}}, \bibinfo {author} {\bibfnamefont {G.}~\bibnamefont {Emperauger}}, \bibinfo {author} {\bibfnamefont {C.}~\bibnamefont {Chen}}, \bibinfo {author} {\bibfnamefont {B.}~\bibnamefont {Ye}}, \bibinfo {author} {\bibfnamefont {M.}~\bibnamefont {Block}}, \bibinfo {author} {\bibfnamefont {M.}~\bibnamefont {Bintz}}, \bibinfo {author} {\bibfnamefont {J.~A.}\ \bibnamefont {Boyd}}, \bibinfo {author} {\bibfnamefont {D.}~\bibnamefont {Barredo}}, \bibinfo {author} {\bibfnamefont {T.}~\bibnamefont {Comparin}}, \bibinfo {author} {\bibfnamefont {F.}~\bibnamefont {Mezzacapo}}, \bibinfo {author} {\bibfnamefont {T.}~\bibnamefont {Roscilde}}, \bibinfo {author} {\bibfnamefont {T.}~\bibnamefont {Lahaye}}, \bibinfo {author} {\bibfnamefont {N.~Y.}\ \bibnamefont {Yao}},\ and\ \bibinfo {author} {\bibfnamefont {A.}~\bibnamefont {Browaeys}},\ }\bibfield  {title} {\bibinfo {title} {Scalable spin squeezing in a dipolar {Rydberg} atom array},\ }\href
  {https://doi.org/10.1038/s41586-023-06414-9} {\bibfield  {journal} {\bibinfo  {journal} {Nature}\ }\textbf {\bibinfo {volume} {621}},\ \bibinfo {pages} {728} (\bibinfo {year} {2023})}\BibitemShut {NoStop}%
\bibitem [{\citenamefont {Bluvstein}\ \emph {et~al.}(2026)\citenamefont {Bluvstein}, \citenamefont {Geim}, \citenamefont {Li}, \citenamefont {Evered}, \citenamefont {Bonilla~Ataides}, \citenamefont {Baranes}, \citenamefont {Gu}, \citenamefont {Manovitz}, \citenamefont {Xu}, \citenamefont {Kalinowski}, \citenamefont {Majidy}, \citenamefont {Kokail}, \citenamefont {Maskara}, \citenamefont {Trapp}, \citenamefont {Stewart}, \citenamefont {Hollerith}, \citenamefont {Zhou}, \citenamefont {Gullans}, \citenamefont {Yelin}, \citenamefont {Greiner}, \citenamefont {Vuleti{\'{c}}}, \citenamefont {Cain},\ and\ \citenamefont {Lukin}}]{Bluvstein2026}%
  \BibitemOpen
  \bibfield  {author} {\bibinfo {author} {\bibfnamefont {D.}~\bibnamefont {Bluvstein}}, \bibinfo {author} {\bibfnamefont {A.~A.}\ \bibnamefont {Geim}}, \bibinfo {author} {\bibfnamefont {S.~H.}\ \bibnamefont {Li}}, \bibinfo {author} {\bibfnamefont {S.~J.}\ \bibnamefont {Evered}}, \bibinfo {author} {\bibfnamefont {J.~P.}\ \bibnamefont {Bonilla~Ataides}}, \bibinfo {author} {\bibfnamefont {G.}~\bibnamefont {Baranes}}, \bibinfo {author} {\bibfnamefont {A.}~\bibnamefont {Gu}}, \bibinfo {author} {\bibfnamefont {T.}~\bibnamefont {Manovitz}}, \bibinfo {author} {\bibfnamefont {M.}~\bibnamefont {Xu}}, \bibinfo {author} {\bibfnamefont {M.}~\bibnamefont {Kalinowski}}, \bibinfo {author} {\bibfnamefont {S.}~\bibnamefont {Majidy}}, \bibinfo {author} {\bibfnamefont {C.}~\bibnamefont {Kokail}}, \bibinfo {author} {\bibfnamefont {N.}~\bibnamefont {Maskara}}, \bibinfo {author} {\bibfnamefont {E.~C.}\ \bibnamefont {Trapp}}, \bibinfo {author} {\bibfnamefont {L.~M.}\ \bibnamefont {Stewart}}, \bibinfo {author} {\bibfnamefont
  {S.}~\bibnamefont {Hollerith}}, \bibinfo {author} {\bibfnamefont {H.}~\bibnamefont {Zhou}}, \bibinfo {author} {\bibfnamefont {M.~J.}\ \bibnamefont {Gullans}}, \bibinfo {author} {\bibfnamefont {S.~F.}\ \bibnamefont {Yelin}}, \bibinfo {author} {\bibfnamefont {M.}~\bibnamefont {Greiner}}, \bibinfo {author} {\bibfnamefont {V.}~\bibnamefont {Vuleti{\'{c}}}}, \bibinfo {author} {\bibfnamefont {M.}~\bibnamefont {Cain}},\ and\ \bibinfo {author} {\bibfnamefont {M.~D.}\ \bibnamefont {Lukin}},\ }\bibfield  {title} {\bibinfo {title} {A fault-tolerant neutral-atom architecture for universal quantum computation},\ }\href {https://doi.org/10.1038/s41586-025-09848-5} {\bibfield  {journal} {\bibinfo  {journal} {Nature}\ }\textbf {\bibinfo {volume} {649}},\ \bibinfo {pages} {39} (\bibinfo {year} {2026})}\BibitemShut {NoStop}%
\bibitem [{\citenamefont {Sales~Rodriguez}\ \emph {et~al.}(2025)\citenamefont {Sales~Rodriguez}, \citenamefont {Robinson}, \citenamefont {Jepsen}, \citenamefont {He}, \citenamefont {Duckering}, \citenamefont {Zhao}, \citenamefont {Wu}, \citenamefont {Campo}, \citenamefont {Bagnall}, \citenamefont {Kwon}, \citenamefont {Karolyshyn}, \citenamefont {Weinberg}, \citenamefont {Cain}, \citenamefont {Evered}, \citenamefont {Geim} \emph {et~al.}}]{sales2025experimental}%
  \BibitemOpen
  \bibfield  {author} {\bibinfo {author} {\bibfnamefont {P.}~\bibnamefont {Sales~Rodriguez}}, \bibinfo {author} {\bibfnamefont {J.~M.}\ \bibnamefont {Robinson}}, \bibinfo {author} {\bibfnamefont {P.~N.}\ \bibnamefont {Jepsen}}, \bibinfo {author} {\bibfnamefont {Z.}~\bibnamefont {He}}, \bibinfo {author} {\bibfnamefont {C.}~\bibnamefont {Duckering}}, \bibinfo {author} {\bibfnamefont {C.}~\bibnamefont {Zhao}}, \bibinfo {author} {\bibfnamefont {K.-H.}\ \bibnamefont {Wu}}, \bibinfo {author} {\bibfnamefont {J.}~\bibnamefont {Campo}}, \bibinfo {author} {\bibfnamefont {K.}~\bibnamefont {Bagnall}}, \bibinfo {author} {\bibfnamefont {M.}~\bibnamefont {Kwon}}, \bibinfo {author} {\bibfnamefont {T.}~\bibnamefont {Karolyshyn}}, \bibinfo {author} {\bibfnamefont {P.}~\bibnamefont {Weinberg}}, \bibinfo {author} {\bibfnamefont {M.}~\bibnamefont {Cain}}, \bibinfo {author} {\bibfnamefont {S.~J.}\ \bibnamefont {Evered}}, \bibinfo {author} {\bibfnamefont {A.~A.}\ \bibnamefont {Geim}}, \emph {et~al.},\ }\bibfield  {title} {\bibinfo
  {title} {Experimental demonstration of logical magic state distillation},\ }\href {https://doi.org/10.1038/s41586-025-09367-3} {\bibfield  {journal} {\bibinfo  {journal} {Nature}\ }\textbf {\bibinfo {volume} {645}},\ \bibinfo {pages} {620} (\bibinfo {year} {2025})}\BibitemShut {NoStop}%
\bibitem [{\citenamefont {Savard}\ \emph {et~al.}(1997)\citenamefont {Savard}, \citenamefont {O'Hara},\ and\ \citenamefont {Thomas}}]{savard1997laser}%
  \BibitemOpen
  \bibfield  {author} {\bibinfo {author} {\bibfnamefont {T.~A.}\ \bibnamefont {Savard}}, \bibinfo {author} {\bibfnamefont {K.~M.}\ \bibnamefont {O'Hara}},\ and\ \bibinfo {author} {\bibfnamefont {J.~E.}\ \bibnamefont {Thomas}},\ }\bibfield  {title} {\bibinfo {title} {Laser-noise-induced heating in far-off resonance optical traps},\ }\href {https://doi.org/10.1103/PhysRevA.56.R1095} {\bibfield  {journal} {\bibinfo  {journal} {Phys. Rev. A}\ }\textbf {\bibinfo {volume} {56}},\ \bibinfo {pages} {R1095} (\bibinfo {year} {1997})}\BibitemShut {NoStop}%
\bibitem [{\citenamefont {Manetsch}\ \emph {et~al.}(2025)\citenamefont {Manetsch}, \citenamefont {Nomura}, \citenamefont {Bataille}, \citenamefont {Lv}, \citenamefont {Leung},\ and\ \citenamefont {Endres}}]{Manetsch2025}%
  \BibitemOpen
  \bibfield  {author} {\bibinfo {author} {\bibfnamefont {H.~J.}\ \bibnamefont {Manetsch}}, \bibinfo {author} {\bibfnamefont {G.}~\bibnamefont {Nomura}}, \bibinfo {author} {\bibfnamefont {E.}~\bibnamefont {Bataille}}, \bibinfo {author} {\bibfnamefont {X.}~\bibnamefont {Lv}}, \bibinfo {author} {\bibfnamefont {K.~H.}\ \bibnamefont {Leung}},\ and\ \bibinfo {author} {\bibfnamefont {M.}~\bibnamefont {Endres}},\ }\bibfield  {title} {\bibinfo {title} {A tweezer array with 6,100 highly coherent atomic qubits},\ }\href {https://doi.org/10.1038/s41586-025-09641-4} {\bibfield  {journal} {\bibinfo  {journal} {Nature}\ }\textbf {\bibinfo {volume} {647}},\ \bibinfo {pages} {60} (\bibinfo {year} {2025})}\BibitemShut {NoStop}%
\bibitem [{\citenamefont {Zhang}\ \emph {et~al.}(2025)\citenamefont {Zhang}, \citenamefont {Hsu}, \citenamefont {Tan}, \citenamefont {Slichter}, \citenamefont {Kaufman}, \citenamefont {Marinelli},\ and\ \citenamefont {Regal}}]{zhang2025high}%
  \BibitemOpen
  \bibfield  {author} {\bibinfo {author} {\bibfnamefont {Z.}~\bibnamefont {Zhang}}, \bibinfo {author} {\bibfnamefont {T.-W.}\ \bibnamefont {Hsu}}, \bibinfo {author} {\bibfnamefont {T.~Y.}\ \bibnamefont {Tan}}, \bibinfo {author} {\bibfnamefont {D.~H.}\ \bibnamefont {Slichter}}, \bibinfo {author} {\bibfnamefont {A.~M.}\ \bibnamefont {Kaufman}}, \bibinfo {author} {\bibfnamefont {M.}~\bibnamefont {Marinelli}},\ and\ \bibinfo {author} {\bibfnamefont {C.~A.}\ \bibnamefont {Regal}},\ }\bibfield  {title} {\bibinfo {title} {High optical access cryogenic system for {Rydberg} atom arrays with a 3000-second trap lifetime},\ }\href {https://doi.org/10.1103/PRXQuantum.6.020337} {\bibfield  {journal} {\bibinfo  {journal} {PRX Quantum}\ }\textbf {\bibinfo {volume} {6}},\ \bibinfo {pages} {020337} (\bibinfo {year} {2025})}\BibitemShut {NoStop}%
\bibitem [{\citenamefont {Schymik}\ \emph {et~al.}(2021)\citenamefont {Schymik}, \citenamefont {Pancaldi}, \citenamefont {Nogrette}, \citenamefont {Barredo}, \citenamefont {Paris}, \citenamefont {Browaeys},\ and\ \citenamefont {Lahaye}}]{schymik2021single}%
  \BibitemOpen
  \bibfield  {author} {\bibinfo {author} {\bibfnamefont {K.-N.}\ \bibnamefont {Schymik}}, \bibinfo {author} {\bibfnamefont {S.}~\bibnamefont {Pancaldi}}, \bibinfo {author} {\bibfnamefont {F.}~\bibnamefont {Nogrette}}, \bibinfo {author} {\bibfnamefont {D.}~\bibnamefont {Barredo}}, \bibinfo {author} {\bibfnamefont {J.}~\bibnamefont {Paris}}, \bibinfo {author} {\bibfnamefont {A.}~\bibnamefont {Browaeys}},\ and\ \bibinfo {author} {\bibfnamefont {T.}~\bibnamefont {Lahaye}},\ }\bibfield  {title} {\bibinfo {title} {Single atoms with 6000-second trapping lifetimes in optical-tweezer arrays at cryogenic temperatures},\ }\href {https://doi.org/10.1103/PhysRevApplied.16.034013} {\bibfield  {journal} {\bibinfo  {journal} {Phys. Rev. Appl.}\ }\textbf {\bibinfo {volume} {16}},\ \bibinfo {pages} {034013} (\bibinfo {year} {2021})}\BibitemShut {NoStop}%
\bibitem [{\citenamefont {Covey}\ \emph {et~al.}(2019)\citenamefont {Covey}, \citenamefont {Madjarov}, \citenamefont {Cooper},\ and\ \citenamefont {Endres}}]{Covey2019A}%
  \BibitemOpen
  \bibfield  {author} {\bibinfo {author} {\bibfnamefont {J.~P.}\ \bibnamefont {Covey}}, \bibinfo {author} {\bibfnamefont {I.~S.}\ \bibnamefont {Madjarov}}, \bibinfo {author} {\bibfnamefont {A.}~\bibnamefont {Cooper}},\ and\ \bibinfo {author} {\bibfnamefont {M.}~\bibnamefont {Endres}},\ }\bibfield  {title} {\bibinfo {title} {2000-times repeated imaging of strontium atoms in clock-magic tweezer arrays},\ }\href {https://doi.org/10.1103/PhysRevLett.122.173201} {\bibfield  {journal} {\bibinfo  {journal} {Phys. Rev. Lett.}\ }\textbf {\bibinfo {volume} {122}},\ \bibinfo {pages} {173201} (\bibinfo {year} {2019})}\BibitemShut {NoStop}%
\bibitem [{\citenamefont {Radnaev}\ \emph {et~al.}(2025)\citenamefont {Radnaev}, \citenamefont {Chung}, \citenamefont {Cole}, \citenamefont {Mason}, \citenamefont {Ballance}, \citenamefont {Bedalov}, \citenamefont {Belknap}, \citenamefont {Berman}, \citenamefont {Blakely}, \citenamefont {Bloomfield} \emph {et~al.}}]{radnaev2025universal}%
  \BibitemOpen
  \bibfield  {author} {\bibinfo {author} {\bibfnamefont {A.}~\bibnamefont {Radnaev}}, \bibinfo {author} {\bibfnamefont {W.}~\bibnamefont {Chung}}, \bibinfo {author} {\bibfnamefont {D.}~\bibnamefont {Cole}}, \bibinfo {author} {\bibfnamefont {D.}~\bibnamefont {Mason}}, \bibinfo {author} {\bibfnamefont {T.}~\bibnamefont {Ballance}}, \bibinfo {author} {\bibfnamefont {M.}~\bibnamefont {Bedalov}}, \bibinfo {author} {\bibfnamefont {D.}~\bibnamefont {Belknap}}, \bibinfo {author} {\bibfnamefont {M.}~\bibnamefont {Berman}}, \bibinfo {author} {\bibfnamefont {M.}~\bibnamefont {Blakely}}, \bibinfo {author} {\bibfnamefont {I.}~\bibnamefont {Bloomfield}}, \emph {et~al.},\ }\bibfield  {title} {\bibinfo {title} {Universal neutral-atom quantum computer with individual optical addressing and nondestructive readout},\ }\href {https://doi.org/10.1103/66s8-jj18} {\bibfield  {journal} {\bibinfo  {journal} {PRX Quantum}\ }\textbf {\bibinfo {volume} {6}},\ \bibinfo {pages} {030334} (\bibinfo {year} {2025})}\BibitemShut {NoStop}%
\bibitem [{\citenamefont {Bluvstein}\ \emph {et~al.}(2024)\citenamefont {Bluvstein}, \citenamefont {Evered}, \citenamefont {Geim}, \citenamefont {Li}, \citenamefont {Zhou}, \citenamefont {Manovitz}, \citenamefont {Ebadi}, \citenamefont {Cain}, \citenamefont {Kalinowski}, \citenamefont {Hangleiter}, \citenamefont {Bonilla~Ataides}, \citenamefont {Maskara}, \citenamefont {Cong}, \citenamefont {Gao}, \citenamefont {Sales~Rodriguez}, \citenamefont {Karolyshyn}, \citenamefont {Semeghini}, \citenamefont {Gullans}, \citenamefont {Greiner}, \citenamefont {Vuleti{\'{c}}},\ and\ \citenamefont {Lukin}}]{Bluvstein2024}%
  \BibitemOpen
  \bibfield  {author} {\bibinfo {author} {\bibfnamefont {D.}~\bibnamefont {Bluvstein}}, \bibinfo {author} {\bibfnamefont {S.~J.}\ \bibnamefont {Evered}}, \bibinfo {author} {\bibfnamefont {A.~A.}\ \bibnamefont {Geim}}, \bibinfo {author} {\bibfnamefont {S.~H.}\ \bibnamefont {Li}}, \bibinfo {author} {\bibfnamefont {H.}~\bibnamefont {Zhou}}, \bibinfo {author} {\bibfnamefont {T.}~\bibnamefont {Manovitz}}, \bibinfo {author} {\bibfnamefont {S.}~\bibnamefont {Ebadi}}, \bibinfo {author} {\bibfnamefont {M.}~\bibnamefont {Cain}}, \bibinfo {author} {\bibfnamefont {M.}~\bibnamefont {Kalinowski}}, \bibinfo {author} {\bibfnamefont {D.}~\bibnamefont {Hangleiter}}, \bibinfo {author} {\bibfnamefont {J.~P.}\ \bibnamefont {Bonilla~Ataides}}, \bibinfo {author} {\bibfnamefont {N.}~\bibnamefont {Maskara}}, \bibinfo {author} {\bibfnamefont {I.}~\bibnamefont {Cong}}, \bibinfo {author} {\bibfnamefont {X.}~\bibnamefont {Gao}}, \bibinfo {author} {\bibfnamefont {P.}~\bibnamefont {Sales~Rodriguez}}, \bibinfo {author} {\bibfnamefont
  {T.}~\bibnamefont {Karolyshyn}}, \bibinfo {author} {\bibfnamefont {G.}~\bibnamefont {Semeghini}}, \bibinfo {author} {\bibfnamefont {M.~J.}\ \bibnamefont {Gullans}}, \bibinfo {author} {\bibfnamefont {M.}~\bibnamefont {Greiner}}, \bibinfo {author} {\bibfnamefont {V.}~\bibnamefont {Vuleti{\'{c}}}},\ and\ \bibinfo {author} {\bibfnamefont {M.~D.}\ \bibnamefont {Lukin}},\ }\bibfield  {title} {\bibinfo {title} {Logical quantum processor based on reconfigurable atom arrays},\ }\href {https://doi.org/10.1038/s41586-023-06927-3} {\bibfield  {journal} {\bibinfo  {journal} {Nature}\ }\textbf {\bibinfo {volume} {626}},\ \bibinfo {pages} {58} (\bibinfo {year} {2024})}\BibitemShut {NoStop}%
\bibitem [{\citenamefont {Anand}\ \emph {et~al.}(2024)\citenamefont {Anand}, \citenamefont {Bradley}, \citenamefont {White}, \citenamefont {Ramesh}, \citenamefont {Singh},\ and\ \citenamefont {Bernien}}]{Anand2024}%
  \BibitemOpen
  \bibfield  {author} {\bibinfo {author} {\bibfnamefont {S.}~\bibnamefont {Anand}}, \bibinfo {author} {\bibfnamefont {C.~E.}\ \bibnamefont {Bradley}}, \bibinfo {author} {\bibfnamefont {R.}~\bibnamefont {White}}, \bibinfo {author} {\bibfnamefont {V.}~\bibnamefont {Ramesh}}, \bibinfo {author} {\bibfnamefont {K.}~\bibnamefont {Singh}},\ and\ \bibinfo {author} {\bibfnamefont {H.}~\bibnamefont {Bernien}},\ }\bibfield  {title} {\bibinfo {title} {A dual-species {Rydberg} array},\ }\href {https://doi.org/10.1038/s41567-024-02638-2} {\bibfield  {journal} {\bibinfo  {journal} {Nat. Phys.}\ }\textbf {\bibinfo {volume} {20}},\ \bibinfo {pages} {1744} (\bibinfo {year} {2024})}\BibitemShut {NoStop}%
\bibitem [{\citenamefont {Muniz}\ \emph {et~al.}(2025)\citenamefont {Muniz}, \citenamefont {Crow}, \citenamefont {Kim}, \citenamefont {Kindem}, \citenamefont {Cairncross}, \citenamefont {Ryou}, \citenamefont {Bohdanowicz}, \citenamefont {Chen}, \citenamefont {Ji}, \citenamefont {Jones}, \citenamefont {Megidish}, \citenamefont {Nishiguchi}, \citenamefont {Urbanek}, \citenamefont {Wadleigh}, \citenamefont {Wilkason} \emph {et~al.}}]{Muniz2025}%
  \BibitemOpen
  \bibfield  {author} {\bibinfo {author} {\bibfnamefont {J.~A.}\ \bibnamefont {Muniz}}, \bibinfo {author} {\bibfnamefont {D.}~\bibnamefont {Crow}}, \bibinfo {author} {\bibfnamefont {H.}~\bibnamefont {Kim}}, \bibinfo {author} {\bibfnamefont {J.~M.}\ \bibnamefont {Kindem}}, \bibinfo {author} {\bibfnamefont {W.~B.}\ \bibnamefont {Cairncross}}, \bibinfo {author} {\bibfnamefont {A.}~\bibnamefont {Ryou}}, \bibinfo {author} {\bibfnamefont {T.~C.}\ \bibnamefont {Bohdanowicz}}, \bibinfo {author} {\bibfnamefont {C.-A.}\ \bibnamefont {Chen}}, \bibinfo {author} {\bibfnamefont {Y.}~\bibnamefont {Ji}}, \bibinfo {author} {\bibfnamefont {A.~M.~W.}\ \bibnamefont {Jones}}, \bibinfo {author} {\bibfnamefont {E.}~\bibnamefont {Megidish}}, \bibinfo {author} {\bibfnamefont {C.}~\bibnamefont {Nishiguchi}}, \bibinfo {author} {\bibfnamefont {M.}~\bibnamefont {Urbanek}}, \bibinfo {author} {\bibfnamefont {L.}~\bibnamefont {Wadleigh}}, \bibinfo {author} {\bibfnamefont {T.}~\bibnamefont {Wilkason}}, \emph {et~al.},\ }\bibfield  {title}
  {\bibinfo {title} {Repeated ancilla reuse for logical computation on a neutral atom quantum computer},\ }\href {https://doi.org/10.1103/v7ny-fg31} {\bibfield  {journal} {\bibinfo  {journal} {Phys. Rev. X}\ }\textbf {\bibinfo {volume} {15}},\ \bibinfo {pages} {041040} (\bibinfo {year} {2025})}\BibitemShut {NoStop}%
\bibitem [{\citenamefont {Machu}\ \emph {et~al.}(2025)\citenamefont {Machu}, \citenamefont {Dur{\'a}n-Hern{\'a}ndez}, \citenamefont {Creutzer}, \citenamefont {Young}, \citenamefont {Raimond}, \citenamefont {Brune},\ and\ \citenamefont {Sayrin}}]{machu2025non}%
  \BibitemOpen
  \bibfield  {author} {\bibinfo {author} {\bibfnamefont {Y.}~\bibnamefont {Machu}}, \bibinfo {author} {\bibfnamefont {A.}~\bibnamefont {Dur{\'a}n-Hern{\'a}ndez}}, \bibinfo {author} {\bibfnamefont {G.}~\bibnamefont {Creutzer}}, \bibinfo {author} {\bibfnamefont {A.~A.}\ \bibnamefont {Young}}, \bibinfo {author} {\bibfnamefont {J.-M.}\ \bibnamefont {Raimond}}, \bibinfo {author} {\bibfnamefont {M.}~\bibnamefont {Brune}},\ and\ \bibinfo {author} {\bibfnamefont {C.}~\bibnamefont {Sayrin}},\ }\bibfield  {title} {\bibinfo {title} {Non-destructive optical read-out and manipulation of circular {Rydberg} atoms},\ }\href {https://doi.org/10.48550/arXiv.2509.24691} {\bibfield  {journal} {\bibinfo  {journal} {arXiv:2509.24691}\ } (\bibinfo {year} {2025})}\BibitemShut {NoStop}%
\bibitem [{\citenamefont {Norcia}\ \emph {et~al.}(2023)\citenamefont {Norcia}, \citenamefont {Cairncross}, \citenamefont {Barnes}, \citenamefont {Battaglino}, \citenamefont {Brown}, \citenamefont {Brown}, \citenamefont {Cassella}, \citenamefont {Chen}, \citenamefont {Coxe}, \citenamefont {Crow}, \citenamefont {Epstein}, \citenamefont {Griger}, \citenamefont {Jones}, \citenamefont {Kim}, \citenamefont {Kindem} \emph {et~al.}}]{Norcia2023A}%
  \BibitemOpen
  \bibfield  {author} {\bibinfo {author} {\bibfnamefont {M.~A.}\ \bibnamefont {Norcia}}, \bibinfo {author} {\bibfnamefont {W.~B.}\ \bibnamefont {Cairncross}}, \bibinfo {author} {\bibfnamefont {K.}~\bibnamefont {Barnes}}, \bibinfo {author} {\bibfnamefont {P.}~\bibnamefont {Battaglino}}, \bibinfo {author} {\bibfnamefont {A.}~\bibnamefont {Brown}}, \bibinfo {author} {\bibfnamefont {M.~O.}\ \bibnamefont {Brown}}, \bibinfo {author} {\bibfnamefont {K.}~\bibnamefont {Cassella}}, \bibinfo {author} {\bibfnamefont {C.-A.}\ \bibnamefont {Chen}}, \bibinfo {author} {\bibfnamefont {R.}~\bibnamefont {Coxe}}, \bibinfo {author} {\bibfnamefont {D.}~\bibnamefont {Crow}}, \bibinfo {author} {\bibfnamefont {J.}~\bibnamefont {Epstein}}, \bibinfo {author} {\bibfnamefont {C.}~\bibnamefont {Griger}}, \bibinfo {author} {\bibfnamefont {A.~M.~W.}\ \bibnamefont {Jones}}, \bibinfo {author} {\bibfnamefont {H.}~\bibnamefont {Kim}}, \bibinfo {author} {\bibfnamefont {J.~M.}\ \bibnamefont {Kindem}}, \emph {et~al.},\ }\bibfield  {title}
  {\bibinfo {title} {Midcircuit qubit measurement and rearrangement in a $^{171}\mathrm{Yb}$ atomic array},\ }\href {https://doi.org/10.1103/PhysRevX.13.041034} {\bibfield  {journal} {\bibinfo  {journal} {Phys. Rev. X}\ }\textbf {\bibinfo {volume} {13}},\ \bibinfo {pages} {041034} (\bibinfo {year} {2023})}\BibitemShut {NoStop}%
\bibitem [{\citenamefont {Deist}\ \emph {et~al.}(2022)\citenamefont {Deist}, \citenamefont {Lu}, \citenamefont {Ho}, \citenamefont {Pasha}, \citenamefont {Zeiher}, \citenamefont {Yan},\ and\ \citenamefont {Stamper-Kurn}}]{Deist2022_midcircuit}%
  \BibitemOpen
  \bibfield  {author} {\bibinfo {author} {\bibfnamefont {E.}~\bibnamefont {Deist}}, \bibinfo {author} {\bibfnamefont {Y.-H.}\ \bibnamefont {Lu}}, \bibinfo {author} {\bibfnamefont {J.}~\bibnamefont {Ho}}, \bibinfo {author} {\bibfnamefont {M.~K.}\ \bibnamefont {Pasha}}, \bibinfo {author} {\bibfnamefont {J.}~\bibnamefont {Zeiher}}, \bibinfo {author} {\bibfnamefont {Z.}~\bibnamefont {Yan}},\ and\ \bibinfo {author} {\bibfnamefont {D.~M.}\ \bibnamefont {Stamper-Kurn}},\ }\bibfield  {title} {\bibinfo {title} {Mid-circuit cavity measurement in a neutral atom array},\ }\href {https://doi.org/10.1103/PhysRevLett.129.203602} {\bibfield  {journal} {\bibinfo  {journal} {Phys. Rev. Lett.}\ }\textbf {\bibinfo {volume} {129}},\ \bibinfo {pages} {203602} (\bibinfo {year} {2022})}\BibitemShut {NoStop}%
\bibitem [{\citenamefont {Graham}\ \emph {et~al.}(2023)\citenamefont {Graham}, \citenamefont {Phuttitarn}, \citenamefont {Chinnarasu}, \citenamefont {Song}, \citenamefont {Poole}, \citenamefont {Jooya}, \citenamefont {Scott}, \citenamefont {Scott}, \citenamefont {Eichler},\ and\ \citenamefont {Saffman}}]{Graham2023}%
  \BibitemOpen
  \bibfield  {author} {\bibinfo {author} {\bibfnamefont {T.~M.}\ \bibnamefont {Graham}}, \bibinfo {author} {\bibfnamefont {L.}~\bibnamefont {Phuttitarn}}, \bibinfo {author} {\bibfnamefont {R.}~\bibnamefont {Chinnarasu}}, \bibinfo {author} {\bibfnamefont {Y.}~\bibnamefont {Song}}, \bibinfo {author} {\bibfnamefont {C.}~\bibnamefont {Poole}}, \bibinfo {author} {\bibfnamefont {K.}~\bibnamefont {Jooya}}, \bibinfo {author} {\bibfnamefont {J.}~\bibnamefont {Scott}}, \bibinfo {author} {\bibfnamefont {A.}~\bibnamefont {Scott}}, \bibinfo {author} {\bibfnamefont {P.}~\bibnamefont {Eichler}},\ and\ \bibinfo {author} {\bibfnamefont {M.}~\bibnamefont {Saffman}},\ }\bibfield  {title} {\bibinfo {title} {Midcircuit measurements on a single-species neutral alkali atom quantum processor},\ }\href {https://doi.org/10.1103/PhysRevX.13.041051} {\bibfield  {journal} {\bibinfo  {journal} {Phys. Rev. X}\ }\textbf {\bibinfo {volume} {13}},\ \bibinfo {pages} {041051} (\bibinfo {year} {2023})}\BibitemShut {NoStop}%
\bibitem [{\citenamefont {Stricker}\ \emph {et~al.}(2020)\citenamefont {Stricker}, \citenamefont {Vodola}, \citenamefont {Erhard}, \citenamefont {Postler}, \citenamefont {Meth}, \citenamefont {Ringbauer}, \citenamefont {Schindler}, \citenamefont {Monz}, \citenamefont {M{\"u}ller},\ and\ \citenamefont {Blatt}}]{Stricker2020}%
  \BibitemOpen
  \bibfield  {author} {\bibinfo {author} {\bibfnamefont {R.}~\bibnamefont {Stricker}}, \bibinfo {author} {\bibfnamefont {D.}~\bibnamefont {Vodola}}, \bibinfo {author} {\bibfnamefont {A.}~\bibnamefont {Erhard}}, \bibinfo {author} {\bibfnamefont {L.}~\bibnamefont {Postler}}, \bibinfo {author} {\bibfnamefont {M.}~\bibnamefont {Meth}}, \bibinfo {author} {\bibfnamefont {M.}~\bibnamefont {Ringbauer}}, \bibinfo {author} {\bibfnamefont {P.}~\bibnamefont {Schindler}}, \bibinfo {author} {\bibfnamefont {T.}~\bibnamefont {Monz}}, \bibinfo {author} {\bibfnamefont {M.}~\bibnamefont {M{\"u}ller}},\ and\ \bibinfo {author} {\bibfnamefont {R.}~\bibnamefont {Blatt}},\ }\bibfield  {title} {\bibinfo {title} {Experimental deterministic correction of qubit loss},\ }\href {https://doi.org/10.1038/s41586-020-2667-0} {\bibfield  {journal} {\bibinfo  {journal} {Nature}\ }\textbf {\bibinfo {volume} {585}},\ \bibinfo {pages} {207} (\bibinfo {year} {2020})}\BibitemShut {NoStop}%
\bibitem [{\citenamefont {Chow}\ \emph {et~al.}(2024)\citenamefont {Chow}, \citenamefont {Buchemmavari}, \citenamefont {Omanakuttan}, \citenamefont {Little}, \citenamefont {Pandey}, \citenamefont {Deutsch},\ and\ \citenamefont {Jau}}]{Chow2024}%
  \BibitemOpen
  \bibfield  {author} {\bibinfo {author} {\bibfnamefont {M.~N.~H.}\ \bibnamefont {Chow}}, \bibinfo {author} {\bibfnamefont {V.}~\bibnamefont {Buchemmavari}}, \bibinfo {author} {\bibfnamefont {S.}~\bibnamefont {Omanakuttan}}, \bibinfo {author} {\bibfnamefont {B.~J.}\ \bibnamefont {Little}}, \bibinfo {author} {\bibfnamefont {S.}~\bibnamefont {Pandey}}, \bibinfo {author} {\bibfnamefont {I.~H.}\ \bibnamefont {Deutsch}},\ and\ \bibinfo {author} {\bibfnamefont {Y.-Y.}\ \bibnamefont {Jau}},\ }\bibfield  {title} {\bibinfo {title} {Circuit-based leakage-to-erasure conversion in a neutral-atom quantum processor},\ }\href {https://doi.org/10.1103/PRXQuantum.5.040343} {\bibfield  {journal} {\bibinfo  {journal} {PRX Quantum}\ }\textbf {\bibinfo {volume} {5}},\ \bibinfo {pages} {040343} (\bibinfo {year} {2024})}\BibitemShut {NoStop}%
\bibitem [{\citenamefont {Scholl}\ \emph {et~al.}(2023)\citenamefont {Scholl}, \citenamefont {Shaw}, \citenamefont {Tsai}, \citenamefont {Finkelstein}, \citenamefont {Choi},\ and\ \citenamefont {Endres}}]{Scholl2023A}%
  \BibitemOpen
  \bibfield  {author} {\bibinfo {author} {\bibfnamefont {P.}~\bibnamefont {Scholl}}, \bibinfo {author} {\bibfnamefont {A.~L.}\ \bibnamefont {Shaw}}, \bibinfo {author} {\bibfnamefont {R.~B.-S.}\ \bibnamefont {Tsai}}, \bibinfo {author} {\bibfnamefont {R.}~\bibnamefont {Finkelstein}}, \bibinfo {author} {\bibfnamefont {J.}~\bibnamefont {Choi}},\ and\ \bibinfo {author} {\bibfnamefont {M.}~\bibnamefont {Endres}},\ }\bibfield  {title} {\bibinfo {title} {Erasure conversion in a high-fidelity {Rydberg} quantum simulator},\ }\href {https://doi.org/10.1038/s41586-023-06516-4} {\bibfield  {journal} {\bibinfo  {journal} {Nature}\ }\textbf {\bibinfo {volume} {622}},\ \bibinfo {pages} {273} (\bibinfo {year} {2023})}\BibitemShut {NoStop}%
\bibitem [{\citenamefont {Muzi~Falconi}\ \emph {et~al.}(2025)\citenamefont {Muzi~Falconi}, \citenamefont {Panza}, \citenamefont {Sbernardori}, \citenamefont {Forti}, \citenamefont {Klemt}, \citenamefont {Abdel~Karim}, \citenamefont {Marinelli},\ and\ \citenamefont {Scazza}}]{Muzi_Falconi_2025}%
  \BibitemOpen
  \bibfield  {author} {\bibinfo {author} {\bibfnamefont {A.}~\bibnamefont {Muzi~Falconi}}, \bibinfo {author} {\bibfnamefont {R.}~\bibnamefont {Panza}}, \bibinfo {author} {\bibfnamefont {S.}~\bibnamefont {Sbernardori}}, \bibinfo {author} {\bibfnamefont {R.}~\bibnamefont {Forti}}, \bibinfo {author} {\bibfnamefont {R.}~\bibnamefont {Klemt}}, \bibinfo {author} {\bibfnamefont {O.}~\bibnamefont {Abdel~Karim}}, \bibinfo {author} {\bibfnamefont {M.}~\bibnamefont {Marinelli}},\ and\ \bibinfo {author} {\bibfnamefont {F.}~\bibnamefont {Scazza}},\ }\bibfield  {title} {\bibinfo {title} {Microsecond-scale high-survival and number-resolved detection of ytterbium atom arrays},\ }\href {https://doi.org/10.1103/n3bg-7yw7} {\bibfield  {journal} {\bibinfo  {journal} {Phys. Rev. Lett.}\ }\textbf {\bibinfo {volume} {135}},\ \bibinfo {pages} {203402} (\bibinfo {year} {2025})}\BibitemShut {NoStop}%
\bibitem [{\citenamefont {Ma}\ \emph {et~al.}(2025)\citenamefont {Ma}, \citenamefont {Dolde}, \citenamefont {Zheng}, \citenamefont {Ganapathy}, \citenamefont {Shtov}, \citenamefont {Chen}, \citenamefont {St\"oltzel}, \citenamefont {Christensen},\ and\ \citenamefont {Kolkowitz}}]{Ma2025}%
  \BibitemOpen
  \bibfield  {author} {\bibinfo {author} {\bibfnamefont {S.}~\bibnamefont {Ma}}, \bibinfo {author} {\bibfnamefont {J.}~\bibnamefont {Dolde}}, \bibinfo {author} {\bibfnamefont {X.}~\bibnamefont {Zheng}}, \bibinfo {author} {\bibfnamefont {D.}~\bibnamefont {Ganapathy}}, \bibinfo {author} {\bibfnamefont {A.}~\bibnamefont {Shtov}}, \bibinfo {author} {\bibfnamefont {J.}~\bibnamefont {Chen}}, \bibinfo {author} {\bibfnamefont {A.}~\bibnamefont {St\"oltzel}}, \bibinfo {author} {\bibfnamefont {B.~J.}\ \bibnamefont {Christensen}},\ and\ \bibinfo {author} {\bibfnamefont {S.}~\bibnamefont {Kolkowitz}},\ }\bibfield  {title} {\bibinfo {title} {Enhancing optical lattice clock coherence times with erasure conversion},\ }\href {https://doi.org/10.1103/2rqf-r9gw} {\bibfield  {journal} {\bibinfo  {journal} {PRX Quantum}\ }\textbf {\bibinfo {volume} {6}},\ \bibinfo {pages} {040340} (\bibinfo {year} {2025})}\BibitemShut {NoStop}%
\bibitem [{\citenamefont {Lis}\ \emph {et~al.}(2023)\citenamefont {Lis}, \citenamefont {Senoo}, \citenamefont {McGrew}, \citenamefont {R\"onchen}, \citenamefont {Jenkins},\ and\ \citenamefont {Kaufman}}]{Lis2023A}%
  \BibitemOpen
  \bibfield  {author} {\bibinfo {author} {\bibfnamefont {J.~W.}\ \bibnamefont {Lis}}, \bibinfo {author} {\bibfnamefont {A.}~\bibnamefont {Senoo}}, \bibinfo {author} {\bibfnamefont {W.~F.}\ \bibnamefont {McGrew}}, \bibinfo {author} {\bibfnamefont {F.}~\bibnamefont {R\"onchen}}, \bibinfo {author} {\bibfnamefont {A.}~\bibnamefont {Jenkins}},\ and\ \bibinfo {author} {\bibfnamefont {A.~M.}\ \bibnamefont {Kaufman}},\ }\bibfield  {title} {\bibinfo {title} {Midcircuit operations using the \textit{omg} architecture in neutral atom arrays},\ }\href {https://doi.org/10.1103/PhysRevX.13.041035} {\bibfield  {journal} {\bibinfo  {journal} {Phys. Rev. X}\ }\textbf {\bibinfo {volume} {13}},\ \bibinfo {pages} {041035} (\bibinfo {year} {2023})}\BibitemShut {NoStop}%
\bibitem [{\citenamefont {Ma}\ \emph {et~al.}(2022)\citenamefont {Ma}, \citenamefont {Burgers}, \citenamefont {Liu}, \citenamefont {Wilson}, \citenamefont {Zhang},\ and\ \citenamefont {Thompson}}]{Ma2022_Yb}%
  \BibitemOpen
  \bibfield  {author} {\bibinfo {author} {\bibfnamefont {S.}~\bibnamefont {Ma}}, \bibinfo {author} {\bibfnamefont {A.~P.}\ \bibnamefont {Burgers}}, \bibinfo {author} {\bibfnamefont {G.}~\bibnamefont {Liu}}, \bibinfo {author} {\bibfnamefont {J.}~\bibnamefont {Wilson}}, \bibinfo {author} {\bibfnamefont {B.}~\bibnamefont {Zhang}},\ and\ \bibinfo {author} {\bibfnamefont {J.~D.}\ \bibnamefont {Thompson}},\ }\bibfield  {title} {\bibinfo {title} {Universal gate operations on nuclear spin qubits in an optical tweezer array of $^{171}\mathrm{Yb}$ atoms},\ }\href {https://doi.org/10.1103/PhysRevX.12.021028} {\bibfield  {journal} {\bibinfo  {journal} {Phys. Rev. X}\ }\textbf {\bibinfo {volume} {12}},\ \bibinfo {pages} {021028} (\bibinfo {year} {2022})}\BibitemShut {NoStop}%
\bibitem [{\citenamefont {Chen}\ \emph {et~al.}(2022)\citenamefont {Chen}, \citenamefont {Li}, \citenamefont {Huie}, \citenamefont {Zhao}, \citenamefont {Vetter}, \citenamefont {Greene},\ and\ \citenamefont {Covey}}]{Chen2022}%
  \BibitemOpen
  \bibfield  {author} {\bibinfo {author} {\bibfnamefont {N.}~\bibnamefont {Chen}}, \bibinfo {author} {\bibfnamefont {L.}~\bibnamefont {Li}}, \bibinfo {author} {\bibfnamefont {W.}~\bibnamefont {Huie}}, \bibinfo {author} {\bibfnamefont {M.}~\bibnamefont {Zhao}}, \bibinfo {author} {\bibfnamefont {I.}~\bibnamefont {Vetter}}, \bibinfo {author} {\bibfnamefont {C.~H.}\ \bibnamefont {Greene}},\ and\ \bibinfo {author} {\bibfnamefont {J.~P.}\ \bibnamefont {Covey}},\ }\bibfield  {title} {\bibinfo {title} {Analyzing the {Rydberg}-based optical-metastable-ground architecture for $^{171}\mathrm{Yb}$ nuclear spins},\ }\href {https://doi.org/10.1103/PhysRevA.105.052438} {\bibfield  {journal} {\bibinfo  {journal} {Phys. Rev. A}\ }\textbf {\bibinfo {volume} {105}},\ \bibinfo {pages} {52438} (\bibinfo {year} {2022})}\BibitemShut {NoStop}%
\bibitem [{SM()}]{SM}%
  \BibitemOpen
  \href@noop {} {\bibinfo {title} {See {Supplemental Material} for experimental, data analysis, and theoretical details, including a discussion of motional shelving errors and the algorithmic cooling pulse sequence.}}\BibitemShut {Stop}%
\bibitem [{\citenamefont {Boykin}\ \emph {et~al.}(2002)\citenamefont {Boykin}, \citenamefont {Mor}, \citenamefont {Roychowdhury}, \citenamefont {Vatan},\ and\ \citenamefont {Vrijen}}]{boykinAlgorithmicCoolingScalable2002}%
  \BibitemOpen
  \bibfield  {author} {\bibinfo {author} {\bibfnamefont {P.~O.}\ \bibnamefont {Boykin}}, \bibinfo {author} {\bibfnamefont {T.}~\bibnamefont {Mor}}, \bibinfo {author} {\bibfnamefont {V.}~\bibnamefont {Roychowdhury}}, \bibinfo {author} {\bibfnamefont {F.}~\bibnamefont {Vatan}},\ and\ \bibinfo {author} {\bibfnamefont {R.}~\bibnamefont {Vrijen}},\ }\bibfield  {title} {\bibinfo {title} {Algorithmic cooling and scalable {{NMR}} quantum computers},\ }\href {https://doi.org/10.1073/pnas.241641898} {\bibfield  {journal} {\bibinfo  {journal} {Proceedings of the National Academy of Sciences}\ }\textbf {\bibinfo {volume} {99}},\ \bibinfo {pages} {3388} (\bibinfo {year} {2002})}\BibitemShut {NoStop}%
\bibitem [{\citenamefont {King}\ \emph {et~al.}(2021)\citenamefont {King}, \citenamefont {Spie\ss{}}, \citenamefont {Micke}, \citenamefont {Wilzewski}, \citenamefont {Leopold}, \citenamefont {Crespo L\'opez-Urrutia},\ and\ \citenamefont {Schmidt}}]{King2021}%
  \BibitemOpen
  \bibfield  {author} {\bibinfo {author} {\bibfnamefont {S.~A.}\ \bibnamefont {King}}, \bibinfo {author} {\bibfnamefont {L.~J.}\ \bibnamefont {Spie\ss{}}}, \bibinfo {author} {\bibfnamefont {P.}~\bibnamefont {Micke}}, \bibinfo {author} {\bibfnamefont {A.}~\bibnamefont {Wilzewski}}, \bibinfo {author} {\bibfnamefont {T.}~\bibnamefont {Leopold}}, \bibinfo {author} {\bibfnamefont {J.~R.}\ \bibnamefont {Crespo L\'opez-Urrutia}},\ and\ \bibinfo {author} {\bibfnamefont {P.~O.}\ \bibnamefont {Schmidt}},\ }\bibfield  {title} {\bibinfo {title} {Algorithmic ground-state cooling of weakly coupled oscillators using quantum logic},\ }\href {https://doi.org/10.1103/PhysRevX.11.041049} {\bibfield  {journal} {\bibinfo  {journal} {Phys. Rev. X}\ }\textbf {\bibinfo {volume} {11}},\ \bibinfo {pages} {041049} (\bibinfo {year} {2021})}\BibitemShut {NoStop}%
\bibitem [{\citenamefont {Blodgett}\ \emph {et~al.}(2023)\citenamefont {Blodgett}, \citenamefont {Peana}, \citenamefont {Phatak}, \citenamefont {Terry}, \citenamefont {Montes},\ and\ \citenamefont {Hood}}]{blodgett2023imaging}%
  \BibitemOpen
  \bibfield  {author} {\bibinfo {author} {\bibfnamefont {K.~N.}\ \bibnamefont {Blodgett}}, \bibinfo {author} {\bibfnamefont {D.}~\bibnamefont {Peana}}, \bibinfo {author} {\bibfnamefont {S.~S.}\ \bibnamefont {Phatak}}, \bibinfo {author} {\bibfnamefont {L.~M.}\ \bibnamefont {Terry}}, \bibinfo {author} {\bibfnamefont {M.~P.}\ \bibnamefont {Montes}},\ and\ \bibinfo {author} {\bibfnamefont {J.~D.}\ \bibnamefont {Hood}},\ }\bibfield  {title} {\bibinfo {title} {Imaging a $^{6}\mathrm{Li}$ atom in an optical tweezer 2000 times with $\mathrm{\ensuremath{\Lambda}}$-enhanced gray molasses},\ }\href {https://doi.org/10.1103/PhysRevLett.131.083001} {\bibfield  {journal} {\bibinfo  {journal} {Phys. Rev. Lett.}\ }\textbf {\bibinfo {volume} {131}},\ \bibinfo {pages} {083001} (\bibinfo {year} {2023})}\BibitemShut {NoStop}%
\bibitem [{\citenamefont {Senoo}\ \emph {et~al.}(2025)\citenamefont {Senoo}, \citenamefont {Baumgärtner}, \citenamefont {Lis}, \citenamefont {Vaidya}, \citenamefont {Zeng}, \citenamefont {Giudici}, \citenamefont {Pichler},\ and\ \citenamefont {Kaufman}}]{Senoo2025}%
  \BibitemOpen
  \bibfield  {author} {\bibinfo {author} {\bibfnamefont {A.}~\bibnamefont {Senoo}}, \bibinfo {author} {\bibfnamefont {A.}~\bibnamefont {Baumgärtner}}, \bibinfo {author} {\bibfnamefont {J.~W.}\ \bibnamefont {Lis}}, \bibinfo {author} {\bibfnamefont {G.~M.}\ \bibnamefont {Vaidya}}, \bibinfo {author} {\bibfnamefont {Z.}~\bibnamefont {Zeng}}, \bibinfo {author} {\bibfnamefont {G.}~\bibnamefont {Giudici}}, \bibinfo {author} {\bibfnamefont {H.}~\bibnamefont {Pichler}},\ and\ \bibinfo {author} {\bibfnamefont {A.~M.}\ \bibnamefont {Kaufman}},\ }\bibfield  {title} {\bibinfo {title} {High-fidelity entanglement and coherent multi-qubit mapping in an atom array},\ }\href {https://arxiv.org/abs/2506.13632} {\bibfield  {journal} {\bibinfo  {journal} {arXiv:2506.13632}\ } (\bibinfo {year} {2025})}\BibitemShut {NoStop}%
\bibitem [{\citenamefont {Su}\ \emph {et~al.}(2025)\citenamefont {Su}, \citenamefont {Douglas}, \citenamefont {Szurek}, \citenamefont {H{\'e}bert}, \citenamefont {Krahn}, \citenamefont {Groth}, \citenamefont {Phelps}, \citenamefont {Markovi{\'c}},\ and\ \citenamefont {Greiner}}]{Su2025}%
  \BibitemOpen
  \bibfield  {author} {\bibinfo {author} {\bibfnamefont {L.}~\bibnamefont {Su}}, \bibinfo {author} {\bibfnamefont {A.}~\bibnamefont {Douglas}}, \bibinfo {author} {\bibfnamefont {M.}~\bibnamefont {Szurek}}, \bibinfo {author} {\bibfnamefont {A.~H.}\ \bibnamefont {H{\'e}bert}}, \bibinfo {author} {\bibfnamefont {A.}~\bibnamefont {Krahn}}, \bibinfo {author} {\bibfnamefont {R.}~\bibnamefont {Groth}}, \bibinfo {author} {\bibfnamefont {G.~A.}\ \bibnamefont {Phelps}}, \bibinfo {author} {\bibfnamefont {O.}~\bibnamefont {Markovi{\'c}}},\ and\ \bibinfo {author} {\bibfnamefont {M.}~\bibnamefont {Greiner}},\ }\bibfield  {title} {\bibinfo {title} {Fast single atom imaging for optical lattice arrays},\ }\href {https://doi.org/10.1038/s41467-025-56305-y} {\bibfield  {journal} {\bibinfo  {journal} {Nat. Commun.}\ }\textbf {\bibinfo {volume} {16}},\ \bibinfo {pages} {1017} (\bibinfo {year} {2025})}\BibitemShut {NoStop}%
\bibitem [{\citenamefont {Braginsky}\ \emph {et~al.}(1980)\citenamefont {Braginsky}, \citenamefont {Vorontsov},\ and\ \citenamefont {Thorne}}]{braginsky1980quantum}%
  \BibitemOpen
  \bibfield  {author} {\bibinfo {author} {\bibfnamefont {V.~B.}\ \bibnamefont {Braginsky}}, \bibinfo {author} {\bibfnamefont {Y.~I.}\ \bibnamefont {Vorontsov}},\ and\ \bibinfo {author} {\bibfnamefont {K.~S.}\ \bibnamefont {Thorne}},\ }\bibfield  {title} {\bibinfo {title} {Quantum nondemolition measurements},\ }\href {https://doi.org/10.1126/science.209.4456.547} {\bibfield  {journal} {\bibinfo  {journal} {Science}\ }\textbf {\bibinfo {volume} {209}},\ \bibinfo {pages} {547} (\bibinfo {year} {1980})}\BibitemShut {NoStop}%
\bibitem [{\citenamefont {Braginsky}\ and\ \citenamefont {Khalili}(1996)}]{braginsky1996quantum}%
  \BibitemOpen
  \bibfield  {author} {\bibinfo {author} {\bibfnamefont {V.~B.}\ \bibnamefont {Braginsky}}\ and\ \bibinfo {author} {\bibfnamefont {F.~Y.}\ \bibnamefont {Khalili}},\ }\bibfield  {title} {\bibinfo {title} {Quantum nondemolition measurements: the route from toys to tools},\ }\href {https://doi.org/10.1103/RevModPhys.68.1} {\bibfield  {journal} {\bibinfo  {journal} {Rev. Mod. Phys.}\ }\textbf {\bibinfo {volume} {68}},\ \bibinfo {pages} {1} (\bibinfo {year} {1996})}\BibitemShut {NoStop}%
\bibitem [{\citenamefont {Shaw}\ \emph {et~al.}(2025)\citenamefont {Shaw}, \citenamefont {Scholl}, \citenamefont {Finkelstein}, \citenamefont {Tsai}, \citenamefont {Choi},\ and\ \citenamefont {Endres}}]{Shaw2025_erasure_cooling}%
  \BibitemOpen
  \bibfield  {author} {\bibinfo {author} {\bibfnamefont {A.~L.}\ \bibnamefont {Shaw}}, \bibinfo {author} {\bibfnamefont {P.}~\bibnamefont {Scholl}}, \bibinfo {author} {\bibfnamefont {R.}~\bibnamefont {Finkelstein}}, \bibinfo {author} {\bibfnamefont {R.~B.-S.}\ \bibnamefont {Tsai}}, \bibinfo {author} {\bibfnamefont {J.}~\bibnamefont {Choi}},\ and\ \bibinfo {author} {\bibfnamefont {M.}~\bibnamefont {Endres}},\ }\bibfield  {title} {\bibinfo {title} {Erasure cooling, control, and hyperentanglement of motion in optical tweezers},\ }\href {https://doi.org/10.1126/science.adn2618} {\bibfield  {journal} {\bibinfo  {journal} {Science}\ }\textbf {\bibinfo {volume} {388}},\ \bibinfo {pages} {845} (\bibinfo {year} {2025})}\BibitemShut {NoStop}%
\bibitem [{\citenamefont {Shaw}\ \emph {et~al.}(2024{\natexlab{b}})\citenamefont {Shaw}, \citenamefont {Finkelstein}, \citenamefont {Tsai}, \citenamefont {Scholl}, \citenamefont {Yoon}, \citenamefont {Choi},\ and\ \citenamefont {Endres}}]{Shaw2024}%
  \BibitemOpen
  \bibfield  {author} {\bibinfo {author} {\bibfnamefont {A.~L.}\ \bibnamefont {Shaw}}, \bibinfo {author} {\bibfnamefont {R.}~\bibnamefont {Finkelstein}}, \bibinfo {author} {\bibfnamefont {R.~B.-S.}\ \bibnamefont {Tsai}}, \bibinfo {author} {\bibfnamefont {P.}~\bibnamefont {Scholl}}, \bibinfo {author} {\bibfnamefont {T.~H.}\ \bibnamefont {Yoon}}, \bibinfo {author} {\bibfnamefont {J.}~\bibnamefont {Choi}},\ and\ \bibinfo {author} {\bibfnamefont {M.}~\bibnamefont {Endres}},\ }\bibfield  {title} {\bibinfo {title} {Multi-ensemble metrology by programming local rotations with atom movements},\ }\href {https://doi.org/10.1038/s41567-023-02323-w} {\bibfield  {journal} {\bibinfo  {journal} {Nat. Phys.}\ }\textbf {\bibinfo {volume} {20}},\ \bibinfo {pages} {195} (\bibinfo {year} {2024}{\natexlab{b}})}\BibitemShut {NoStop}%
\bibitem [{\citenamefont {Levine}\ \emph {et~al.}(2019)\citenamefont {Levine}, \citenamefont {Keesling}, \citenamefont {Semeghini}, \citenamefont {Omran}, \citenamefont {Wang}, \citenamefont {Ebadi}, \citenamefont {Bernien}, \citenamefont {Greiner}, \citenamefont {Vuletić}, \citenamefont {Pichler},\ and\ \citenamefont {Lukin}}]{Levine2019}%
  \BibitemOpen
  \bibfield  {author} {\bibinfo {author} {\bibfnamefont {H.}~\bibnamefont {Levine}}, \bibinfo {author} {\bibfnamefont {A.}~\bibnamefont {Keesling}}, \bibinfo {author} {\bibfnamefont {G.}~\bibnamefont {Semeghini}}, \bibinfo {author} {\bibfnamefont {A.}~\bibnamefont {Omran}}, \bibinfo {author} {\bibfnamefont {T.~T.}\ \bibnamefont {Wang}}, \bibinfo {author} {\bibfnamefont {S.}~\bibnamefont {Ebadi}}, \bibinfo {author} {\bibfnamefont {H.}~\bibnamefont {Bernien}}, \bibinfo {author} {\bibfnamefont {M.}~\bibnamefont {Greiner}}, \bibinfo {author} {\bibfnamefont {V.}~\bibnamefont {Vuletić}}, \bibinfo {author} {\bibfnamefont {H.}~\bibnamefont {Pichler}},\ and\ \bibinfo {author} {\bibfnamefont {M.~D.}\ \bibnamefont {Lukin}},\ }\bibfield  {title} {\bibinfo {title} {Parallel implementation of high-fidelity multiqubit gates with neutral atoms},\ }\href {https://doi.org/10.1103/PhysRevLett.123.170503} {\bibfield  {journal} {\bibinfo  {journal} {Phys. Rev. Lett.}\ }\textbf {\bibinfo {volume} {123}},\ \bibinfo {pages}
  {170503} (\bibinfo {year} {2019})}\BibitemShut {NoStop}%
\bibitem [{\citenamefont {Graham}\ \emph {et~al.}(2019)\citenamefont {Graham}, \citenamefont {Kwon}, \citenamefont {Grinkemeyer}, \citenamefont {Marra}, \citenamefont {Jiang}, \citenamefont {Lichtman}, \citenamefont {Sun}, \citenamefont {Ebert},\ and\ \citenamefont {Saffman}}]{Graham2019}%
  \BibitemOpen
  \bibfield  {author} {\bibinfo {author} {\bibfnamefont {T.~M.}\ \bibnamefont {Graham}}, \bibinfo {author} {\bibfnamefont {M.}~\bibnamefont {Kwon}}, \bibinfo {author} {\bibfnamefont {B.}~\bibnamefont {Grinkemeyer}}, \bibinfo {author} {\bibfnamefont {Z.}~\bibnamefont {Marra}}, \bibinfo {author} {\bibfnamefont {X.}~\bibnamefont {Jiang}}, \bibinfo {author} {\bibfnamefont {M.~T.}\ \bibnamefont {Lichtman}}, \bibinfo {author} {\bibfnamefont {Y.}~\bibnamefont {Sun}}, \bibinfo {author} {\bibfnamefont {M.}~\bibnamefont {Ebert}},\ and\ \bibinfo {author} {\bibfnamefont {M.}~\bibnamefont {Saffman}},\ }\bibfield  {title} {\bibinfo {title} {Rydberg-mediated entanglement in a two-dimensional neutral atom qubit array},\ }\href {https://doi.org/10.1103/PhysRevLett.123.230501} {\bibfield  {journal} {\bibinfo  {journal} {Phys. Rev. Lett.}\ }\textbf {\bibinfo {volume} {123}},\ \bibinfo {pages} {230501} (\bibinfo {year} {2019})}\BibitemShut {NoStop}%
\bibitem [{\citenamefont {Evered}\ \emph {et~al.}(2023)\citenamefont {Evered}, \citenamefont {Bluvstein}, \citenamefont {Kalinowski}, \citenamefont {Ebadi}, \citenamefont {Manovitz}, \citenamefont {Zhou}, \citenamefont {Li}, \citenamefont {Geim}, \citenamefont {Wang}, \citenamefont {Maskara}, \citenamefont {Levine}, \citenamefont {Semeghini}, \citenamefont {Greiner}, \citenamefont {Vuleti{\'{c}}},\ and\ \citenamefont {Lukin}}]{Evered2023}%
  \BibitemOpen
  \bibfield  {author} {\bibinfo {author} {\bibfnamefont {S.~J.}\ \bibnamefont {Evered}}, \bibinfo {author} {\bibfnamefont {D.}~\bibnamefont {Bluvstein}}, \bibinfo {author} {\bibfnamefont {M.}~\bibnamefont {Kalinowski}}, \bibinfo {author} {\bibfnamefont {S.}~\bibnamefont {Ebadi}}, \bibinfo {author} {\bibfnamefont {T.}~\bibnamefont {Manovitz}}, \bibinfo {author} {\bibfnamefont {H.}~\bibnamefont {Zhou}}, \bibinfo {author} {\bibfnamefont {S.~H.}\ \bibnamefont {Li}}, \bibinfo {author} {\bibfnamefont {A.~A.}\ \bibnamefont {Geim}}, \bibinfo {author} {\bibfnamefont {T.~T.}\ \bibnamefont {Wang}}, \bibinfo {author} {\bibfnamefont {N.}~\bibnamefont {Maskara}}, \bibinfo {author} {\bibfnamefont {H.}~\bibnamefont {Levine}}, \bibinfo {author} {\bibfnamefont {G.}~\bibnamefont {Semeghini}}, \bibinfo {author} {\bibfnamefont {M.}~\bibnamefont {Greiner}}, \bibinfo {author} {\bibfnamefont {V.}~\bibnamefont {Vuleti{\'{c}}}},\ and\ \bibinfo {author} {\bibfnamefont {M.~D.}\ \bibnamefont {Lukin}},\ }\bibfield  {title} {\bibinfo
  {title} {High-fidelity parallel entangling gates on a neutral-atom quantum computer},\ }\href {https://doi.org/10.1038/s41586-023-06481-y} {\bibfield  {journal} {\bibinfo  {journal} {Nature}\ }\textbf {\bibinfo {volume} {622}},\ \bibinfo {pages} {268} (\bibinfo {year} {2023})}\BibitemShut {NoStop}%
\bibitem [{\citenamefont {Ma}\ \emph {et~al.}(2023)\citenamefont {Ma}, \citenamefont {Liu}, \citenamefont {Peng}, \citenamefont {Zhang}, \citenamefont {Jandura}, \citenamefont {Claes}, \citenamefont {Burgers}, \citenamefont {Pupillo}, \citenamefont {Puri},\ and\ \citenamefont {Thompson}}]{Ma2023}%
  \BibitemOpen
  \bibfield  {author} {\bibinfo {author} {\bibfnamefont {S.}~\bibnamefont {Ma}}, \bibinfo {author} {\bibfnamefont {G.}~\bibnamefont {Liu}}, \bibinfo {author} {\bibfnamefont {P.}~\bibnamefont {Peng}}, \bibinfo {author} {\bibfnamefont {B.}~\bibnamefont {Zhang}}, \bibinfo {author} {\bibfnamefont {S.}~\bibnamefont {Jandura}}, \bibinfo {author} {\bibfnamefont {J.}~\bibnamefont {Claes}}, \bibinfo {author} {\bibfnamefont {A.~P.}\ \bibnamefont {Burgers}}, \bibinfo {author} {\bibfnamefont {G.}~\bibnamefont {Pupillo}}, \bibinfo {author} {\bibfnamefont {S.}~\bibnamefont {Puri}},\ and\ \bibinfo {author} {\bibfnamefont {J.~D.}\ \bibnamefont {Thompson}},\ }\bibfield  {title} {\bibinfo {title} {High-fidelity gates and mid-circuit erasure conversion in an atomic qubit},\ }\href {https://doi.org/10.1038/s41586-023-06438-1} {\bibfield  {journal} {\bibinfo  {journal} {Nature}\ }\textbf {\bibinfo {volume} {622}},\ \bibinfo {pages} {279} (\bibinfo {year} {2023})}\BibitemShut {NoStop}%
\bibitem [{\citenamefont {Tsai}\ \emph {et~al.}(2025)\citenamefont {Tsai}, \citenamefont {Sun}, \citenamefont {Shaw}, \citenamefont {Finkelstein},\ and\ \citenamefont {Endres}}]{Tsai2025}%
  \BibitemOpen
  \bibfield  {author} {\bibinfo {author} {\bibfnamefont {R.~B.-S.}\ \bibnamefont {Tsai}}, \bibinfo {author} {\bibfnamefont {X.}~\bibnamefont {Sun}}, \bibinfo {author} {\bibfnamefont {A.~L.}\ \bibnamefont {Shaw}}, \bibinfo {author} {\bibfnamefont {R.}~\bibnamefont {Finkelstein}},\ and\ \bibinfo {author} {\bibfnamefont {M.}~\bibnamefont {Endres}},\ }\bibfield  {title} {\bibinfo {title} {Benchmarking and fidelity response theory of high-fidelity {Rydberg} entangling gates},\ }\href {https://doi.org/10.1103/PRXQuantum.6.010331} {\bibfield  {journal} {\bibinfo  {journal} {PRX Quantum}\ }\textbf {\bibinfo {volume} {6}},\ \bibinfo {pages} {010331} (\bibinfo {year} {2025})}\BibitemShut {NoStop}%
\bibitem [{\citenamefont {Bluvstein}\ \emph {et~al.}(2022)\citenamefont {Bluvstein}, \citenamefont {Levine}, \citenamefont {Semeghini}, \citenamefont {Wang}, \citenamefont {Ebadi}, \citenamefont {Kalinowski}, \citenamefont {Keesling}, \citenamefont {Maskara}, \citenamefont {Pichler}, \citenamefont {Greiner}, \citenamefont {Vuletić},\ and\ \citenamefont {Lukin}}]{Bluvstein2022}%
  \BibitemOpen
  \bibfield  {author} {\bibinfo {author} {\bibfnamefont {D.}~\bibnamefont {Bluvstein}}, \bibinfo {author} {\bibfnamefont {H.}~\bibnamefont {Levine}}, \bibinfo {author} {\bibfnamefont {G.}~\bibnamefont {Semeghini}}, \bibinfo {author} {\bibfnamefont {T.~T.}\ \bibnamefont {Wang}}, \bibinfo {author} {\bibfnamefont {S.}~\bibnamefont {Ebadi}}, \bibinfo {author} {\bibfnamefont {M.}~\bibnamefont {Kalinowski}}, \bibinfo {author} {\bibfnamefont {A.}~\bibnamefont {Keesling}}, \bibinfo {author} {\bibfnamefont {N.}~\bibnamefont {Maskara}}, \bibinfo {author} {\bibfnamefont {H.}~\bibnamefont {Pichler}}, \bibinfo {author} {\bibfnamefont {M.}~\bibnamefont {Greiner}}, \bibinfo {author} {\bibfnamefont {V.}~\bibnamefont {Vuletić}},\ and\ \bibinfo {author} {\bibfnamefont {M.~D.}\ \bibnamefont {Lukin}},\ }\bibfield  {title} {\bibinfo {title} {A quantum processor based on coherent transport of entangled atom arrays},\ }\href {https://doi.org/10.1038/s41586-022-04592-6} {\bibfield  {journal} {\bibinfo  {journal} {Nature}\ }\textbf
  {\bibinfo {volume} {604}},\ \bibinfo {pages} {451} (\bibinfo {year} {2022})}\BibitemShut {NoStop}%
\bibitem [{\citenamefont {Madjarov}(2021)}]{Madjarovthesis}%
  \BibitemOpen
  \bibfield  {author} {\bibinfo {author} {\bibfnamefont {I.~S.}\ \bibnamefont {Madjarov}},\ }\bibfield  {title} {\bibinfo {title} {Ph.{D}. thesis},\ }\href {https://resolver.caltech.edu/CaltechTHESIS:01292021-001639979} {\bibfield  {journal} {\bibinfo  {journal} {California Institute of Technology}\ } (\bibinfo {year} {2021})}\BibitemShut {NoStop}%
\bibitem [{\citenamefont {Madjarov}\ \emph {et~al.}(2019)\citenamefont {Madjarov}, \citenamefont {Cooper}, \citenamefont {Shaw}, \citenamefont {Covey}, \citenamefont {Schkolnik}, \citenamefont {Yoon}, \citenamefont {Williams},\ and\ \citenamefont {Endres}}]{Madjarov2019}%
  \BibitemOpen
  \bibfield  {author} {\bibinfo {author} {\bibfnamefont {I.~S.}\ \bibnamefont {Madjarov}}, \bibinfo {author} {\bibfnamefont {A.}~\bibnamefont {Cooper}}, \bibinfo {author} {\bibfnamefont {A.~L.}\ \bibnamefont {Shaw}}, \bibinfo {author} {\bibfnamefont {J.~P.}\ \bibnamefont {Covey}}, \bibinfo {author} {\bibfnamefont {V.}~\bibnamefont {Schkolnik}}, \bibinfo {author} {\bibfnamefont {T.~H.}\ \bibnamefont {Yoon}}, \bibinfo {author} {\bibfnamefont {J.~R.}\ \bibnamefont {Williams}},\ and\ \bibinfo {author} {\bibfnamefont {M.}~\bibnamefont {Endres}},\ }\bibfield  {title} {\bibinfo {title} {An atomic-array optical clock with single-atom readout},\ }\href {https://doi.org/10.1103/PhysRevX.9.041052} {\bibfield  {journal} {\bibinfo  {journal} {Phys. Rev. X}\ }\textbf {\bibinfo {volume} {9}},\ \bibinfo {pages} {41052} (\bibinfo {year} {2019})}\BibitemShut {NoStop}%
\bibitem [{\citenamefont {Chin}\ \emph {et~al.}(2017)\citenamefont {Chin}, \citenamefont {Steiner},\ and\ \citenamefont {Kurtsiefer}}]{chin2017polarization}%
  \BibitemOpen
  \bibfield  {author} {\bibinfo {author} {\bibfnamefont {Y.-S.}\ \bibnamefont {Chin}}, \bibinfo {author} {\bibfnamefont {M.}~\bibnamefont {Steiner}},\ and\ \bibinfo {author} {\bibfnamefont {C.}~\bibnamefont {Kurtsiefer}},\ }\bibfield  {title} {\bibinfo {title} {Polarization gradient cooling of single atoms in optical dipole traps},\ }\href {https://doi.org/10.1103/PhysRevA.96.033406} {\bibfield  {journal} {\bibinfo  {journal} {Phys. Rev. A}\ }\textbf {\bibinfo {volume} {96}},\ \bibinfo {pages} {033406} (\bibinfo {year} {2017})}\BibitemShut {NoStop}%
\bibitem [{\citenamefont {Brown}\ \emph {et~al.}(2019)\citenamefont {Brown}, \citenamefont {Thiele}, \citenamefont {Kiehl}, \citenamefont {Hsu},\ and\ \citenamefont {Regal}}]{Brown2019}%
  \BibitemOpen
  \bibfield  {author} {\bibinfo {author} {\bibfnamefont {M.~O.}\ \bibnamefont {Brown}}, \bibinfo {author} {\bibfnamefont {T.}~\bibnamefont {Thiele}}, \bibinfo {author} {\bibfnamefont {C.}~\bibnamefont {Kiehl}}, \bibinfo {author} {\bibfnamefont {T.-W.}\ \bibnamefont {Hsu}},\ and\ \bibinfo {author} {\bibfnamefont {C.~A.}\ \bibnamefont {Regal}},\ }\bibfield  {title} {\bibinfo {title} {Gray-molasses optical-tweezer loading: Controlling collisions for scaling atom-array assembly},\ }\href {https://doi.org/10.1103/PhysRevX.9.011057} {\bibfield  {journal} {\bibinfo  {journal} {Phys. Rev. X}\ }\textbf {\bibinfo {volume} {9}},\ \bibinfo {pages} {11057} (\bibinfo {year} {2019})}\BibitemShut {NoStop}%
\bibitem [{\citenamefont {Cooper}\ \emph {et~al.}(2018)\citenamefont {Cooper}, \citenamefont {Covey}, \citenamefont {Madjarov}, \citenamefont {Porsev}, \citenamefont {Safronova},\ and\ \citenamefont {Endres}}]{Cooper2018}%
  \BibitemOpen
  \bibfield  {author} {\bibinfo {author} {\bibfnamefont {A.}~\bibnamefont {Cooper}}, \bibinfo {author} {\bibfnamefont {J.~P.}\ \bibnamefont {Covey}}, \bibinfo {author} {\bibfnamefont {I.~S.}\ \bibnamefont {Madjarov}}, \bibinfo {author} {\bibfnamefont {S.~G.}\ \bibnamefont {Porsev}}, \bibinfo {author} {\bibfnamefont {M.~S.}\ \bibnamefont {Safronova}},\ and\ \bibinfo {author} {\bibfnamefont {M.}~\bibnamefont {Endres}},\ }\bibfield  {title} {\bibinfo {title} {Alkaline-earth atoms in optical tweezers},\ }\href {https://doi.org/10.1103/PhysRevX.8.041055} {\bibfield  {journal} {\bibinfo  {journal} {Phys. Rev. X}\ }\textbf {\bibinfo {volume} {8}},\ \bibinfo {pages} {41055} (\bibinfo {year} {2018})}\BibitemShut {NoStop}%
\bibitem [{\citenamefont {Urech}\ \emph {et~al.}(2022)\citenamefont {Urech}, \citenamefont {Knottnerus}, \citenamefont {Spreeuw},\ and\ \citenamefont {Schreck}}]{urech2022narrow}%
  \BibitemOpen
  \bibfield  {author} {\bibinfo {author} {\bibfnamefont {A.}~\bibnamefont {Urech}}, \bibinfo {author} {\bibfnamefont {I.~H.~A.}\ \bibnamefont {Knottnerus}}, \bibinfo {author} {\bibfnamefont {R.~J.~C.}\ \bibnamefont {Spreeuw}},\ and\ \bibinfo {author} {\bibfnamefont {F.}~\bibnamefont {Schreck}},\ }\bibfield  {title} {\bibinfo {title} {Narrow-line imaging of single strontium atoms in shallow optical tweezers},\ }\href {https://doi.org/10.1103/PhysRevResearch.4.023245} {\bibfield  {journal} {\bibinfo  {journal} {Phys. Rev. Res.}\ }\textbf {\bibinfo {volume} {4}},\ \bibinfo {pages} {023245} (\bibinfo {year} {2022})}\BibitemShut {NoStop}%
\bibitem [{\citenamefont {H\"olzl}\ \emph {et~al.}(2023)\citenamefont {H\"olzl}, \citenamefont {G\"otzelmann}, \citenamefont {Wirth}, \citenamefont {Safronova}, \citenamefont {Weber},\ and\ \citenamefont {Meinert}}]{holzl2023motional}%
  \BibitemOpen
  \bibfield  {author} {\bibinfo {author} {\bibfnamefont {C.}~\bibnamefont {H\"olzl}}, \bibinfo {author} {\bibfnamefont {A.}~\bibnamefont {G\"otzelmann}}, \bibinfo {author} {\bibfnamefont {M.}~\bibnamefont {Wirth}}, \bibinfo {author} {\bibfnamefont {M.~S.}\ \bibnamefont {Safronova}}, \bibinfo {author} {\bibfnamefont {S.}~\bibnamefont {Weber}},\ and\ \bibinfo {author} {\bibfnamefont {F.}~\bibnamefont {Meinert}},\ }\bibfield  {title} {\bibinfo {title} {Motional ground-state cooling of single atoms in state-dependent optical tweezers},\ }\href {https://doi.org/10.1103/PhysRevResearch.5.033093} {\bibfield  {journal} {\bibinfo  {journal} {Phys. Rev. Res.}\ }\textbf {\bibinfo {volume} {5}},\ \bibinfo {pages} {033093} (\bibinfo {year} {2023})}\BibitemShut {NoStop}%
\bibitem [{\citenamefont {Blodgett}\ \emph {et~al.}(2025)\citenamefont {Blodgett}, \citenamefont {Phatak}, \citenamefont {Chen}, \citenamefont {Peana}, \citenamefont {Pritts},\ and\ \citenamefont {Hood}}]{blodgett2025narrow}%
  \BibitemOpen
  \bibfield  {author} {\bibinfo {author} {\bibfnamefont {K.~N.}\ \bibnamefont {Blodgett}}, \bibinfo {author} {\bibfnamefont {S.~S.}\ \bibnamefont {Phatak}}, \bibinfo {author} {\bibfnamefont {M.~R.}\ \bibnamefont {Chen}}, \bibinfo {author} {\bibfnamefont {D.}~\bibnamefont {Peana}}, \bibinfo {author} {\bibfnamefont {C.}~\bibnamefont {Pritts}},\ and\ \bibinfo {author} {\bibfnamefont {J.~D.}\ \bibnamefont {Hood}},\ }\bibfield  {title} {\bibinfo {title} {Narrow-line electric quadrupole cooling and background-free imaging of a single {Cs} atom with spatially structured light},\ }\href {https://doi.org/10.1103/vr4g-h995} {\bibfield  {journal} {\bibinfo  {journal} {Phys. Rev. A}\ }\textbf {\bibinfo {volume} {112}},\ \bibinfo {pages} {043109} (\bibinfo {year} {2025})}\BibitemShut {NoStop}%
\bibitem [{\citenamefont {Kaufman}\ \emph {et~al.}(2012)\citenamefont {Kaufman}, \citenamefont {Lester},\ and\ \citenamefont {Regal}}]{Kaufman2012}%
  \BibitemOpen
  \bibfield  {author} {\bibinfo {author} {\bibfnamefont {A.~M.}\ \bibnamefont {Kaufman}}, \bibinfo {author} {\bibfnamefont {B.~J.}\ \bibnamefont {Lester}},\ and\ \bibinfo {author} {\bibfnamefont {C.~A.}\ \bibnamefont {Regal}},\ }\bibfield  {title} {\bibinfo {title} {Cooling a single atom in an optical tweezer to its quantum ground state},\ }\href {https://doi.org/10.1103/PhysRevX.2.041014} {\bibfield  {journal} {\bibinfo  {journal} {Phys. Rev. X}\ }\textbf {\bibinfo {volume} {2}},\ \bibinfo {pages} {41014} (\bibinfo {year} {2012})}\BibitemShut {NoStop}%
\bibitem [{\citenamefont {Thompson}\ \emph {et~al.}(2013)\citenamefont {Thompson}, \citenamefont {Tiecke}, \citenamefont {Zibrov}, \citenamefont {Vuletić},\ and\ \citenamefont {Lukin}}]{Thompson2013}%
  \BibitemOpen
  \bibfield  {author} {\bibinfo {author} {\bibfnamefont {J.~D.}\ \bibnamefont {Thompson}}, \bibinfo {author} {\bibfnamefont {T.~G.}\ \bibnamefont {Tiecke}}, \bibinfo {author} {\bibfnamefont {A.~S.}\ \bibnamefont {Zibrov}}, \bibinfo {author} {\bibfnamefont {V.}~\bibnamefont {Vuletić}},\ and\ \bibinfo {author} {\bibfnamefont {M.~D.}\ \bibnamefont {Lukin}},\ }\bibfield  {title} {\bibinfo {title} {Coherence and {Raman} sideband cooling of a single atom in an optical tweezer},\ }\href {https://doi.org/10.1103/PhysRevLett.110.133001} {\bibfield  {journal} {\bibinfo  {journal} {Phys. Rev. Lett.}\ }\textbf {\bibinfo {volume} {110}},\ \bibinfo {pages} {133001} (\bibinfo {year} {2013})}\BibitemShut {NoStop}%
\bibitem [{\citenamefont {Yu}\ \emph {et~al.}(2018)\citenamefont {Yu}, \citenamefont {Hutzler}, \citenamefont {Zhang}, \citenamefont {Liu}, \citenamefont {Hood}, \citenamefont {Rosenband},\ and\ \citenamefont {Ni}}]{yu2018motional}%
  \BibitemOpen
  \bibfield  {author} {\bibinfo {author} {\bibfnamefont {Y.}~\bibnamefont {Yu}}, \bibinfo {author} {\bibfnamefont {N.~R.}\ \bibnamefont {Hutzler}}, \bibinfo {author} {\bibfnamefont {J.~T.}\ \bibnamefont {Zhang}}, \bibinfo {author} {\bibfnamefont {L.~R.}\ \bibnamefont {Liu}}, \bibinfo {author} {\bibfnamefont {J.~D.}\ \bibnamefont {Hood}}, \bibinfo {author} {\bibfnamefont {T.}~\bibnamefont {Rosenband}},\ and\ \bibinfo {author} {\bibfnamefont {K.-K.}\ \bibnamefont {Ni}},\ }\bibfield  {title} {\bibinfo {title} {Motional-ground-state cooling outside the {Lamb}-{Dicke} regime},\ }\href {https://doi.org/10.1103/PhysRevA.97.063423} {\bibfield  {journal} {\bibinfo  {journal} {Phys. Rev. A}\ }\textbf {\bibinfo {volume} {97}},\ \bibinfo {pages} {063423} (\bibinfo {year} {2018})}\BibitemShut {NoStop}%
\bibitem [{\citenamefont {Norcia}\ \emph {et~al.}(2018)\citenamefont {Norcia}, \citenamefont {Young},\ and\ \citenamefont {Kaufman}}]{Norcia2018}%
  \BibitemOpen
  \bibfield  {author} {\bibinfo {author} {\bibfnamefont {M.~A.}\ \bibnamefont {Norcia}}, \bibinfo {author} {\bibfnamefont {A.~W.}\ \bibnamefont {Young}},\ and\ \bibinfo {author} {\bibfnamefont {A.~M.}\ \bibnamefont {Kaufman}},\ }\bibfield  {title} {\bibinfo {title} {Microscopic control and detection of ultracold strontium in optical-tweezer arrays},\ }\href {https://doi.org/10.1103/PhysRevX.8.041054} {\bibfield  {journal} {\bibinfo  {journal} {Phys. Rev. X}\ }\textbf {\bibinfo {volume} {8}},\ \bibinfo {pages} {41054} (\bibinfo {year} {2018})}\BibitemShut {NoStop}%
\bibitem [{\citenamefont {Jenkins}\ \emph {et~al.}(2022)\citenamefont {Jenkins}, \citenamefont {Lis}, \citenamefont {Senoo}, \citenamefont {McGrew},\ and\ \citenamefont {Kaufman}}]{Jenkins2022}%
  \BibitemOpen
  \bibfield  {author} {\bibinfo {author} {\bibfnamefont {A.}~\bibnamefont {Jenkins}}, \bibinfo {author} {\bibfnamefont {J.~W.}\ \bibnamefont {Lis}}, \bibinfo {author} {\bibfnamefont {A.}~\bibnamefont {Senoo}}, \bibinfo {author} {\bibfnamefont {W.~F.}\ \bibnamefont {McGrew}},\ and\ \bibinfo {author} {\bibfnamefont {A.~M.}\ \bibnamefont {Kaufman}},\ }\bibfield  {title} {\bibinfo {title} {Ytterbium nuclear-spin qubits in an optical tweezer array},\ }\href {https://doi.org/10.1103/PhysRevX.12.021027} {\bibfield  {journal} {\bibinfo  {journal} {Phys. Rev. X}\ }\textbf {\bibinfo {volume} {12}},\ \bibinfo {pages} {21027} (\bibinfo {year} {2022})}\BibitemShut {NoStop}%
\bibitem [{\citenamefont {Wineland}\ and\ \citenamefont {Itano}(1979)}]{Wineland1979}%
  \BibitemOpen
  \bibfield  {author} {\bibinfo {author} {\bibfnamefont {D.~J.}\ \bibnamefont {Wineland}}\ and\ \bibinfo {author} {\bibfnamefont {W.~M.}\ \bibnamefont {Itano}},\ }\bibfield  {title} {\bibinfo {title} {Laser cooling of atoms},\ }\href {https://doi.org/10.1103/PhysRevA.20.1521} {\bibfield  {journal} {\bibinfo  {journal} {Phys. Rev. A}\ }\textbf {\bibinfo {volume} {20}},\ \bibinfo {pages} {1521} (\bibinfo {year} {1979})}\BibitemShut {NoStop}%
\bibitem [{\citenamefont {Phatak}\ \emph {et~al.}(2024)\citenamefont {Phatak}, \citenamefont {Blodgett}, \citenamefont {Peana}, \citenamefont {Chen},\ and\ \citenamefont {Hood}}]{phatak2024generalized}%
  \BibitemOpen
  \bibfield  {author} {\bibinfo {author} {\bibfnamefont {S.~S.}\ \bibnamefont {Phatak}}, \bibinfo {author} {\bibfnamefont {K.~N.}\ \bibnamefont {Blodgett}}, \bibinfo {author} {\bibfnamefont {D.}~\bibnamefont {Peana}}, \bibinfo {author} {\bibfnamefont {M.~R.}\ \bibnamefont {Chen}},\ and\ \bibinfo {author} {\bibfnamefont {J.~D.}\ \bibnamefont {Hood}},\ }\bibfield  {title} {\bibinfo {title} {Generalized theory for optical cooling of a trapped atom with spin},\ }\href {https://doi.org/10.1103/PhysRevA.110.043116} {\bibfield  {journal} {\bibinfo  {journal} {Phys. Rev. A}\ }\textbf {\bibinfo {volume} {110}},\ \bibinfo {pages} {043116} (\bibinfo {year} {2024})}\BibitemShut {NoStop}%
\bibitem [{\citenamefont {Shaw}\ \emph {et~al.}(2023)\citenamefont {Shaw}, \citenamefont {Scholl}, \citenamefont {Finklestein}, \citenamefont {Madjarov}, \citenamefont {Grinkemeyer},\ and\ \citenamefont {Endres}}]{Shaw2023A}%
  \BibitemOpen
  \bibfield  {author} {\bibinfo {author} {\bibfnamefont {A.~L.}\ \bibnamefont {Shaw}}, \bibinfo {author} {\bibfnamefont {P.}~\bibnamefont {Scholl}}, \bibinfo {author} {\bibfnamefont {R.}~\bibnamefont {Finklestein}}, \bibinfo {author} {\bibfnamefont {I.~S.}\ \bibnamefont {Madjarov}}, \bibinfo {author} {\bibfnamefont {B.}~\bibnamefont {Grinkemeyer}},\ and\ \bibinfo {author} {\bibfnamefont {M.}~\bibnamefont {Endres}},\ }\bibfield  {title} {\bibinfo {title} {Dark-state enhanced loading of an optical tweezer array},\ }\href {https://doi.org/10.1103/PhysRevLett.130.193402} {\bibfield  {journal} {\bibinfo  {journal} {Phys. Rev. Lett.}\ }\textbf {\bibinfo {volume} {130}},\ \bibinfo {pages} {193402} (\bibinfo {year} {2023})}\BibitemShut {NoStop}%
\bibitem [{\citenamefont {Belyansky}\ \emph {et~al.}(2019)\citenamefont {Belyansky}, \citenamefont {Young}, \citenamefont {Bienias}, \citenamefont {Eldredge}, \citenamefont {Kaufman}, \citenamefont {Zoller},\ and\ \citenamefont {Gorshkov}}]{Belyansky2019}%
  \BibitemOpen
  \bibfield  {author} {\bibinfo {author} {\bibfnamefont {R.}~\bibnamefont {Belyansky}}, \bibinfo {author} {\bibfnamefont {J.~T.}\ \bibnamefont {Young}}, \bibinfo {author} {\bibfnamefont {P.}~\bibnamefont {Bienias}}, \bibinfo {author} {\bibfnamefont {Z.}~\bibnamefont {Eldredge}}, \bibinfo {author} {\bibfnamefont {A.~M.}\ \bibnamefont {Kaufman}}, \bibinfo {author} {\bibfnamefont {P.}~\bibnamefont {Zoller}},\ and\ \bibinfo {author} {\bibfnamefont {A.~V.}\ \bibnamefont {Gorshkov}},\ }\bibfield  {title} {\bibinfo {title} {Nondestructive cooling of an atomic quantum register via state-insensitive {Rydberg} interactions},\ }\href {https://doi.org/10.1103/PhysRevLett.123.213603} {\bibfield  {journal} {\bibinfo  {journal} {Phys. Rev. Lett.}\ }\textbf {\bibinfo {volume} {123}},\ \bibinfo {pages} {213603} (\bibinfo {year} {2019})}\BibitemShut {NoStop}%
\bibitem [{\citenamefont {Monroe}\ \emph {et~al.}(1995)\citenamefont {Monroe}, \citenamefont {Meekhof}, \citenamefont {King}, \citenamefont {Jefferts}, \citenamefont {Itano}, \citenamefont {Wineland},\ and\ \citenamefont {Gould}}]{Monroe1995}%
  \BibitemOpen
  \bibfield  {author} {\bibinfo {author} {\bibfnamefont {C.}~\bibnamefont {Monroe}}, \bibinfo {author} {\bibfnamefont {D.~M.}\ \bibnamefont {Meekhof}}, \bibinfo {author} {\bibfnamefont {B.~E.}\ \bibnamefont {King}}, \bibinfo {author} {\bibfnamefont {S.~R.}\ \bibnamefont {Jefferts}}, \bibinfo {author} {\bibfnamefont {W.~M.}\ \bibnamefont {Itano}}, \bibinfo {author} {\bibfnamefont {D.~J.}\ \bibnamefont {Wineland}},\ and\ \bibinfo {author} {\bibfnamefont {P.}~\bibnamefont {Gould}},\ }\bibfield  {title} {\bibinfo {title} {Resolved-sideband {Raman} cooling of a bound atom to the {3D} zero-point energy},\ }\href {https://doi.org/10.1103/PhysRevLett.75.4011} {\bibfield  {journal} {\bibinfo  {journal} {Phys. Rev. Lett.}\ }\textbf {\bibinfo {volume} {75}},\ \bibinfo {pages} {4011} (\bibinfo {year} {1995})}\BibitemShut {NoStop}%
\bibitem [{\citenamefont {Gyger}\ \emph {et~al.}(2024)\citenamefont {Gyger}, \citenamefont {Ammenwerth}, \citenamefont {Tao}, \citenamefont {Timme}, \citenamefont {Snigirev}, \citenamefont {Bloch},\ and\ \citenamefont {Zeiher}}]{Gyger2024}%
  \BibitemOpen
  \bibfield  {author} {\bibinfo {author} {\bibfnamefont {F.}~\bibnamefont {Gyger}}, \bibinfo {author} {\bibfnamefont {M.}~\bibnamefont {Ammenwerth}}, \bibinfo {author} {\bibfnamefont {R.}~\bibnamefont {Tao}}, \bibinfo {author} {\bibfnamefont {H.}~\bibnamefont {Timme}}, \bibinfo {author} {\bibfnamefont {S.}~\bibnamefont {Snigirev}}, \bibinfo {author} {\bibfnamefont {I.}~\bibnamefont {Bloch}},\ and\ \bibinfo {author} {\bibfnamefont {J.}~\bibnamefont {Zeiher}},\ }\bibfield  {title} {\bibinfo {title} {Continuous operation of large-scale atom arrays in optical lattices},\ }\href {https://doi.org/10.1103/PhysRevResearch.6.033104} {\bibfield  {journal} {\bibinfo  {journal} {Phys. Rev. Res.}\ }\textbf {\bibinfo {volume} {6}},\ \bibinfo {pages} {033104} (\bibinfo {year} {2024})}\BibitemShut {NoStop}%
\bibitem [{\citenamefont {Li}\ \emph {et~al.}(2025)\citenamefont {Li}, \citenamefont {Bao}, \citenamefont {Peper}, \citenamefont {Li},\ and\ \citenamefont {Thompson}}]{Li2025}%
  \BibitemOpen
  \bibfield  {author} {\bibinfo {author} {\bibfnamefont {Y.}~\bibnamefont {Li}}, \bibinfo {author} {\bibfnamefont {Y.}~\bibnamefont {Bao}}, \bibinfo {author} {\bibfnamefont {M.}~\bibnamefont {Peper}}, \bibinfo {author} {\bibfnamefont {C.}~\bibnamefont {Li}},\ and\ \bibinfo {author} {\bibfnamefont {J.~D.}\ \bibnamefont {Thompson}},\ }\bibfield  {title} {\bibinfo {title} {Fast, continuous and coherent atom replacement in a neutral atom qubit array},\ }\href {https://arxiv.org/abs/2506.15633} {\bibfield  {journal} {\bibinfo  {journal} {arXiv:2506.15633}\ } (\bibinfo {year} {2025})}\BibitemShut {NoStop}%
\bibitem [{\citenamefont {Chiu}\ \emph {et~al.}(2025)\citenamefont {Chiu}, \citenamefont {Trapp}, \citenamefont {Guo}, \citenamefont {Abobeih}, \citenamefont {Stewart}, \citenamefont {Hollerith}, \citenamefont {Stroganov}, \citenamefont {Kalinowski}, \citenamefont {Geim}, \citenamefont {Evered}, \citenamefont {Li}, \citenamefont {Lyu}, \citenamefont {Peters}, \citenamefont {Bluvstein}, \citenamefont {Wang}, \citenamefont {Greiner}, \citenamefont {Vuleti{\'{c}}},\ and\ \citenamefont {Lukin}}]{Chiu2025}%
  \BibitemOpen
  \bibfield  {author} {\bibinfo {author} {\bibfnamefont {N.-C.}\ \bibnamefont {Chiu}}, \bibinfo {author} {\bibfnamefont {E.~C.}\ \bibnamefont {Trapp}}, \bibinfo {author} {\bibfnamefont {J.}~\bibnamefont {Guo}}, \bibinfo {author} {\bibfnamefont {M.~H.}\ \bibnamefont {Abobeih}}, \bibinfo {author} {\bibfnamefont {L.~M.}\ \bibnamefont {Stewart}}, \bibinfo {author} {\bibfnamefont {S.}~\bibnamefont {Hollerith}}, \bibinfo {author} {\bibfnamefont {P.~L.}\ \bibnamefont {Stroganov}}, \bibinfo {author} {\bibfnamefont {M.}~\bibnamefont {Kalinowski}}, \bibinfo {author} {\bibfnamefont {A.~A.}\ \bibnamefont {Geim}}, \bibinfo {author} {\bibfnamefont {S.~J.}\ \bibnamefont {Evered}}, \bibinfo {author} {\bibfnamefont {S.~H.}\ \bibnamefont {Li}}, \bibinfo {author} {\bibfnamefont {X.}~\bibnamefont {Lyu}}, \bibinfo {author} {\bibfnamefont {L.~M.}\ \bibnamefont {Peters}}, \bibinfo {author} {\bibfnamefont {D.}~\bibnamefont {Bluvstein}}, \bibinfo {author} {\bibfnamefont {T.~T.}\ \bibnamefont {Wang}}, \bibinfo {author} {\bibfnamefont
  {M.}~\bibnamefont {Greiner}}, \bibinfo {author} {\bibfnamefont {V.}~\bibnamefont {Vuleti{\'{c}}}},\ and\ \bibinfo {author} {\bibfnamefont {M.~D.}\ \bibnamefont {Lukin}},\ }\bibfield  {title} {\bibinfo {title} {Continuous operation of a coherent 3,000-qubit system},\ }\href {https://doi.org/10.1038/s41586-025-09596-6} {\bibfield  {journal} {\bibinfo  {journal} {Nature}\ }\textbf {\bibinfo {volume} {646}},\ \bibinfo {pages} {1075} (\bibinfo {year} {2025})}\BibitemShut {NoStop}%
\bibitem [{\citenamefont {Hu}\ \emph {et~al.}(2025)\citenamefont {Hu}, \citenamefont {Sinclair}, \citenamefont {Bytyqi}, \citenamefont {Chong}, \citenamefont {Rudelis}, \citenamefont {Ramette}, \citenamefont {Vendeiro},\ and\ \citenamefont {Vuleti\ifmmode~\acute{c}\else \'{c}\fi{}}}]{Hu2025}%
  \BibitemOpen
  \bibfield  {author} {\bibinfo {author} {\bibfnamefont {B.}~\bibnamefont {Hu}}, \bibinfo {author} {\bibfnamefont {J.}~\bibnamefont {Sinclair}}, \bibinfo {author} {\bibfnamefont {E.}~\bibnamefont {Bytyqi}}, \bibinfo {author} {\bibfnamefont {M.}~\bibnamefont {Chong}}, \bibinfo {author} {\bibfnamefont {A.}~\bibnamefont {Rudelis}}, \bibinfo {author} {\bibfnamefont {J.}~\bibnamefont {Ramette}}, \bibinfo {author} {\bibfnamefont {Z.}~\bibnamefont {Vendeiro}},\ and\ \bibinfo {author} {\bibfnamefont {V.}~\bibnamefont {Vuleti\ifmmode~\acute{c}\else \'{c}\fi{}}},\ }\bibfield  {title} {\bibinfo {title} {Site-selective cavity readout and classical error correction of a 5-bit atomic register},\ }\href {https://doi.org/10.1103/PhysRevLett.134.120801} {\bibfield  {journal} {\bibinfo  {journal} {Phys. Rev. Lett.}\ }\textbf {\bibinfo {volume} {134}},\ \bibinfo {pages} {120801} (\bibinfo {year} {2025})}\BibitemShut {NoStop}%
\bibitem [{\citenamefont {Zhao}\ \emph {et~al.}(2024)\citenamefont {Zhao}, \citenamefont {Xu}, \citenamefont {Shi}, \citenamefont {Chen}, \citenamefont {Kong}, \citenamefont {Yang}, \citenamefont {Wang}, \citenamefont {Ye}, \citenamefont {Yu}, \citenamefont {Wang}, \citenamefont {Xie}, \citenamefont {Shi},\ and\ \citenamefont {Du}}]{Zhao2024_repetitive}%
  \BibitemOpen
  \bibfield  {author} {\bibinfo {author} {\bibfnamefont {Z.}~\bibnamefont {Zhao}}, \bibinfo {author} {\bibfnamefont {S.}~\bibnamefont {Xu}}, \bibinfo {author} {\bibfnamefont {Q.}~\bibnamefont {Shi}}, \bibinfo {author} {\bibfnamefont {Y.}~\bibnamefont {Chen}}, \bibinfo {author} {\bibfnamefont {X.}~\bibnamefont {Kong}}, \bibinfo {author} {\bibfnamefont {Z.}~\bibnamefont {Yang}}, \bibinfo {author} {\bibfnamefont {M.}~\bibnamefont {Wang}}, \bibinfo {author} {\bibfnamefont {X.}~\bibnamefont {Ye}}, \bibinfo {author} {\bibfnamefont {P.}~\bibnamefont {Yu}}, \bibinfo {author} {\bibfnamefont {Y.}~\bibnamefont {Wang}}, \bibinfo {author} {\bibfnamefont {T.}~\bibnamefont {Xie}}, \bibinfo {author} {\bibfnamefont {F.}~\bibnamefont {Shi}},\ and\ \bibinfo {author} {\bibfnamefont {J.}~\bibnamefont {Du}},\ }\bibfield  {title} {\bibinfo {title} {Optimal repetitive readout of single solid-state spins determined by {Fisher} information},\ }\href {https://doi.org/10.1126/sciadv.adp9228} {\bibfield  {journal} {\bibinfo  {journal}
  {Sci. Adv.}\ }\textbf {\bibinfo {volume} {10}},\ \bibinfo {pages} {eadp9228} (\bibinfo {year} {2024})}\BibitemShut {NoStop}%
\bibitem [{\citenamefont {Norcia}\ \emph {et~al.}(2019)\citenamefont {Norcia}, \citenamefont {Young}, \citenamefont {Eckner}, \citenamefont {Oelker}, \citenamefont {Ye},\ and\ \citenamefont {Kaufman}}]{Norcia2019}%
  \BibitemOpen
  \bibfield  {author} {\bibinfo {author} {\bibfnamefont {M.~A.}\ \bibnamefont {Norcia}}, \bibinfo {author} {\bibfnamefont {A.~W.}\ \bibnamefont {Young}}, \bibinfo {author} {\bibfnamefont {W.~J.}\ \bibnamefont {Eckner}}, \bibinfo {author} {\bibfnamefont {E.}~\bibnamefont {Oelker}}, \bibinfo {author} {\bibfnamefont {J.}~\bibnamefont {Ye}},\ and\ \bibinfo {author} {\bibfnamefont {A.~M.}\ \bibnamefont {Kaufman}},\ }\bibfield  {title} {\bibinfo {title} {Seconds-scale coherence on an optical clock transition in a tweezer array},\ }\href {https://doi.org/10.1126/science.aay0644} {\bibfield  {journal} {\bibinfo  {journal} {Science}\ }\textbf {\bibinfo {volume} {366}},\ \bibinfo {pages} {93} (\bibinfo {year} {2019})}\BibitemShut {NoStop}%
\bibitem [{\citenamefont {Katori}(2021)}]{Katori_2021}%
  \BibitemOpen
  \bibfield  {author} {\bibinfo {author} {\bibfnamefont {H.}~\bibnamefont {Katori}},\ }\bibfield  {title} {\bibinfo {title} {Longitudinal {Ramsey} spectroscopy of atoms for continuous operation of optical clocks},\ }\href {https://doi.org/10.35848/1882-0786/ac0e16} {\bibfield  {journal} {\bibinfo  {journal} {Applied Physics Express}\ }\textbf {\bibinfo {volume} {14}},\ \bibinfo {pages} {072006} (\bibinfo {year} {2021})}\BibitemShut {NoStop}%
\bibitem [{\citenamefont {Liu}\ \emph {et~al.}(2025)\citenamefont {Liu}, \citenamefont {Liu}, \citenamefont {Li}, \citenamefont {Zhang}, \citenamefont {Wang}, \citenamefont {Jia}, \citenamefont {Zhang}, \citenamefont {Zhu}, \citenamefont {Kong}, \citenamefont {Song}, \citenamefont {Niu}, \citenamefont {Yang}, \citenamefont {Feng}, \citenamefont {Liu}, \citenamefont {Cui}, \citenamefont {Xu}, \citenamefont {Jiang}, \citenamefont {Yin}, \citenamefont {Liao}, \citenamefont {Peng}, \citenamefont {Dai}, \citenamefont {Chen},\ and\ \citenamefont {Pan}}]{Liu2025_zerodeadtimeclock}%
  \BibitemOpen
  \bibfield  {author} {\bibinfo {author} {\bibfnamefont {X.-Y.}\ \bibnamefont {Liu}}, \bibinfo {author} {\bibfnamefont {P.}~\bibnamefont {Liu}}, \bibinfo {author} {\bibfnamefont {J.}~\bibnamefont {Li}}, \bibinfo {author} {\bibfnamefont {Y.-C.}\ \bibnamefont {Zhang}}, \bibinfo {author} {\bibfnamefont {Y.-B.}\ \bibnamefont {Wang}}, \bibinfo {author} {\bibfnamefont {Z.-P.}\ \bibnamefont {Jia}}, \bibinfo {author} {\bibfnamefont {X.}~\bibnamefont {Zhang}}, \bibinfo {author} {\bibfnamefont {X.-Q.}\ \bibnamefont {Zhu}}, \bibinfo {author} {\bibfnamefont {D.-Q.}\ \bibnamefont {Kong}}, \bibinfo {author} {\bibfnamefont {W.-L.}\ \bibnamefont {Song}}, \bibinfo {author} {\bibfnamefont {G.-Z.}\ \bibnamefont {Niu}}, \bibinfo {author} {\bibfnamefont {Y.-M.}\ \bibnamefont {Yang}}, \bibinfo {author} {\bibfnamefont {P.-J.}\ \bibnamefont {Feng}}, \bibinfo {author} {\bibfnamefont {X.-P.}\ \bibnamefont {Liu}}, \bibinfo {author} {\bibfnamefont {X.-Y.}\ \bibnamefont {Cui}}, \bibinfo {author} {\bibfnamefont {P.}~\bibnamefont {Xu}},
  \bibinfo {author} {\bibfnamefont {X.}~\bibnamefont {Jiang}}, \bibinfo {author} {\bibfnamefont {J.}~\bibnamefont {Yin}}, \bibinfo {author} {\bibfnamefont {S.-K.}\ \bibnamefont {Liao}}, \bibinfo {author} {\bibfnamefont {C.-Z.}\ \bibnamefont {Peng}}, \bibinfo {author} {\bibfnamefont {H.-N.}\ \bibnamefont {Dai}}, \bibinfo {author} {\bibfnamefont {Y.-A.}\ \bibnamefont {Chen}},\ and\ \bibinfo {author} {\bibfnamefont {J.-W.}\ \bibnamefont {Pan}},\ }\bibfield  {title} {\bibinfo {title} {Zero-dead-time strontium lattice clock with a stability at ${10}^{\ensuremath{-}19}$ level},\ }\href {https://doi.org/10.1103/zbpb-6qxb} {\bibfield  {journal} {\bibinfo  {journal} {Phys. Rev. Lett.}\ }\textbf {\bibinfo {volume} {135}},\ \bibinfo {pages} {263402} (\bibinfo {year} {2025})}\BibitemShut {NoStop}%
\bibitem [{\citenamefont {Allen}\ \emph {et~al.}(2025)\citenamefont {Allen}, \citenamefont {Machado}, \citenamefont {Chuang}, \citenamefont {Huang},\ and\ \citenamefont {Choi}}]{allen2025quantumcomputingenhancedsensing}%
  \BibitemOpen
  \bibfield  {author} {\bibinfo {author} {\bibfnamefont {R.~R.}\ \bibnamefont {Allen}}, \bibinfo {author} {\bibfnamefont {F.}~\bibnamefont {Machado}}, \bibinfo {author} {\bibfnamefont {I.~L.}\ \bibnamefont {Chuang}}, \bibinfo {author} {\bibfnamefont {H.-Y.}\ \bibnamefont {Huang}},\ and\ \bibinfo {author} {\bibfnamefont {S.}~\bibnamefont {Choi}},\ }\bibfield  {title} {\bibinfo {title} {Quantum computing enhanced sensing},\ }\href {https://arxiv.org/abs/2501.07625} {\bibfield  {journal} {\bibinfo  {journal} {arXiv:2501.07625}\ } (\bibinfo {year} {2025})}\BibitemShut {NoStop}%
\bibitem [{\citenamefont {Direkci}\ \emph {et~al.}(2026)\citenamefont {Direkci}, \citenamefont {Finkelstein}, \citenamefont {Endres},\ and\ \citenamefont {Gefen}}]{Direkci2026}%
  \BibitemOpen
  \bibfield  {author} {\bibinfo {author} {\bibfnamefont {S.}~\bibnamefont {Direkci}}, \bibinfo {author} {\bibfnamefont {R.}~\bibnamefont {Finkelstein}}, \bibinfo {author} {\bibfnamefont {M.}~\bibnamefont {Endres}},\ and\ \bibinfo {author} {\bibfnamefont {T.}~\bibnamefont {Gefen}},\ }\bibfield  {title} {\bibinfo {title} {Heisenberg-limited bayesian phase estimation with low-depth digital quantum circuits},\ }\href {https://doi.org/10.1038/s41534-025-01177-9} {\bibfield  {journal} {\bibinfo  {journal} {npj Quantum Information}\ }\textbf {\bibinfo {volume} {12}},\ \bibinfo {pages} {31} (\bibinfo {year} {2026})}\BibitemShut {NoStop}%
\bibitem [{\citenamefont {Dudinets}\ \emph {et~al.}(2025)\citenamefont {Dudinets}, \citenamefont {Straupe}, \citenamefont {Fedorov},\ and\ \citenamefont {Lychkovskiy}}]{dudinets2025}%
  \BibitemOpen
  \bibfield  {author} {\bibinfo {author} {\bibfnamefont {I.~V.}\ \bibnamefont {Dudinets}}, \bibinfo {author} {\bibfnamefont {S.~S.}\ \bibnamefont {Straupe}}, \bibinfo {author} {\bibfnamefont {A.~K.}\ \bibnamefont {Fedorov}},\ and\ \bibinfo {author} {\bibfnamefont {O.~V.}\ \bibnamefont {Lychkovskiy}},\ }\bibfield  {title} {\bibinfo {title} {All-to-all connectivity of {Rydberg}-atom-based quantum processors with messenger qubits},\ }\href {https://arxiv.org/abs/2504.05087} {\bibfield  {journal} {\bibinfo  {journal} {arXiv:2504.05087}\ } (\bibinfo {year} {2025})}\BibitemShut {NoStop}%
\end{thebibliography}%


\begin{thebibliography}{12}%
\makeatletter
\providecommand \@ifxundefined [1]{%
 \@ifx{#1\undefined}
}%
\providecommand \@ifnum [1]{%
 \ifnum #1\expandafter \@firstoftwo
 \else \expandafter \@secondoftwo
 \fi
}%
\providecommand \@ifx [1]{%
 \ifx #1\expandafter \@firstoftwo
 \else \expandafter \@secondoftwo
 \fi
}%
\providecommand \natexlab [1]{#1}%
\providecommand \enquote  [1]{``#1''}%
\providecommand \bibnamefont  [1]{#1}%
\providecommand \bibfnamefont [1]{#1}%
\providecommand \citenamefont [1]{#1}%
\providecommand \href@noop [0]{\@secondoftwo}%
\providecommand \href [0]{\begingroup \@sanitize@url \@href}%
\providecommand \@href[1]{\@@startlink{#1}\@@href}%
\providecommand \@@href[1]{\endgroup#1\@@endlink}%
\providecommand \@sanitize@url [0]{\catcode `\\12\catcode `\$12\catcode `\&12\catcode `\#12\catcode `\^12\catcode `\_12\catcode `\%12\relax}%
\providecommand \@@startlink[1]{}%
\providecommand \@@endlink[0]{}%
\providecommand \url  [0]{\begingroup\@sanitize@url \@url }%
\providecommand \@url [1]{\endgroup\@href {#1}{\urlprefix }}%
\providecommand \urlprefix  [0]{URL }%
\providecommand \Eprint [0]{\href }%
\providecommand \doibase [0]{https://doi.org/}%
\providecommand \selectlanguage [0]{\@gobble}%
\providecommand \bibinfo  [0]{\@secondoftwo}%
\providecommand \bibfield  [0]{\@secondoftwo}%
\providecommand \translation [1]{[#1]}%
\providecommand \BibitemOpen [0]{}%
\providecommand \bibitemStop [0]{}%
\providecommand \bibitemNoStop [0]{.\EOS\space}%
\providecommand \EOS [0]{\spacefactor3000\relax}%
\providecommand \BibitemShut  [1]{\csname bibitem#1\endcsname}%
\let\auto@bib@innerbib\@empty
\bibitem [{\citenamefont {Scholl}\ \emph {et~al.}(2023)\citenamefont {Scholl}, \citenamefont {Shaw}, \citenamefont {Tsai}, \citenamefont {Finkelstein}, \citenamefont {Choi},\ and\ \citenamefont {Endres}}]{Scholl2023A}%
  \BibitemOpen
  \bibfield  {author} {\bibinfo {author} {\bibfnamefont {P.}~\bibnamefont {Scholl}}, \bibinfo {author} {\bibfnamefont {A.~L.}\ \bibnamefont {Shaw}}, \bibinfo {author} {\bibfnamefont {R.~B.-S.}\ \bibnamefont {Tsai}}, \bibinfo {author} {\bibfnamefont {R.}~\bibnamefont {Finkelstein}}, \bibinfo {author} {\bibfnamefont {J.}~\bibnamefont {Choi}},\ and\ \bibinfo {author} {\bibfnamefont {M.}~\bibnamefont {Endres}},\ }\bibfield  {title} {\bibinfo {title} {Erasure conversion in a high-fidelity {Rydberg} quantum simulator},\ }\href {https://doi.org/10.1038/s41586-023-06516-4} {\bibfield  {journal} {\bibinfo  {journal} {Nature}\ }\textbf {\bibinfo {volume} {622}},\ \bibinfo {pages} {273} (\bibinfo {year} {2023})}\BibitemShut {NoStop}%
\bibitem [{\citenamefont {Shaw}\ \emph {et~al.}(2024)\citenamefont {Shaw}, \citenamefont {Finkelstein}, \citenamefont {Tsai}, \citenamefont {Scholl}, \citenamefont {Yoon}, \citenamefont {Choi},\ and\ \citenamefont {Endres}}]{Shaw2024}%
  \BibitemOpen
  \bibfield  {author} {\bibinfo {author} {\bibfnamefont {A.~L.}\ \bibnamefont {Shaw}}, \bibinfo {author} {\bibfnamefont {R.}~\bibnamefont {Finkelstein}}, \bibinfo {author} {\bibfnamefont {R.~B.-S.}\ \bibnamefont {Tsai}}, \bibinfo {author} {\bibfnamefont {P.}~\bibnamefont {Scholl}}, \bibinfo {author} {\bibfnamefont {T.~H.}\ \bibnamefont {Yoon}}, \bibinfo {author} {\bibfnamefont {J.}~\bibnamefont {Choi}},\ and\ \bibinfo {author} {\bibfnamefont {M.}~\bibnamefont {Endres}},\ }\bibfield  {title} {\bibinfo {title} {Multi-ensemble metrology by programming local rotations with atom movements},\ }\href {https://doi.org/10.1038/s41567-023-02323-w} {\bibfield  {journal} {\bibinfo  {journal} {Nat. Phys.}\ }\textbf {\bibinfo {volume} {20}},\ \bibinfo {pages} {195} (\bibinfo {year} {2024})}\BibitemShut {NoStop}%
\bibitem [{\citenamefont {Lis}\ \emph {et~al.}(2023)\citenamefont {Lis}, \citenamefont {Senoo}, \citenamefont {McGrew}, \citenamefont {R\"onchen}, \citenamefont {Jenkins},\ and\ \citenamefont {Kaufman}}]{Lis2023A}%
  \BibitemOpen
  \bibfield  {author} {\bibinfo {author} {\bibfnamefont {J.~W.}\ \bibnamefont {Lis}}, \bibinfo {author} {\bibfnamefont {A.}~\bibnamefont {Senoo}}, \bibinfo {author} {\bibfnamefont {W.~F.}\ \bibnamefont {McGrew}}, \bibinfo {author} {\bibfnamefont {F.}~\bibnamefont {R\"onchen}}, \bibinfo {author} {\bibfnamefont {A.}~\bibnamefont {Jenkins}},\ and\ \bibinfo {author} {\bibfnamefont {A.~M.}\ \bibnamefont {Kaufman}},\ }\bibfield  {title} {\bibinfo {title} {Midcircuit operations using the \textit{omg} architecture in neutral atom arrays},\ }\href {https://doi.org/10.1103/PhysRevX.13.041035} {\bibfield  {journal} {\bibinfo  {journal} {Phys. Rev. X}\ }\textbf {\bibinfo {volume} {13}},\ \bibinfo {pages} {041035} (\bibinfo {year} {2023})}\BibitemShut {NoStop}%
\bibitem [{\citenamefont {Ma}\ \emph {et~al.}(2022)\citenamefont {Ma}, \citenamefont {Burgers}, \citenamefont {Liu}, \citenamefont {Wilson}, \citenamefont {Zhang},\ and\ \citenamefont {Thompson}}]{Ma2022_Yb}%
  \BibitemOpen
  \bibfield  {author} {\bibinfo {author} {\bibfnamefont {S.}~\bibnamefont {Ma}}, \bibinfo {author} {\bibfnamefont {A.~P.}\ \bibnamefont {Burgers}}, \bibinfo {author} {\bibfnamefont {G.}~\bibnamefont {Liu}}, \bibinfo {author} {\bibfnamefont {J.}~\bibnamefont {Wilson}}, \bibinfo {author} {\bibfnamefont {B.}~\bibnamefont {Zhang}},\ and\ \bibinfo {author} {\bibfnamefont {J.~D.}\ \bibnamefont {Thompson}},\ }\bibfield  {title} {\bibinfo {title} {Universal gate operations on nuclear spin qubits in an optical tweezer array of $^{171}\mathrm{Yb}$ atoms},\ }\href {https://doi.org/10.1103/PhysRevX.12.021028} {\bibfield  {journal} {\bibinfo  {journal} {Phys. Rev. X}\ }\textbf {\bibinfo {volume} {12}},\ \bibinfo {pages} {021028} (\bibinfo {year} {2022})}\BibitemShut {NoStop}%
\bibitem [{\citenamefont {Chen}\ \emph {et~al.}(2022)\citenamefont {Chen}, \citenamefont {Li}, \citenamefont {Huie}, \citenamefont {Zhao}, \citenamefont {Vetter}, \citenamefont {Greene},\ and\ \citenamefont {Covey}}]{Chen2022}%
  \BibitemOpen
  \bibfield  {author} {\bibinfo {author} {\bibfnamefont {N.}~\bibnamefont {Chen}}, \bibinfo {author} {\bibfnamefont {L.}~\bibnamefont {Li}}, \bibinfo {author} {\bibfnamefont {W.}~\bibnamefont {Huie}}, \bibinfo {author} {\bibfnamefont {M.}~\bibnamefont {Zhao}}, \bibinfo {author} {\bibfnamefont {I.}~\bibnamefont {Vetter}}, \bibinfo {author} {\bibfnamefont {C.~H.}\ \bibnamefont {Greene}},\ and\ \bibinfo {author} {\bibfnamefont {J.~P.}\ \bibnamefont {Covey}},\ }\bibfield  {title} {\bibinfo {title} {Analyzing the {Rydberg}-based optical-metastable-ground architecture for $^{171}\mathrm{Yb}$ nuclear spins},\ }\href {https://doi.org/10.1103/PhysRevA.105.052438} {\bibfield  {journal} {\bibinfo  {journal} {Phys. Rev. A}\ }\textbf {\bibinfo {volume} {105}},\ \bibinfo {pages} {52438} (\bibinfo {year} {2022})}\BibitemShut {NoStop}%
\bibitem [{\citenamefont {Shaw}\ \emph {et~al.}(2025)\citenamefont {Shaw}, \citenamefont {Scholl}, \citenamefont {Finkelstein}, \citenamefont {Tsai}, \citenamefont {Choi},\ and\ \citenamefont {Endres}}]{Shaw2025_erasure_cooling}%
  \BibitemOpen
  \bibfield  {author} {\bibinfo {author} {\bibfnamefont {A.~L.}\ \bibnamefont {Shaw}}, \bibinfo {author} {\bibfnamefont {P.}~\bibnamefont {Scholl}}, \bibinfo {author} {\bibfnamefont {R.}~\bibnamefont {Finkelstein}}, \bibinfo {author} {\bibfnamefont {R.~B.-S.}\ \bibnamefont {Tsai}}, \bibinfo {author} {\bibfnamefont {J.}~\bibnamefont {Choi}},\ and\ \bibinfo {author} {\bibfnamefont {M.}~\bibnamefont {Endres}},\ }\bibfield  {title} {\bibinfo {title} {Erasure cooling, control, and hyperentanglement of motion in optical tweezers},\ }\href {https://doi.org/10.1126/science.adn2618} {\bibfield  {journal} {\bibinfo  {journal} {Science}\ }\textbf {\bibinfo {volume} {388}},\ \bibinfo {pages} {845} (\bibinfo {year} {2025})}\BibitemShut {NoStop}%
\bibitem [{\citenamefont {Madjarov}(2021)}]{Madjarovthesis}%
  \BibitemOpen
  \bibfield  {author} {\bibinfo {author} {\bibfnamefont {I.~S.}\ \bibnamefont {Madjarov}},\ }\bibfield  {title} {\bibinfo {title} {Ph.{D}. thesis},\ }\href {https://resolver.caltech.edu/CaltechTHESIS:01292021-001639979} {\bibfield  {journal} {\bibinfo  {journal} {California Institute of Technology}\ } (\bibinfo {year} {2021})}\BibitemShut {NoStop}%
\bibitem [{\citenamefont {Tsai}\ \emph {et~al.}(2025)\citenamefont {Tsai}, \citenamefont {Sun}, \citenamefont {Shaw}, \citenamefont {Finkelstein},\ and\ \citenamefont {Endres}}]{Tsai2025}%
  \BibitemOpen
  \bibfield  {author} {\bibinfo {author} {\bibfnamefont {R.~B.-S.}\ \bibnamefont {Tsai}}, \bibinfo {author} {\bibfnamefont {X.}~\bibnamefont {Sun}}, \bibinfo {author} {\bibfnamefont {A.~L.}\ \bibnamefont {Shaw}}, \bibinfo {author} {\bibfnamefont {R.}~\bibnamefont {Finkelstein}},\ and\ \bibinfo {author} {\bibfnamefont {M.}~\bibnamefont {Endres}},\ }\bibfield  {title} {\bibinfo {title} {Benchmarking and fidelity response theory of high-fidelity {Rydberg} entangling gates},\ }\href {https://doi.org/10.1103/PRXQuantum.6.010331} {\bibfield  {journal} {\bibinfo  {journal} {PRX Quantum}\ }\textbf {\bibinfo {volume} {6}},\ \bibinfo {pages} {010331} (\bibinfo {year} {2025})}\BibitemShut {NoStop}%
\bibitem [{\citenamefont {Finkelstein}\ \emph {et~al.}(2024)\citenamefont {Finkelstein}, \citenamefont {Tsai}, \citenamefont {Sun}, \citenamefont {Scholl}, \citenamefont {Direkci}, \citenamefont {Gefen}, \citenamefont {Choi}, \citenamefont {Shaw},\ and\ \citenamefont {Endres}}]{Finkelstein2024}%
  \BibitemOpen
  \bibfield  {author} {\bibinfo {author} {\bibfnamefont {R.}~\bibnamefont {Finkelstein}}, \bibinfo {author} {\bibfnamefont {R.~B.-S.}\ \bibnamefont {Tsai}}, \bibinfo {author} {\bibfnamefont {X.}~\bibnamefont {Sun}}, \bibinfo {author} {\bibfnamefont {P.}~\bibnamefont {Scholl}}, \bibinfo {author} {\bibfnamefont {S.}~\bibnamefont {Direkci}}, \bibinfo {author} {\bibfnamefont {T.}~\bibnamefont {Gefen}}, \bibinfo {author} {\bibfnamefont {J.}~\bibnamefont {Choi}}, \bibinfo {author} {\bibfnamefont {A.~L.}\ \bibnamefont {Shaw}},\ and\ \bibinfo {author} {\bibfnamefont {M.}~\bibnamefont {Endres}},\ }\bibfield  {title} {\bibinfo {title} {Universal quantum operations and ancilla-based read-out for tweezer clocks},\ }\href {https://doi.org/10.1038/s41586-024-08005-8} {\bibfield  {journal} {\bibinfo  {journal} {Nature}\ }\textbf {\bibinfo {volume} {634}},\ \bibinfo {pages} {321} (\bibinfo {year} {2024})}\BibitemShut {NoStop}%
\bibitem [{\citenamefont {Nielsen}\ and\ \citenamefont {Chuang}(2010)}]{Nielsen2010}%
  \BibitemOpen
  \bibfield  {author} {\bibinfo {author} {\bibfnamefont {M.~A.}\ \bibnamefont {Nielsen}}\ and\ \bibinfo {author} {\bibfnamefont {I.~L.}\ \bibnamefont {Chuang}},\ }\href {https://doi.org/10.1017/CBO9780511976667} {\emph {\bibinfo {title} {Quantum Computation and Quantum Information}}}\ (\bibinfo  {publisher} {Cambridge University Press},\ \bibinfo {year} {2010})\BibitemShut {NoStop}%
\bibitem [{\citenamefont {Monroe}\ \emph {et~al.}(1995)\citenamefont {Monroe}, \citenamefont {Meekhof}, \citenamefont {King}, \citenamefont {Jefferts}, \citenamefont {Itano}, \citenamefont {Wineland},\ and\ \citenamefont {Gould}}]{Monroe1995}%
  \BibitemOpen
  \bibfield  {author} {\bibinfo {author} {\bibfnamefont {C.}~\bibnamefont {Monroe}}, \bibinfo {author} {\bibfnamefont {D.~M.}\ \bibnamefont {Meekhof}}, \bibinfo {author} {\bibfnamefont {B.~E.}\ \bibnamefont {King}}, \bibinfo {author} {\bibfnamefont {S.~R.}\ \bibnamefont {Jefferts}}, \bibinfo {author} {\bibfnamefont {W.~M.}\ \bibnamefont {Itano}}, \bibinfo {author} {\bibfnamefont {D.~J.}\ \bibnamefont {Wineland}},\ and\ \bibinfo {author} {\bibfnamefont {P.}~\bibnamefont {Gould}},\ }\bibfield  {title} {\bibinfo {title} {Resolved-sideband {Raman} cooling of a bound atom to the {3D} zero-point energy},\ }\href {https://doi.org/10.1103/PhysRevLett.75.4011} {\bibfield  {journal} {\bibinfo  {journal} {Phys. Rev. Lett.}\ }\textbf {\bibinfo {volume} {75}},\ \bibinfo {pages} {4011} (\bibinfo {year} {1995})}\BibitemShut {NoStop}%
\bibitem [{\citenamefont {Wineland}\ \emph {et~al.}(1998)\citenamefont {Wineland}, \citenamefont {Monroe}, \citenamefont {Itano}, \citenamefont {Leibfried}, \citenamefont {King},\ and\ \citenamefont {Meekhof}}]{Wineland1998}%
  \BibitemOpen
  \bibfield  {author} {\bibinfo {author} {\bibfnamefont {D.}~\bibnamefont {Wineland}}, \bibinfo {author} {\bibfnamefont {C.}~\bibnamefont {Monroe}}, \bibinfo {author} {\bibfnamefont {W.}~\bibnamefont {Itano}}, \bibinfo {author} {\bibfnamefont {D.}~\bibnamefont {Leibfried}}, \bibinfo {author} {\bibfnamefont {B.}~\bibnamefont {King}},\ and\ \bibinfo {author} {\bibfnamefont {D.}~\bibnamefont {Meekhof}},\ }\bibfield  {title} {\bibinfo {title} {Experimental issues in coherent quantum-state manipulation of trapped atomic ions},\ }\href {https://doi.org/10.6028/jres.103.019} {\bibfield  {journal} {\bibinfo  {journal} {Journal of Research of the National Institute of Standards and Technology}\ }\textbf {\bibinfo {volume} {103}},\ \bibinfo {pages} {259} (\bibinfo {year} {1998})}\BibitemShut {NoStop}%
\end{thebibliography}%

\end{document}


\title{Gate-based Readout and Cooling of Neutral Atoms: Supplemental Material}

\author{Richard Bing-Shiun Tsai}
\altaffiliation{These authors contributed equally to this work.}
\author{Lewis R. B. Picard}
\altaffiliation{These authors contributed equally to this work.}
\author{Xiangkai Sun}
\altaffiliation{These authors contributed equally to this work.}
\author{Yuan Le}
\author{Kon H. Leung}
\author{Manuel Endres}
\email{mendres@caltech.edu}
\affiliation{\Caltech}


\clearpage

\setcounter{equation}{0}
\setcounter{figure}{0}
\setcounter{table}{0}
\setcounter{section}{0}
\setcounter{page}{1}

\renewcommand{\theequation}{S\arabic{equation}}
\renewcommand{\thefigure}{S\arabic{figure}}
\renewcommand{\thetable}{S\Roman{table}}
\renewcommand{\thesection}{\Roman{section}}
\renewcommand{\bibnumfmt}[1]{[S#1]}
\renewcommand{\citenumfont}[1]{S#1}


\onecolumngrid
\begin{center}
    \textbf{\large Gate-based Readout and Cooling of Neutral Atoms: Supplemental Material} \\
    \vspace{10pt}
    Richard Bing-Shiun Tsai,\textsuperscript{1,*} Lewis R. B. Picard,\textsuperscript{1,*} Xiangkai Sun,\textsuperscript{1,*} \\Yuan Le,\textsuperscript{1} Kon H. Leung,\textsuperscript{1} and Manuel Endres\textsuperscript{1,$\dagger$}\\
    \vspace{4pt}
    \textit{\small \textsuperscript{1}California Institute of Technology, Pasadena, CA 91125, USA} \\
    \textsuperscript{*} These authors contributed equally to this work. \\
    \textsuperscript{$\dagger$} mendres@caltech.edu
    \vspace{2pt}
\end{center}

\section{Experimental setup and motional \textit{omg}-architecture}\label{Appendix:Setup}
Our experimental setup has been detailed in previous works~\cite{Scholl2023A, Shaw2024}. Relevant electronic states are addressed with global beams, as shown in the simplified setup diagram [Fig.1(a,b)]. In this work, we consider a motional \textit{omg}-architecture~\cite{Lis2023A,Ma2022_Yb,Chen2022}, specific to $^{88}\text{Sr}$ atoms without nuclear spin. The optical qubit is defined on the \ground($\ket{\downarrow}$) $\leftrightarrow$ \clock($\ket{\uparrow}$) transition. Using the lowest two quantized motional states $\{\ket{0_m}, \ket{1_m} \}$ of the atoms in tweezers, the metastable-state motional qubit is defined on $\ket{\uparrow, 1_{m}}$ and $\ket{\uparrow, 0_{m}}$, whereas the ground-state motional qubit can be defined on $\ket{\downarrow, 1_{m}}$ and $\ket{\downarrow, 0_{m}}$. The latter is not explicitly shown in the level diagram since we are not using it in this work. Single-site tweezer control with an acousto-optic deflector (AOD) allows us to implement local $Z$~gates~\cite{Shaw2024}. Transitions between different motional states rely on sideband pulses after deterministic motional state initialization with erasure-cooling~\cite{Shaw2025_erasure_cooling}.

Ancilla signals from fast imaging shown in Fig.2 and Fig.3 are derived from the signals collected on the EMCCD camera. We first apply a binarization on the electron signals to obtain either 0 or 1 on each pixel~\cite{Scholl2023A}. From this, we obtain a Gaussian-weighted average signal over the pixel area, assigned to each tweezer, and quote this as the signal. Assuming the point spread function of a given tweezer site does not change over the course of data-taking, this signal scales monotonously with the number of photons collected. The aggregated ancilla signal is simply the sum of ancilla signals from considered detection rounds.

\section{Data analysis on low atomic temperature} \label{Appendix:Data_analysis}
To extract the atomic temperature from sideband spectroscopy spectra in Fig.2(e), extra care is warranted as the cooling sideband peak (left peaks) is barely resolvable due to the low atomic temperature after erasure cooling~\cite{Shaw2025_erasure_cooling}. We first perform a standard least-squares fit of the heating sideband peak (right peaks) to a Gaussian. From the fit, we obtain the peak height, peak location, and the width of the heating sideband.

To further extract the cooling sideband peak height and its associated uncertainty, we perform maximum-likelihood estimation with two free parameters: the offset and the peak height. By the symmetry of sideband features, we constrain the width and position by taking the width and center position (with opposite sign) obtained from the fitting of the heating sideband peak.

With the peak geometry constrained, we construct a weighted least-squares statistic, $\chi^2(a_1, d)$, dependent on the peak amplitude $a_1$ and the background offset $d$. To rigorously account for the correlation between the background level and the weak signal amplitude, we calculate the profile likelihood for $a_1$ by treating $d$ as a nuisance parameter; for each stepped value of $a_1$, $d$ was optimized to minimize $\chi^2$. The best-fit value for $a_1$ was identified at the global minimum $\chi^2_{\text{min}}$, and the $1\sigma$ confidence interval was determined by the boundary values satisfying $\Delta \chi^2(a_1) = \chi^2(a_1) - \chi^2_{\text{min}} \leq 1$. The atomic temperature (average motional occupation number) is further computed with the standard conversion formula~\cite{Madjarovthesis}, assuming a thermal distribution over motional states, and propagating the error accordingly. 

\section{Limitation of shelving into motional states}\label{Appendix:limitation_shelving}
One of the major limitations on coherence-preserving ancilla-based atom loss detection [Fig.3] is the motional shelving efficiency. Discussed and identified in our previous work~\cite{Shaw2025_erasure_cooling}, the two major technical limitations are the laser frequency noise and trap frequency inhomogeneity and fluctuations. Here, we provide a more quantitative perspective to discuss how future technical upgrades will improve the shelving efficiency, based on the fidelity response formalism in our previous work~\cite{Tsai2025}. Our goal is to obtain the noiseless Hamiltonian for blue sideband pulse, the corresponding noise operators, and the associated fidelity response functions.

We start with the atom-laser interaction Hamiltonian with fixed laser intensity and frequency in the lab frame, under the dipole approximation, with two electronic states (ground $\ket{g}$ and excited $\ket{e}$) and quantized motional states (associated with the annihilation/creation operators $\hat{a}$, $\hat{a}^\dagger$) without noise:
\begin{equation}
\hat{H}_{\text{lab}} = \hbar \omega_0 \frac{\hat{\sigma}_z}{2} + \hbar \omega_{t,0} \left(\hat{a}^\dagger \hat{a} + \tfrac{1}{2}\right) + \frac{\hbar \Omega_0}{2} \left[e^{i \left(k \hat{x} - \omega_L t \right)} \hat{\sigma}_+ + \text{h.c.} \right].
\end{equation}
where $\omega_0$ is the atom transition (angular) frequency, $\omega_{t,0}$ is the tweezer trap frequency, $\Omega_0$ is the fixed carrier Rabi frequency, $\hat{\sigma}_z = \ket{e}\bra{e} - \ket{g}\bra{g}$, $\hat{\sigma}_+ = \ket{e}\bra{g}$. The laser phase factor $\left(k \hat{x} - \omega_L t\right)$ consists of the fixed laser frequency $\omega_L$ and the position dependence $k \hat{x}$, where $\hat{x} = \sqrt{\frac{\hbar}{2m \omega_{t,0}}}\left(\hat{a}+\hat{a}^\dagger \right)$ and $k=\frac{\omega_L}{c}$.

We then add some classical stochastic noise to the Hamiltonian, in particular the trap frequency noise $\delta\omega_t(t)$, laser phase noise $\phi(t)$, and laser amplitude noise $\delta\Omega(t)$. For simplicity, we also drop the offset $\tfrac{1}{2}\hbar\omega_{t,0}$, which only adds as global phase and clearly doesn't enter into the dynamics.
\begin{equation}
    \hat{H}_{\text{lab}} = \hbar \omega_0 \frac{\hat{\sigma}_z}{2} + \hbar (\omega_{t,0}+\delta\omega_t(t)) \hat{a}^\dagger \hat{a} + \frac{\hbar (\Omega_0+\delta\Omega(t))}{2} \left[e^{i \left(k \hat{x} - \omega_L t + \phi(t)\right)} \hat{\sigma}_+ + \text{h.c.} \right].
\end{equation}

We define the Lamb-Dicke parameter $\eta = k\sqrt{\frac{\hbar}{2m \omega_{t,0}}}\approx 0.36$ (it is judicious to see $k$ as a fixed parameter despite the frequency noise, since we should be comparing the frequency noise, on the order of $\sim100$~Hz to the optical transition frequency on the order of 400~THz) the laser detuning $\delta=\omega_L-\omega_0$. We move to the interaction picture with the reference Hamiltonian,

\begin{equation}
\begin{split}
    \hat{H}_{\text{ref}}=&\hbar\omega_{t,0}\hat{a}^\dagger \hat{a}+\frac{\hbar\omega_0}{2} \hat{\sigma}_z,
\end{split}
\end{equation}

such that the noisy atom-laser interaction Hamiltonian transforms as:
\begin{equation}
\hat{H}_{\text{I, noisy}} =  \hbar \delta\omega_t(t) \hat{a}^\dagger \hat{a} + \frac{\hbar (\Omega_0+\delta\Omega(t))}{2} \left[e^{i \eta\left(\hat{a}e^{-i\omega_{t,0}t}+\hat{a}^\dagger e^{i\omega_{t,0}t}\right) - i\delta t + i\phi(t)} \hat{\sigma}_+ + \text{h.c.} \right].
\end{equation}

We expand to the first order in $\eta$.
\begin{equation}
\hat{H}_{\text{I, noisy}} \approx  \hbar \delta\omega_t(t) \hat{a}^\dagger \hat{a} + \frac{\hbar (\Omega_0+\delta\Omega(t))}{2} \{\left[1+i\eta\left(\hat{a}e^{-i\omega_{t,0}t}+\hat{a}^\dagger e^{i\omega_{t,0}t}\right) \right]e^{- i\delta t + i\phi(t)} \hat{\sigma}_+ + \text{h.c.} \}.
\end{equation}

We are particularly interested in the case of $\delta = \omega_{t,0}$ (blue sideband). We substitute this in, perform a frame change with $\hat{U}(t)=e^{i\phi(t)/2}\ket{g}\bra{g}+e^{-i\phi(t)/2}\ket{e}\bra{e}$, and obtain:

\begin{equation}
\hat{H}'_{\text{noisy}} \approx  \hbar \delta\omega_t(t) \hat{a}^\dagger \hat{a} +\frac{\hbar\dot{\phi}(t)}{2}\hat{\sigma}_z+ \frac{\hbar (\Omega_0+\delta\Omega(t))}{2} \{\left[1+i\eta\left(\hat{a}e^{-i\omega_{t,0}t}+\hat{a}^\dagger e^{i\omega_{t,0}t}\right) \right]e^{- i\omega_{t,0} t} \hat{\sigma}_+ + \text{h.c.} \}.
\end{equation}
After the standard rotating-wave approximation
\begin{equation}
\hat{H}'_{\text{noisy}} \approx  \hbar \delta\omega_t(t) \hat{a}^\dagger \hat{a} +\frac{\hbar\dot{\phi}(t)}{2}\hat{\sigma}_z+ \frac{\hbar (\Omega_0+\delta\Omega(t))}{2} \{i\eta\hat{a}^\dagger \hat{\sigma}_+ + \text{h.c.} \}.
\end{equation}

It is instructive to see that in the noiseless scenario, this reduces to a simple two-level Hamiltonian describing the coupling between an isolated pair of states within an infinite motional ladder.
\begin{equation}
    \hat{H}'_{\text{noiseless}} \approx   \frac{\hbar \Omega_0}{2} \{i\eta\hat{a}^\dagger \hat{\sigma}_+ + \text{h.c.} \}.
\end{equation}

Assuming we start with the state $\ket{g,0_m}$, the only state that $\hat{H}'_{\text{noiseless}}$ couples to is $\ket{e,1_m}$. Let us now restrict the states that we consider to $\{\ket{g'}=\ket{g,0_m}, \ket{e'}=\ket{e,1_m}\}$ and rewrite $\hat{H}'_{\text{noisy}}$ in this restricted space with the usual Pauli operators $\hat{\sigma}'_z=\ket{e'}\bra{e'}-\ket{g'}\bra{g'}$  and $\hat{\sigma}'_y = i\ket{e'}\bra{g'}-i\ket{g'}\bra{e'}$

\begin{equation}
\begin{split}
    \hat{H}'_{\text{noisy}} &\approx  \hbar \delta\omega_t(t) \ket{e'}\bra{e'} +\frac{\hbar\dot{\phi}(t)}{2}\left(\ket{e'}\bra{e'}-\ket{g'}\bra{g'}\right)+ \frac{\hbar (\Omega_0+\delta\Omega(t))}{2} \left(i\eta\ket{e'}\bra{g'} + \text{h.c.} \right)\\
    &=  \hbar \delta\omega_t(t) \ket{e'}\bra{e'} +\frac{\hbar\dot{\phi}(t)}{2}\hat{\sigma}'_z+ \frac{\hbar (\Omega_0+\delta\Omega(t))}{2} \eta\hat{\sigma}'_y
\end{split}
\end{equation}

Now we identify the noise operators and the associated power spectral density, based on the fidelity response theory~\cite{Tsai2025}:
\begin{equation}
    \hat{H}(t) = \hat{H}_0(t) + \sum_j h_j(t) \hat{O}_j(t),
\end{equation}
where $\hat{H}_0(t)$ is the target, programmed time-dependent drive, and each $h_j(t) \hat{O}_j(t)$ is an independent noise term on the Hamiltonian, with $h_j(t)$ described by a power spectral density $S_j(f)$, and $\hat{O}_j(t)$ a noise operator that only depends on the noise type and is independent of the noise spectrum. Then we evaluate the fidelity response for each noise term in the Heisenberg picture with the unitary transformation $\hat{U}(t) = e^{-\frac{i\eta\Omega_0}{2}\hat{\sigma}'_y t}$. We also take $T=T_{\pi}=\frac{\pi}{\eta\Omega_0}$, corresponding to a $\pi$-pulse on the blue sideband.

Let us take trap frequency noise as an example: $\hbar \delta\omega_t(t) \ket{e'}\bra{e'}$

\begin{equation}
\begin{split}
    h_t(t) &= \delta\omega_t(t)\\
    \hat{O}_t &= \hbar \ket{e'}\bra{e'}\\
\end{split}  
\end{equation}

We note that the laser phase (frequency) noise actually has the same noise operator with a different offset: $\frac{\hbar}{2}\dot{\phi}(t)\hat{\sigma}'_z$
\begin{equation}
\begin{split}
    h_{\nu}(t) &= \dot{\phi}(t)\\
    \hat{O}_{\nu} &= \frac{\hbar}{2}\hat{\sigma}'_z
\end{split}  
\end{equation}

We evaluate the fidelity response associated to the trap frequency noise with the closed analytical form.

\begin{equation}
\begin{split}
    I_t(f) &= \int_0^T dt \int_0^T d\tau \cos (2\pi f(t-\tau)) \langle \hat{O}_{t,H}(t) \hat{O}_{t,H}(\tau) \rangle_c\\
    &=\int_0^T dt \int_0^T d\tau \cos (2\pi f(t-\tau)) \frac{1}{4}\sin(\eta\Omega_0t) \sin(\eta\Omega_0\tau)\\
    &= {\left[ \frac{\eta\Omega_0\cos(\pi fT_{\pi})}{\eta^2\Omega_0^2-4\pi^2f^2}\right]}^2 (\text{for}f>0, f\neq\frac{\eta\Omega_0}{2\pi})
\end{split}  
\end{equation}
The same fidelity response function applies to laser frequency noise $I_{\nu}(f)$. Evaluating $I_{\nu}(f)$ and $I_t(f)$ with our carrier Rabi frequency $\Omega_0 = 2\pi\times2$~kHz and $\eta$ gives the values for $I_{\nu}$. The contribution from our clock laser frequency noise (see ED Fig. 2f of Ref.~\cite{Finkelstein2024}) is evaluated to be $1.7\times10^{-2}$. For trap frequency noise, most of the contributions come from the DC trap intensity noise and disuniformity across the array. With ${\sim}0.5\%$ variation on the trap frequency of 35~kHz, the infidelity contribution is calculated to be $5.6\times10^{-2}$. These two contributions account for the infidelity that we see on our experimental blue sideband transfer fidelity of ${\sim}0.92$ over the course of data taking for this work.

\section{Algorithmic cooling sequence and data analysis}
\begin{figure*}[ht!]
    \centering
    \includegraphics[width=\textwidth]{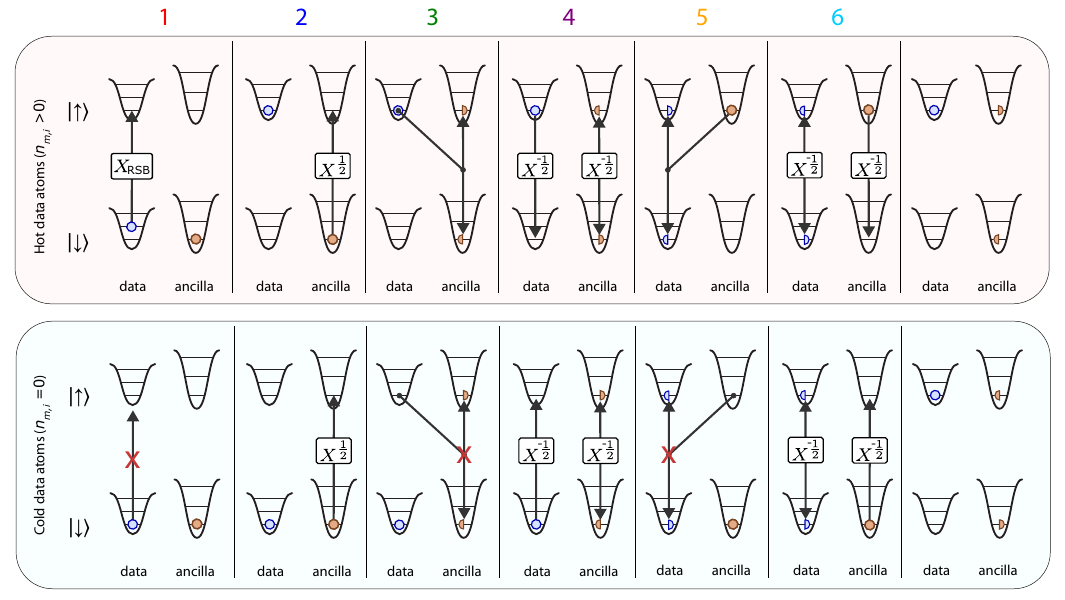}
    \caption{\textbf{Algorithmic cooling pulse sequence.}
    Illustration of the algorithmic cooling circuit depicted in Fig.4(a) explicitly showing the electronic and motional states of data and ancilla atoms at each step. The top row shows the sequence when the data atoms are initially in a motional excited state of the tweezer, with an atom initially in $n_m = 1$ used for illustrative purposes. The bottom row shows the sequence when the atom is initially in the motional ground state. Red crosses are used to illustrate the pulses that have no effect on the atoms in this case. The ancilla atoms are assumed to be in the motional ground state, although this is not strictly required for the success of the cooling cycle. In this figure specifically, solid colors of atoms are only for distinguishing between data and ancilla atoms.
    } 
    \vspace{-0.3cm}
    \label{fig:AC_motion}
\end{figure*}

In Fig.~\ref{fig:AC_motion}, we explicitly show the motional and electronic states of data and ancilla atoms during one round of algorithmic cooling. The scheme relies on both the data and ancilla atoms initially being in the same electronic ground state $\ket{\downarrow}$. The motional state of the data atoms is then coupled to their electronic state using a red sideband pulse on the clock transition, which transfers atoms from $\ket{\downarrow,n_{m,i}}$ to $\ket{\uparrow,n_{m,i}-1}$ only if they are not in the motional ground state to begin with. If they are in the motional ground state then the pulse is off-resonant and the atom remains in $\ket{\downarrow,0}$. At the start of the sequence, the tweezers holding the data atoms are ramped to a different depth than the ancilla tweezers, allowing the red-sideband pulse to be selectively applied to the data atoms due to the strong separation of the trap frequencies between data and ancilla. This difference of trap depths is also what allows us to selectively initialize data atoms at a range of initial motional temperatures in Fig.4(c,d) through laser-induced heating on the narrow intercombination line. After the initial red sideband pulse, all subsequent single-qubit rotations are applied on the clock \textit{carrier transition}, coupling $\ket{\downarrow,n_m} \leftrightarrow \ket{\uparrow,n_m}$. We represent the ancilla atoms as being in the motional ground state throughout the sequence because we initialize them using erasure cooling to an average motional state of $\bar{n} = 0.002^{+5}_{-2}$. Using such motionally cold ancilla affects the overall performance of algorithmic cooling because it improves the fidelity of the single-qubit gates on the clock transition~\cite{Finkelstein2024}. However, we emphasize that this is not a strict requirement for the success of the cooling sequence. In fact, as long as the electronic state of the ancilla can be reliably initialized and single- and two-qubit gates can be executed, motional cooling of the data atoms can be achieved regardless of the temperature of the ancilla.

Following the sideband pulse, a selective $X^{1/2}$ gate is applied to the ancilla atoms using programmable local tweezer movement \cite{Shaw2024}. A controlled-$Z$ Rydberg gate~\cite{Tsai2025} is then applied, which flips the phase of the ancilla qubits only if the data were excited in the first step. Another $X^{-1/2}$ gate then drives the ancilla to either $\ket{\downarrow}$ or $\ket{\uparrow}$, and prepares the data qubits in the $X$-basis. A second controlled-$Z$ gate flips the phase of the data atoms depending on the state of the ancilla. As a result, the final $X^{-1/2}$ gate deterministically leaves the data atom in the $\ket{\uparrow}$ state. Steps 2-6 operate much like the last two CNOTs in the standard compilation of a SWAP gate \cite{Nielsen2010}, coherently transferring the electronic state (in this case effectively a mixed state) information from the data to the ancilla qubit.

\begin{figure*}[ht!]
    \centering
    \includegraphics[width=\textwidth]{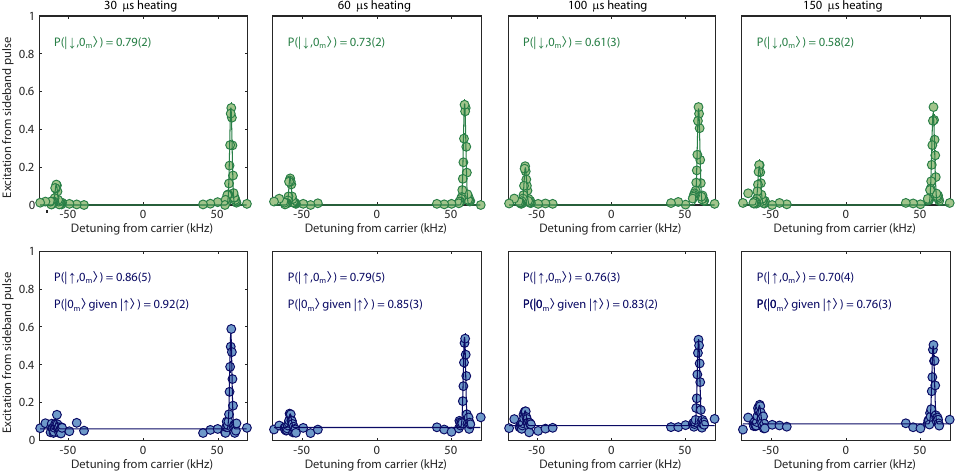}
    \caption{\textbf{Sideband spectra before (upper) and after (lower) algorithmic cooling for various initial temperatures.}
    (a-d) Baseline data atom sideband spectra for different durations of site-selective laser-induced Sisyphus heating. After heating, atoms remain initially entirely in the electronic ground state $\ket{\downarrow}$. Sideband spectra are fit to a sum of two Gaussians with the same width, and the motional ground state population and its uncertainty are determined using the heights of the heating and cooling sidebands. (e-h) Sideband spectra taken after one round of algorithmic cooling. Atoms are nominally prepared in the clock state $\ket{\uparrow}$, but some residual population remains in $\ket{\downarrow}$ due to gate imperfections. The spectra are thus fit to the same double-Gaussian function as the baseline, with an additional global offset, which reflects the fraction of the population in $\ket{\downarrow}$. Based on the fit to the sideband spectra, we report the fraction of the total population which is in both the motional ground state and the correct electronic state, as well as the motional ground state fraction of the atoms, conditioned on them being in the correct electronic state.
    } 
    \vspace{-0.3cm}
    \label{fig:AC_sideband}
\end{figure*}

We evaluate the performance of the algorithmic cooling sequence using sideband thermometry (spectroscopy) \cite{Monroe1995} on the clock transition. Fig.~\ref{fig:AC_sideband} shows the sideband spectra from which the motional ground state fraction data in Fig.4 is extracted. For each duration of laser-induced heating applied to the data atoms, we measure the data atom temperature before algorithmic cooling. The spectra for these initial temperature measurements are shown in the top row of the figure. The sidebands are fit to a sum of two Gaussian peaks with the same width parameter but independent heights. Since the population begins entirely in the $\ket{\downarrow}$ state, the global offset of the excitation is fixed at zero for this fit. We then take sideband spectra after one round of algorithmic cooling, which are shown in the bottom row. In this case the atoms begin in the clock state, and the sideband pulse drives them back to the ground state. We apply a pushout pulse before the sideband thermometry to remove any remaining population in the electronic ground state, which would otherwise interfere with the sideband thermometry signal. We then apply another pushout pulse after the thermometry, which allows us to measure the sideband spectrum via population loss. The spectra are fit to the same function as the top row, with the addition of a global offset parameter. For all spectra the motional ground state fraction is extracted simply from the ratio of the heights of the fit peaks, and the incorrect electronic state population after algorithmic cooling is extracted from the offset of the fit.

Upon removal of one motional quantum, the original Boltzmann distribution of motional state occupancy no longer strictly holds, which can affect the interpretation of the sideband thermometry results. Here, we estimate the scale of any error in the motional ground state population estimates incurred by treating the atoms as if they are in thermal equilibrium after algorithmic cooling. Let's assume that the initial Boltzmann distribution is characterized by the ratio $q=\frac{E^{\text{init}}_{\text{red}}}{E^{\text{init}}_{\text{blue}}}=\frac{\bar{n}_\text{init}}{\bar{n}_\text{init}+1}$. The first few motional state populations are $p_0=1-q$, $p_1=q-q^2$, $p_2=q^2-q^3$, etc. If the system undergoes an ideal process of one motional quantum removal (as algorithmic cooling is designed to do), the motional ground state occupancy is $p_0^\prime=1-q^2$ with the excited state occupancies being $p_1^\prime=q^2-q^3$, $p_2^\prime=q^3-q^4$. We denote as $t_{n_i, n_i+1}$ the transfer probability for a thermometry pulse on the sideband bridging $\ket{n=n_i}$ and $\ket{n'=n_i+1}$. In a typical sideband thermometry experiment, we maximize the population transfer on the heating sideband, which roughly equates to setting $t_{0, 1}=1$. This allows us to write the sideband excitation ratio as $r=\frac{p_1^\prime t_{0,1}+p_2^\prime t_{1,2}+\mathcal{O}(q^4)}{p_0^\prime t_{0,1}+p_1^\prime t_{1,2}+\mathcal{O}(q^3)}=\frac{q^2-q^3(1-t_{1,2})+\mathcal{O}(q^4)}{1-q^2(1-t_{1,2})+\mathcal{O}(q^3)}=q^2-q^3(1-t_{1,2})+\mathcal{O}(q^4)$. Here, if we estimate the motional ground state occupancy $p_{0,\text{est}}^\prime=1-r=1-q^2+q^3(1-t_{1,2})+\mathcal{O}(q^4)$, there would be an additional error term $q^3(1-t_{1,2})+\mathcal{O}(q^4)=r^{\frac{3}{2}}(1-t_{1,2})+\mathcal{O}(r^2)$, resulting in a slight overestimation of the ground state population. In our experimental settings $1-t_{1,2}=0.24$, which follows from the fact that we set $t_{0,1} = 1$~\cite{Wineland1998}. Examining the data in Fig.~\ref{fig:AC_sideband}, our measured sideband excitation ratio $r_{\text{est}}$ ranges from $0.08$ to $0.24$, which suggests a correction term of $0.005$ to $0.03$, which is smaller than the experimental uncertainty in our population measurements. In reality, the ground state populations we measure are also smaller than one would achieve if the algorithmic cooling population transfer were perfect, indicating that the motional states of the atoms are likely at some intermediate point between a fully thermal distribution and the ideal non-thermal distribution from one round of cooling. We thus report the estimated quantity $p_{0,\text{est}}^\prime=1-r_{\text{est}}$ without further corrections in this work.

\bibliography{library-endreslab}